\documentclass[letter,iop]{emulateapj}
\usepackage{amssymb}
\usepackage{natbib}
\usepackage{graphicx}

\def\ltsima{$\; \buildrel < \over \sim \;$}
\def\simlt{\lower.5ex\hbox{\ltsima}}
\def\gtsima{$\; \buildrel > \over \sim \;$}
\def\simgt{\lower.5ex\hbox{\gtsima}}
\def\simless{\mathbin{\lower 3pt\hbox
   {$\rlap{\raise 5pt\hbox{$\char'074$}}\mathchar"7218$}}}   
\def\simgreat{\mathbin{\lower 3pt\hbox
   {$\rlap{\raise 5pt\hbox{$\char'076$}}\mathchar"7218$}}}   

\def\rosat{{\it ROSAT}}
\def\xmm{{\it XMM-Newton}}
\def\chandra{{\it Chandra}}

\newcommand{\asca}{{\it ASCA}}

\long\def\symbolfootnote[#1]#2{\begingroup%
\def\thefootnote{\fnsymbol{footnote}}\footnote[#1]{#2}\endgroup} 

\shorttitle{Physical Properties of ACT SZE Clusters}
\shortauthors{Menanteau et al.}
\submitted{submitted 2010 June 25; accepted 2010 September 8}

\begin{document}

\title{The Atacama Cosmology Telescope: Physical Properties and Purity
  of a Galaxy Cluster Sample Selected via the Sunyaev-Zel'dovich Effect}

\author{
Felipe Menanteau\altaffilmark{1,\dag,\ddag}, 
Jorge~Gonz\'alez\altaffilmark{2},
Jean-Baptiste~Juin\altaffilmark{2},
Tobias~A.~Marriage\altaffilmark{3},
Erik~D.~Reese\altaffilmark{4},
Viviana~Acquaviva\altaffilmark{1,3},
Paula~Aguirre\altaffilmark{2},
John~William~Appel\altaffilmark{5},
Andrew~J.~Baker\altaffilmark{1},
L.~Felipe~Barrientos\altaffilmark{2},
Elia~S.~Battistelli\altaffilmark{6},
J.~Richard~Bond\altaffilmark{7},
Sudeep~Das\altaffilmark{8},
Amruta J.~Deshpande\altaffilmark{1},
Mark~J.~Devlin\altaffilmark{4},
Simon~Dicker\altaffilmark{4},
Joanna~Dunkley\altaffilmark{9},
Rolando~D\"{u}nner\altaffilmark{2},
Thomas~Essinger-Hileman\altaffilmark{5},
Joseph~W.~Fowler\altaffilmark{5},
Amir~Hajian\altaffilmark{7},
Mark~Halpern\altaffilmark{10},
Matthew~Hasselfield\altaffilmark{10},
Carlos~Hern\'andez-Monteagudo\altaffilmark{11},
Matt~Hilton\altaffilmark{12},
Adam~D.~Hincks\altaffilmark{5},
Ren\'ee~Hlozek\altaffilmark{9},
Kevin~M.~Huffenberger\altaffilmark{13},
John~P. Hughes\altaffilmark{1,\ddag}, 
Leopoldo~Infante\altaffilmark{2},
Kent~D.~Irwin\altaffilmark{14},
Jeff~Klein\altaffilmark{4},
Arthur~Kosowsky\altaffilmark{15},
Yen-Ting~Lin\altaffilmark{16},
Danica~Marsden\altaffilmark{4},
Kavilan~Moodley\altaffilmark{12},
Michael~D.~Niemack\altaffilmark{14},
Michael~R.~Nolta\altaffilmark{7},
Lyman~A.~Page\altaffilmark{5},
Lucas~Parker\altaffilmark{5},
Bruce~Partridge\altaffilmark{17},
Neelima Sehgal\altaffilmark{18},
Jon~Sievers\altaffilmark{7},
David~N.~Spergel\altaffilmark{3},
Suzanne~T.~Staggs\altaffilmark{5},
Daniel~Swetz\altaffilmark{4},
Eric~Switzer\altaffilmark{19},
Robert~Thornton\altaffilmark{20},
Hy~Trac\altaffilmark{21},
Ryan~Warne\altaffilmark{12} and
Ed~Wollack\altaffilmark{22},
}

\altaffiltext{1}{Rutgers University, Department of Physics \& Astronomy, 136 Frelinghuysen Rd, Piscataway, NJ 08854, USA }
\altaffiltext{2}{Departamento de Astronom{\'{i}}a y Astrof{\'{i}}sica,
  Facultad de F{\'{i}}sica, Pontific\'{i}a Universidad Cat\'{o}lica de
  Chile, Casilla 306, Santiago 22, Chile}
\altaffiltext{3}{Department of Astrophysical Sciences, Peyton Hall, Princeton University, Princeton, NJ 08544, USA}
\altaffiltext{4}{University of Pennsylvania, Physics and Astronomy, 209 South 33rd Street, Philadelphia, PA 19104, USA}
\altaffiltext{5}{Joseph Henry Laboratories of Physics, Jadwin Hall, Princeton University, Princeton, NJ, USA 08544}
\altaffiltext{6}{Department of Physics, University of Rome ``La Sapienza'', Piazzale Aldo Moro 5, I-00185 Rome, Italy}
\altaffiltext{7}{Canadian Institute for Theoretical Astrophysics,  University of Toronto, Toronto, ON, Canada M5S 3H8}
\altaffiltext{8}{Berkeley Center for Cosmological Physics, UC Berkeley, 366 Le Conte Hall, Berkeley, CA 94720}
\altaffiltext{9}{Department of Astrophysics, Oxford University, Oxford, UK OX1 3RH}
\altaffiltext{10}{Department of Physics and Astronomy, University of British Columbia, Vancouver, BC, Canada V6T 1Z4}
\altaffiltext{11}{Max Planck Institut f\"ur Astrophysik, Postfach 1317, D-85741 Garching bei M\"unchen, Germany}
\altaffiltext{12}{University of KwaZulu-Natal, Astrophysics \& Cosmology Research Unit, School of Mathematical Sciences, Durban, 4041, South Africa.}
\altaffiltext{13}{Department of Physics, University of Miami, 1320 Campo Sano Avenue, Coral Gables, FL 33124}
\altaffiltext{14}{NIST Quantum Devices Group, 325 Broadway Mailcode 817.03, Boulder, CO, USA 80305}
\altaffiltext{15}{University of Pittsburgh, Physics \& Astronomy Department, 100 Allen Hall, 3941 O'Hara Street, Pittsburgh, PA 15260, USA}
\altaffiltext{16}{Institute for the Physics and Mathematics of the Universe, The University of Tokyo, Kashiwa, Chiba 277-8568, Japan}
\altaffiltext{17}{Department of Physics and Astronomy, Haverford College, Haverford, PA, USA 19041}
\altaffiltext{18}{Kavli Institute for Particle Astrophysics and Cosmology, Stanford University, Stanford, CA 94305, USA}
\altaffiltext{19}{Kavli Institute for Cosmological Physics, Laboratory for Astrophysics and Space Research, 5620 South Ellis Ave.,
Chicago, IL, USA 60637}
\altaffiltext{20}{Department of Physics , West Chester University of Pennsylvania, West Chester, PA, USA 19383}
\altaffiltext{21}{Harvard-Smithsonian Center for Astrophysics, Harvard University, Cambridge, MA 02138}
\altaffiltext{22}{Code 553/665, NASA/Goddard Space Flight Center, Greenbelt, MD, USA 20771}
\altaffiltext{\dag}{Based on observations made with ESO Telescopes at the La Silla Observatories under programme ID 084.A-0619.}
\altaffiltext{\ddag}{Visiting astronomer, Cerro Tololo Inter-American Observatory, National
Optical Astronomy Observatory, which are operated by the Association
of Universities for Research in Astronomy, under contract with the
National Science Foundation.}

\begin{abstract}

We present optical and X-ray properties for the first confirmed galaxy
cluster sample selected by the Sunyaev-Zel'dovich Effect from 148~GHz
maps over 455 square degrees of sky made with the Atacama Cosmology
Telescope.  These maps, coupled with multi-band imaging on
4-meter-class optical telescopes, have yielded a sample of 23 galaxy
clusters with redshifts between 0.118 and 1.066. Of these 23 clusters,
10 are newly discovered. The selection of this sample is approximately
mass limited and essentially independent of redshift.  We provide
optical positions, images, redshifts and X-ray fluxes and luminosities for the
full sample, and X-ray temperatures of an important subset.  The mass
limit of the full sample is around $8.0\times10^{14}\,M_\odot$, with a
number distribution that peaks around a redshift of 0.4.  For the 10
highest significance SZE-selected cluster candidates, all of which are
optically confirmed, the mass threshold is $1\times10^{15}\,M_\odot$
and the redshift range is 0.167 to 1.066.  Archival observations from
\chandra, \xmm, and \rosat\ provide X-ray luminosities and
temperatures that are broadly consistent with this mass threshold.
Our optical follow-up procedure also allowed us to assess the purity
of the ACT cluster sample. Eighty (one hundred) percent of the 148~GHz
candidates with signal-to-noise ratios greater than 5.1 (5.7) are
confirmed as massive clusters.  The reported sample represents one of
the largest SZE-selected sample of massive clusters over all redshifts
within a cosmologically-significant survey volume, which will enable
cosmological studies as well as future studies on the evolution,
morphology, and stellar populations in the most massive clusters in
the Universe.

\end{abstract}


\section{Introduction}

Clusters of galaxies mark the largest virialized structures in the
Universe, and their formation and evolution rates depend on
cosmological parameters and the kinematics of the dark matter. The
cluster mass function constrains the normalization of the matter power
spectrum, typically expressed as $\sigma_8$, the root-mean-square
(rms) mass fluctuation on a scale of 8 $h^{-1}$Mpc. As a result, the
number of clusters as a function of redshift provides a powerful
constraint on both the expansion history of the Universe and the
gravitational growth of structure within it \citep[see][and references
 therein]{Carlstrom-02}.  Both functions are strongly influenced by
the matter content $\Omega_M$ and the equation of state of the
so-called ``dark energy'' that comprises most of the effective energy
density in the Universe. Galaxy clusters also harbor a significant
fraction of the visible baryons in the Universe, in the form of a hot
intracluster medium that leaves an imprint on the Cosmic Microwave
Background (CMB) radiation though the Sunyaev Zel'dovich Effect (SZE)
\citep{SZ72}.  At frequencies below the SZ null frequency around
218~GHz, clusters appear as small ($\la1$mK) decrements in the
temperature of the CMB.

Selecting clusters through the strength of their SZE signal has key
advantages over other methods of selection: (1) it is relatively independent
of redshift, (2) according to simulations, the SZE flux
($Y\equiv \int y\, d\Omega$, where $y$ is the usual Compton
$y$-parameter) should be cleanly related to the cluster mass with low
scatter \citep{Motl-05, Nagai-06, Reid-Spergel-06}, and (3) 
according to simulations the SZE signal 
is more robust to baryonic astrophysics (cooling, AGN
feedback). Thus an SZE-selected sample could offer a reliable
correlation with cluster mass, which is the key cosmological property,
particularly at high redshift. However, optical and X-ray observations
are crucial to confirm the SZE detections and calibrate the $Y$-mass
relation predicted by simulations.

The SZE has been the focus of a considerable number of observational
efforts over the past three decades. Some of the earliest convincing
measurements were made by \cite{Birkinshaw-84} observing at 20~GHz on
the 40-m telescope of the Owens Valley Radio Observatory. Each
individual cluster detection required many hours of telescope time,
typically spread out over several years.  By the late 1990's there
were still only a dozen or so clusters for which reliable detections
had been published \citep[see the review by][]{Birkinshaw-99}.  

Large-area cosmological SZE surveys have only now become a reality,
thanks to the development of mm-band cameras composed of arrays of
order a thousand bolometers located at the focus of
large-diameter telescopes (6 to 10 m) offering diffraction-limited
imaging at arcminute resolution (a factor of five finer than the {\it
  Planck} satellite). Two such experiments, the Atacama Cosmology
Telescope \citep[ACT,][]{Fowler-07} and the South Pole Telescope
\citep[SPT,][]{SPTref} have surveyed hundreds of square degrees of the
southern sky at arcminute resolutions and are reporting the first
cluster detections \citep{Stan09, Hincks-09, Vanderlinde10}.
ACT is conducting a wide-area mm-band survey at arcminute resolution
with $\simeq$30 $\mu$K~s$^{1/2}$ array sensitivity to efficiently search
for massive galaxy clusters using the SZE. The first ACT SZE clusters
detected in the 2008 observation season are reported in
\citet{Hincks-09} and the first analysis of the ACT 148~GHz CMB power
spectrum reaching to arcminute scales is presented in \citet{Fowler-10}.

This paper introduces the first sample of optically-confirmed SZE
clusters from the ACT 2008 148~GHz data \citep[see][for the
recent SPT cluster follow-up study]{High-10} and, for the first time,
characterizes observationally the purity of an SZE survey.
For this purpose we selected candidate galaxy clusters from ACT maps
covering 455 square degrees (a comoving volume of 2.46~Gpc$^3$ at
$z<1.2$) using a rigorous and uniformly controlled procedure (see
Marriage et al. 2010). Subsequently, during the 2009B semester, we
imaged candidates in the bands $g$, $r$, and $i$ on 4-m class
telescopes to confirm the presence of a brightest cluster galaxy (BCG)
and an accompanying red sequence of cluster members.

In two companion papers we describe in detail the SZE cluster
selection from the ACT data (Marriage et~al.\ 2010) and cosmological
constraints from the sample presented here (Sehgal
et~al.\ 2010). Throughout this paper we assume a flat cosmology with
$H_0=100 h$~km~s$^{-1}$~Mpc$^{-1}$, $h=0.7$, and matter density
$\Omega_M=0.3$. The reported magnitudes are always in the AB system.
Cluster masses are quoted as $M_{200}$, which corresponds to
integrating the cluster mass within a spherical volume such that
the density is 200 times the average density of the Universe at the
cluster's redshift.

\section{Observational Setup}

\begin{figure}
\centerline{\includegraphics[width=3.9in]{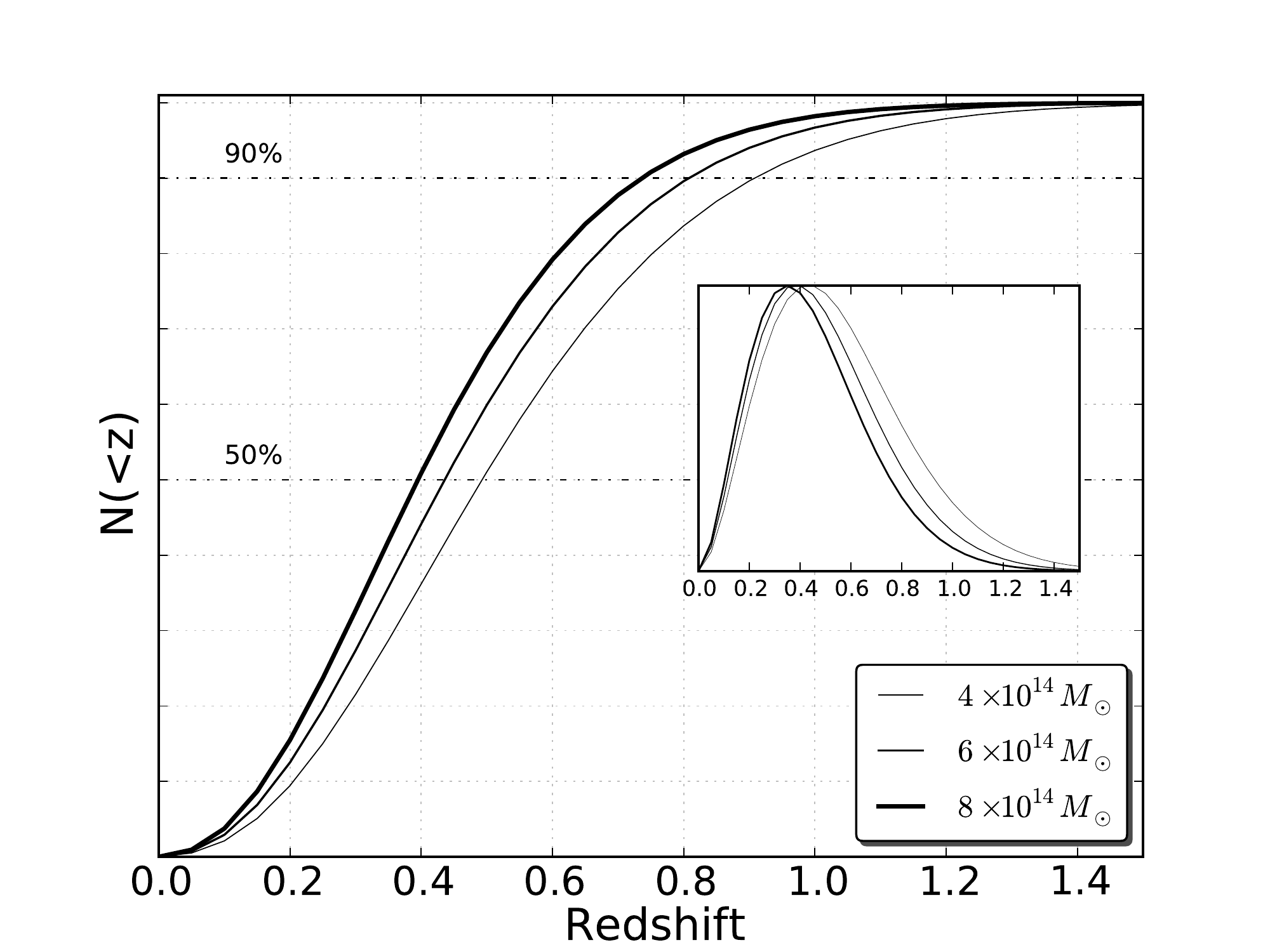}}
\caption{The predicted number of clusters more massive than $4$, $6$ and
  $8\times10^{14}M_{\sun}$ (solid lines) as a function of redshift and
  normalized to unity from the mass function of \cite{Tinker-08}. For
  reference we show the 50\% and 90\% fractional completeness levels as dashed
  lines. The inset shows the normalized differential $dN/dz$
  distribution over the same redshift range and mass limits.}
\label{fig:Nc}
\end{figure}

Both ACT and SPT aim to provide a unique sample of massive clusters of
galaxies, in principle selected by mass, nearly independently of
redshift and over a large area of the sky covering a cosmologically
significant volume.  Utilizing these cluster samples to constrain
cosmology and cluster physics requires empirical characterization of the
SZE selection function.

As the first step in this process we measure the ``purity'' of the ACT
sample, i.e., the fraction of real clusters as a function of the SZE
detection significance, using a signal-to-noise limited sample of
cluster candidates extracted from the 148~GHz maps.
We search for an optical cluster associated with each candidate's
SZE decrement.  This is relatively straightforward,
since for a wide range of mass limits in a $\Lambda$CDM cosmology 90\%
of clusters will lie below $z\approx0.8$ and are therefore accessible
in modest exposures on 4-m class telescopes. To illustrate this point,
in Figure~\ref{fig:Nc} we show the expected number of clusters of
galaxies more massive than $4$, $6$ and $8\times10^{14}M_{\sun}$ as a
function of redshift from the mass function of \cite{Tinker-08}. This
range in mass limit covers the current relevant range for ACT.
A key requirement of our observing plan is efficient multi-band
optical imaging of ACT cluster candidates. Unfortunately, the current
generation of large and wide format CCD mosaic cameras are not ideally
suited for this task as their long read-out times (typically between
$100-200$~s) often become comparable to the required exposure times to
find $L^*$ (the characteristic \citeauthor{Schechter-76} luminosity)
ellipticals at $z\simeq1$, resulting in significant time overheads. For
our observations, we decided instead to trade the large field-of-view
for smaller but faster readout CCD cameras (around 10~s) that allow
us to move rapidly from target to target and secure multi-band imaging
for the central $5'$ of the clusters, sufficient for the
identification process because of the accurate astrometry of the
mm-band data (see below).

The number of nights required for this measurement is set by several
factors, but the basic requirement is unambiguously detecting all
optical clusters out to a redshift of $z\approx0.8$ (as shown in
Figure~\ref{fig:Nc}), where we can detect a high fraction ($\simeq$90\%)
of {\it all} massive clusters even for a conservative mass threshold
of $4\times10^{14}\,M_\odot$.
The optical cluster detection requires seeing a cluster red sequence
(i.e., early-type galaxies with luminosities less than $L^*$).  Our
exposure times are therefore determined by the observed $i-$band
magnitudes of low-luminosity ($\simeq0.4L^*$) red-sequence galaxies at
$z\simeq0.8$, so that we can confidently identify clusters using the
overdensity of red galaxies.
In addition to being dominated by a red sequence of galaxies, massive
clusters exhibit another distinctive signature, the presence of
a BCG (typically with luminosity of a few $L^*$).
Our observation plan is hence designed to detect both of these features of
real clusters (BCG and red-sequence) for 90\% of the SZE candidates
(i.e., $z<0.8$).  For most of the remaining clusters at $z>0.8$ 
(estimated to be only 10\% of the total number) we will be able to 
detect the presence of a BCG as well as some of the brightest cluster members.

\section{Analysis}

\subsection{Selection of ACT SZE Cluster Candidates}
\label{sec:sz}

ACT is a dedicated arcminute resolution CMB experiment consisting of a
6-m off-axis Gregorian telescope located at 5190m on Cerro Toco, near
the Chajnantor Plateau, in the Atacama Desert in northern Chile
\citep[see][for a complete
  description]{Fowler-07,Hincks-09,Fowler-10,Swetz-10}.
The SZE cluster selection for this work is based on 148~GHz data from
the ACT 2008 southern survey using the Millimeter Bolometer Array
Camera \citep[MBAC;][]{Niemack-08} on the ACT focal plane and with
band centers at 148~GHz, 218~GHz and 277~GHz.
The 2008 southern survey area is $9^{\circ}$ wide, centered on
declination $-53.5^{\circ}$, and extends from right ascension
$19^{\mathrm h}$ to $24^{\mathrm h}$ and $00^{\mathrm h}$ to
$07^{\mathrm h}30^{\mathrm m}$. A total of 850 hours of data with an
average of 680 working detectors at 148\,GHz (yielding a data volume
of 3200 GB) are used to make the microwave sky map. Relative detector
pointings are established through observations of Saturn which has a
high enough signal-to-noise ratio to allow a pointing fit for each
detector. Absolute pointings for the survey are iteratively solved by
comparing ACT radio source detections.  The resulting rms uncertainty
in the reported positions is $5\arcsec$ \citep{Fowler-10}.

The data are calibrated with 6\% precision to Uranus by extrapolating
the lower frequency WMAP7 Uranus calibration \citep{Weiland-10}. A
highly-parallelized preconditioned conjugate gradient code is used to
simultaneously solve for the maximum likelihood millimeter-wave map
and correlated noise (e.g., from large-scale atmospheric emission).

Cluster candidates were detected in a 148\,GHz, 455 square-degree
submap from the ACT 2008 observing season. The submap lies between
right ascensions $00^{\mathrm h}12^{\mathrm m}$ and $07^{\mathrm
  h}08^{\mathrm m}$ and declinations $-56^{\circ}11\arcmin$ and
$-49^{\circ}$00$\arcmin$. The median noise level in this submap in
terms of the decrement amplitude is $31\mu$K.  After masking bright
sources from the ACT 148 GHz source catalog, the map is match-filtered
in the Fourier domain using the isothermal $\beta$-model ($\beta=2/3$)
convolved with the ACT 148\,GHz beam and models for the ACT noise,
primary CMB fluctuations, and undetected sources.
The form of the filter \citep{Melin-06,Haehnelt-96} is
\begin{equation}
\Phi(\vec{k}) = \frac{  \tilde{B_{\theta_{\rm c}}}^*(\vec{k}) \mid \widetilde{T}_{\rm other}(\vec{k}) \mid^{-2}} { \int   \tilde{B_{\theta_{\rm c}}}^*(\vec{k'}) \mid \widetilde{T}_{\rm other}(\vec{k'}) \mid^{-2} \tilde{B_{\theta_{\rm c}}}(\vec{k'}) {\rm d}\vec{k'}}
\label{equ:filter}
\end{equation}
where $\tilde{B_{\theta_{\rm c}}}(\vec{k'})$ is the Fourier transform
of the beam-convolved $\beta$-model with critical radius $\theta_{\rm
  c}$, and $\widetilde{T}_{\rm other}(\vec{k'})$ is the transform of
the noise obtained from a combination of ACT difference maps, the
WMAP5 primary CMB signal, and the undetected sources (and clusters)
based on the amplitudes of \cite{Fowler-10}.  The map is filtered
using $\beta$-models with critical radii from $0.25\arcmin$ to
$4.0\arcmin$ in $0.25\arcmin$ steps.  For a given filter, the SNR of a
cluster detection is defined as the ratio of decrement amplitude to
total rms. The reported significance of a cluster detection
corresponds to the filter which maximizes this SNR. Furthermore, any
detection that was not detected at greater than SNR$>$3.5 in at least
four conjoined pixels was discarded.

During 2009B the ACT data reduction pipeline and cluster extraction
algorithms evolved such that everything from pointing to data
selection to the matched filter for the final candidate list changed
from one observing run to the next. Thus we observed more SZE
candidates than used in the purity study (although we report all
confirmed cluster detections regardless of its origin). Despite this
inefficiency the optical study completely vetted a SNR-ordered list of
candidates selected in the manner described above. A final caveat is
that after the optical observing season ended, the ACT pipeline
continued to evolve such that the finalized candidate list tested here
is not identical to the list presented in Marriage et
al.~(2010). While the SNR of the high significance candidates remains
similar between this study and Marriage et al.~(2010), the fainter
candidates differ in SNR with some of the faintest candidates falling
below the SNR threshold of the Marriage et al.\ list.  This is
expected: different map-making configurations weight the data in
different ways, and a faint candidate which is decrement-boosted by
noise in one realization of map may not be boosted in the next.

\subsection{2009B Optical Observing Campaign}

\begin{deluxetable}{clllcl}
\tablecaption{2009B Approved Observing Runs}
\tablehead{
\colhead{Run} &
\colhead{Tel.} &
\colhead{Camera} &
\colhead{Dates} &
\colhead{Nights} & 
\colhead{TAC} 
}
\startdata
1 & SOAR  & SOI   & 2009, Aug 22-23  & 0/2 & CNTAC  \\
2 & NTT  & EFOSC  & 2009, Oct 22-26  & 3/4 & ESO/Chile   \\
3 & SOAR & SOI    & 2009, Dec  9-12  & 4/4 & NOAO   \\
\enddata
\label{tab:runs}
\tablecomments{For run 1 we applied for SOAR time but were reassigned
  by the CNTAC to the CTIO/Blanco 4-m telescope due to
  re-aluminization of the SOAR mirror.}
\end{deluxetable}

During the 2009B semester we were awarded a
total of 10 dark nights for this project in three observing runs on 4-m
class telescopes obtained through different Telescope Allocation
Committees (TACs). Unfortunately our first run (2 nights), which
occurred during the Chilean winter, was completely lost due to snow and
no data were obtained (09B-1007, PI:Infante).  
The next two observing runs had optimal conditions and enabled the
successful completion of our program. In Table~\ref{tab:runs} we
summarize the telescopes, instruments, dates, number of good
(photometric) nights, and the TACs that awarded the time.

In the run during October 22-26, 2009 we used the ESO Faint Object Spectrograph
and Camera (EFOSC) on the 3.6-m NTT on La Silla (084.A-0619,
PI:Barrientos) and observed SZE candidates using the Gunn $gri$
filter, with exposure times of 270~s ($3\times90$~s), 360~s
($3\times120$~s) and 1100~s ($4\times275$~s) in $g$, $r$ and $i$
respectively.  EFOSC consists of a $2048\times2048$~pixel thinned, AR
coated CCD\#40 detector projecting a $4.1'\times4.1'$ field-of-view at
$0.2408''$/pixel ($2\times2$ binning) on the sky. For the three first
nights of this run atmospheric conditions were photometric with seeing
$0.\arcsec7-1\arcsec$. The last and fourth night of the run was lost
due to clouds and no data were taken.  Throughout the run we also observed
photometric standards from the Southern Hemisphere Standards Stars
Catalog \citep{Smith-07} to calibrate our data.

For our final run, we used the SOAR Optical Imager (SOI) on the 4.1-m
SOAR Telescope on Cerro Pach\'on (09B-0355, PI: Menanteau) for four
nights from December 9-12, 2009. SOI is a bent-Cassegrain mounted
optical imager using a mini-mosaic of two E2V $2048\times4096$~pixel
CCDs to cover a $5.26'\times5.26'$ field-of-view at a scale of
$0.\arcsec154$/pixel ($2\times2$ binning). For our cluster search we
used the same exposure times as in the NTT using the SDSS $gri$ filter
set. For three cluster candidates presumed to be at high redshift we
obtained observations twice as long as the nominal exposure times
and additional deep 2200s ($8\times275$s) $z$-band observations. The
run conditions were optimal and all four nights were photometric, with
seeing in the range $0.\arcsec5-1\arcsec$. We observed photometric
standard stars using the same set and method as we did for the earlier
NTT run.

\subsection{Optical Data Reduction}

\begin{figure}
\centerline{\includegraphics[width=3.9in]{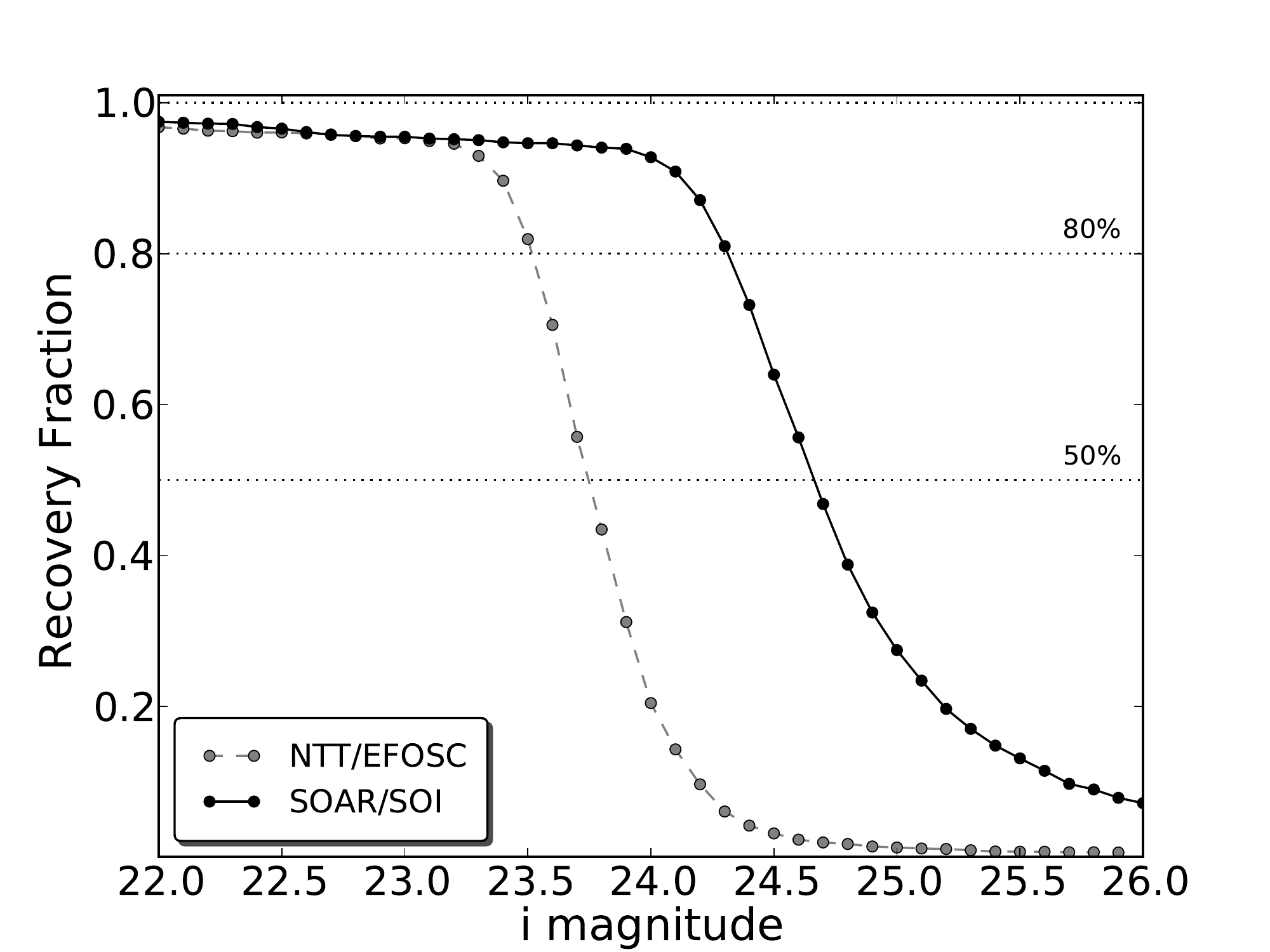}}
\caption{The recovery fraction of simulated elliptical galaxies, using
  a \citeauthor{deVauc48} profile, as a function of the observed $i$-band
  magnitude for the NTT (dashed line) and SOAR (solid line)
  observations. We show for reference the 50\% and 80\% recovery
  fraction levels.}
\label{fig:recovery}
\end{figure}

For the optical analysis we take advantage of an existing imaging
data pipeline that takes us from raw data to galaxy
catalogs. We use a modified version of the Python-based Rutgers
Southern Cosmology image pipeline \citep[][]{SCSI,SCSII} to process
the data from the NTT and SOAR runs. The initial standard CCD
processing steps are performed using IRAF\footnote{IRAF is distributed
  by the National Optical Astronomy Observatory, which is operated by
  the AURA under cooperative agreement with NSF}/{\tt ccdproc} tasks via the
STScI/PyRAF\footnote{PyRAF is a product of the Space Telescope Science
  Institute, which is operated by AURA for NASA} interface.
The fast read-out time of the CCDs allowed us to accommodate multiple
dithered exposures for each targeted cluster candidate. In our cluster
observations we followed a $10''$ off-center 3--, 3--, and 4--point
dither pattern in the $g$, $r$, and $i$ bands, respectively, which helped 
to avoid saturated stars, remove chip defects and artifacts, and
cover the $10''$ gap between CCDs in the SOI camera.
Individual science frames were astrometrically recalibrated before
stacking by matching sources with stars from the US Naval Observatory
Catalog. Observed fields with photometric standard stars from the
\cite{Smith-07} catalog were processed in the same fashion and we
obtained a photometric zero-point solution in AB magnitudes for each
observing night. 
For each targeted cluster we associated all science frames in all
filters and aligned and median combined them into a common pixel
reference frame using SWarp \citep{SWarp} to a plate scale of
$0.''2408$/pixel for the NTT and $0.''1533$/pixel for SOAR. Source
detection and photometry for the science catalogs were performed using
SExtractor \citep{SEx} in dual-image mode in which sources were
identified on the $i-$band images using a $1.5\sigma$ detection
threshold, while magnitudes were extracted at matching locations from
all other bands.
The computed magnitudes in our catalogs were corrected for Galactic
dust absorption in each observed band utilizing the infrared maps and
C~routines provided by \cite{Schlegel98}. This step is particularly
important for obtaining unbiased colors, required for accurate photometric
redshifts, as significant variations occur over the large area covered
by the ACT 2008 survey region.

The final step in our pipeline involves computing the photometric
redshift and redshift probability distributions, $p_{\rm BPZ}(z)$, for
each object using the spectral energy distribution (SED) based BPZ
code \citep{BPZ} with a flat prior.
We compute photometric redshifts using the dust-corrected $g,r,i$
isophotal magnitudes, as defined by the $i$-band detection and the BPZ
set of template spectra described in \cite{Benitez-04} which is based
on the templates from \cite{C80} and \cite{K96}. This set consists of
El, Sbc, Scd, Im, SB3, and SB2 and represent the typical SEDs of
elliptical, early/intermediate type spiral, late-type spiral,
irregular, and two types of starburst galaxies respectively.

\renewcommand{\arraystretch}{1.05} 
\begin{deluxetable*}{ccclrl} 
\tablecaption{ACT 2008 clusters} 
\tablehead{
\colhead{ACT Descriptor} & 
\colhead{R.A. (J2000)} & 
\colhead{Dec. (J2000)} &
\colhead{Redshift} &
\colhead{SNR} &
\colhead{Alt Name}
}
\startdata
ACT-CL~J0145-5301  &  01:45:03.6  & -53:01:23.4  &   0.118 \tablenotemark{a} &   4.7 (4.0) &  Abell 2941 \\ 
ACT-CL~J0641-4949  &  06:41:37.8  & -49:46:55.0  &   0.146 \tablenotemark{b} &   4.9 (4.9) &  Abell 3402 \\ 
ACT-CL~J0645-5413  &  06:45:29.5  & -54:13:37.0  &   0.167 \tablenotemark{a} &   7.1 (7.1) &  Abell 3404 \\ 
ACT-CL~J0638-5358  &  06:38:49.4  & -53:58:40.8  &   0.222 \tablenotemark{a} &  10.6 (10.0) & Abell S0592 \\ 
ACT-CL~J0516-5430  &  05:16:37.4  & -54:30:01.5  &   0.294 \tablenotemark{c} &   5.2  (4.7) &  Abell S0520/SPT-CL~J0516-5430\\ 
ACT-CL~J0658-5557  &  06:58:33.1  & -55:57:07.2  &   0.296 \tablenotemark{d} &  11.6  (11.5) &  1E0657-56 (Bullet)\\ 
ACT-CL~J0245-5302  &  02:45:35.8  & -53:02:16.8  &   0.300 \tablenotemark{e} &   8.3  (9.1) &  Abell S0295 \\ 
ACT-CL~J0217-5245  &  02:17:12.6  & -52:44:49.0  &   0.343 \tablenotemark{f} &   4.5  (4.1) &  RXC~J0217.2-5244 \\ 
ACT-CL~J0237-4939  &  02:37:01.7  & -49:38:10.0  &  $0.40\pm0.05$   &   4.9 (3.9) &  \\ 
ACT-CL~J0707-5522  &  07:07:04.7  & -55:23:08.5  &  $0.43\pm0.06$   &   4.2 ($\ldots$) &  \\ 
ACT-CL~J0235-5121  &  02:35:45.3  & -51:21:05.2  &  $0.43\pm0.07$   &   5.7 (6.2) &  \\ 
ACT-CL~J0330-5227  &  03:30:56.8  & -52:28:13.7  &  0.440 \tablenotemark{g} &   7.4 (6.1) &  Abell 3128(NE) \\ 
ACT-CL~J0509-5341  &  05:09:21.4  & -53:42:12.3  &  0.461 \tablenotemark{h} &   4.4 (4.8) &  SPT-CL~J0509-5342 \\ 
ACT-CL~J0304-4921  &  03:04:16.0  & -49:21:26.3  &  $0.47\pm0.05$   &   5.0 (3.9) &  \\ 
ACT-CL~J0215-5212  &  02:15:12.3  & -52:12:25.3  &  $0.51\pm0.05$   &   4.8 (4.9) &  \\ 
ACT-CL~J0438-5419  &  04:38:17.7  & -54:19:20.7  &  $0.54\pm0.05$   &   8.8 (8.0) &  \\ 
ACT-CL~J0346-5438  &  03:46:55.5  & -54:38:54.8  &  $0.55\pm0.05$   &   4.4 (4.4) &  \\ 
ACT-CL~J0232-5257  &  02:32:46.2  & -52:57:50.0  &  $0.59\pm0.07$   &   5.2 (4.7) &  \\ 
ACT-CL~J0559-5249  &  05:59:43.2  & -52:49:27.1  &  0.611 \tablenotemark{i} &   5.1 (5.1) &  SPT-CL~J0559-5249 \\ 
ACT-CL~J0616-5227  &  06:16:34.2  & -52:27:13.3  &  $0.71\pm0.10$   &   6.3 (5.9) &  \\ 
ACT-CL~J0102-4915  &  01:02:52.5  & -49:14:58.0  &  $0.75\pm0.04$   &   8.8 (9.0) &  \\ 
ACT-CL~J0528-5259  &  05:28:05.3  & -52:59:52.8  &  0.768  \tablenotemark{h} &   4.7 ($\ldots$) &  SPT-CL~J0528-5300 \\ 
ACT-CL~J0546-5345  &  05:46:37.7  & -53:45:31.1  &  1.066  \tablenotemark{h} &   7.2 (6.5) &  SPT-CL~J0546-5345 \\ 
\enddata
\label{tab:clusters}
\tablecomments{R.A.\ and Dec.\ positions denote the BCG position in the
  optical images of the cluster. The SZE position was used to construct the
  ACT descriptor identifiers. Clusters with uncertainty estimates on 
  their redshifts are those systems for which only photometric redshifts are
  available from the NTT and SOAR $gri$ imaging. The values of SNR in parentheses 
  are the current values from Marriage et al.~(2010), while the others
  are the ones used in this study}
\tablenotetext{a}{spec-z from \cite{deGrandi-99}}
\tablenotetext{b}{spec-z from \cite{6dF}} 
\tablenotetext{c}{spec-z from \cite{Guzzo-99}} 
\tablenotetext{d}{spec-z from \cite{Bullet-z}}
\tablenotetext{e}{spec-z from \cite{Edge-94}} 
\tablenotetext{f}{spec-z from \cite{Bohringer-04}} 
\tablenotetext{g}{spec-z from \cite{Werner-07}}
\tablenotetext{h}{spec-z from \cite{Infante-10}, \cite{Brodwin-10}}
\tablenotetext{i}{spec-z from \cite{High-10}}
\end{deluxetable*}

\subsection{Galaxy Completeness and Recovery Fraction}

\begin{figure}
\centerline{\includegraphics[width=3.9in]{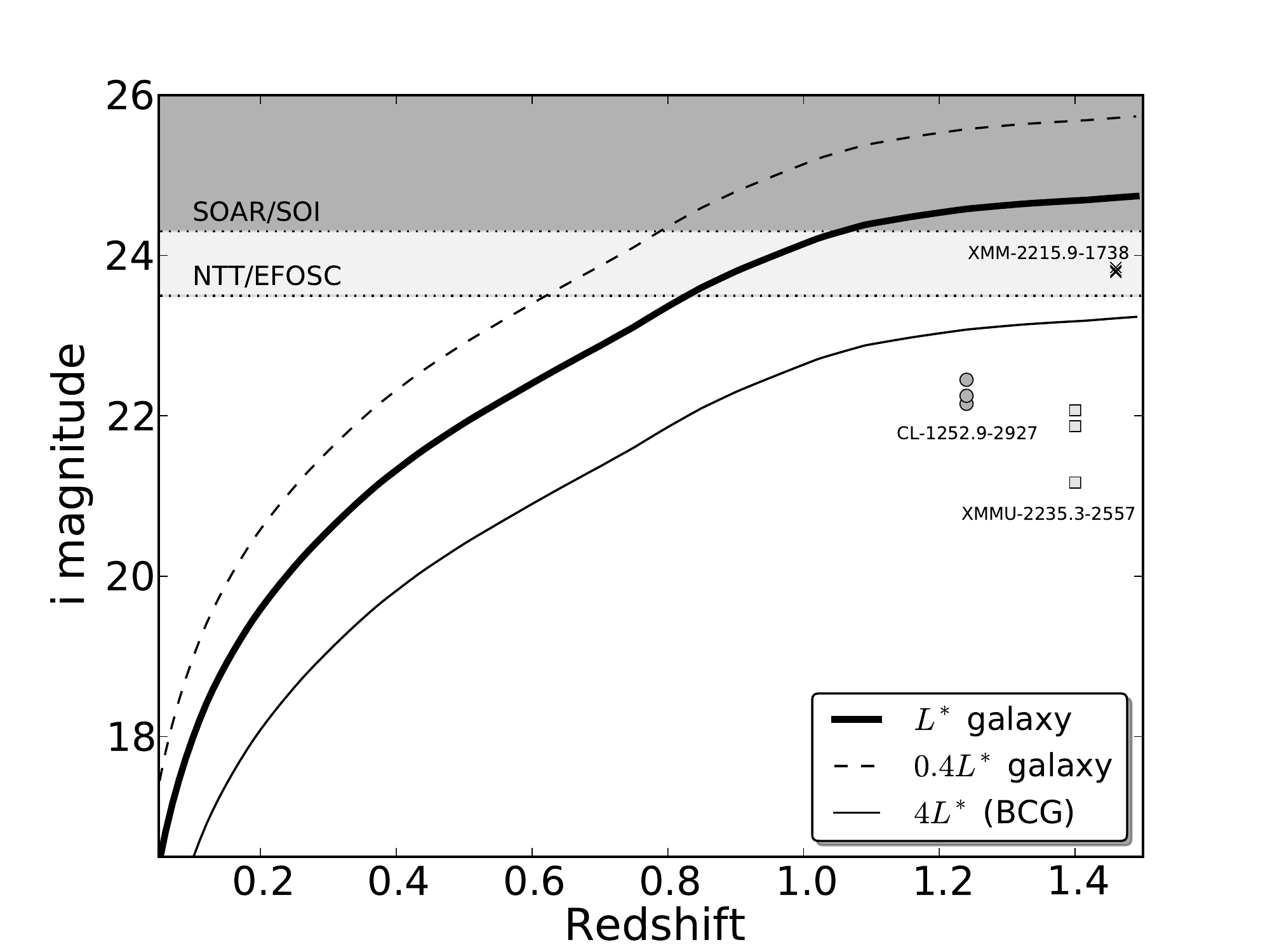}}
\caption{The observed $i$-band magnitudes of $L^*$, $0.4L^*$ and $4L^*$
  (BCG) early-type galaxies as a function of redshift. We use $L^*$ as
  defined \cite{Blanton-03} for the population of red galaxies at
  $z=0.1$ and allow it to passively evolve with redshift. We show in
  different gray levels the 80\% completeness limits for the NTT and
  SOAR observations.  In other words, at magnitudes below the gray
  band we detect elliptical galaxies with at least 80\%
  completeness. We also show the observed magnitudes for the three
  brightest galaxies in three high redshift clusters.}
 \label{fig:mi_z}
\end{figure}

The exposure times for our observations were designed to confidently detect
sub-$L^*$ early-type galaxies up to $z\approx0.8$ and BCGs at $z>1.0$ in
order to confirm the presence of a galaxy cluster. In order to assess
the depth reached by the observations as well as the differences between
telescopes, we performed Monte Carlo simulations to estimate the
recovery fraction of early-type galaxies as a function of magnitude in
each of our observing runs.
We used the IRAF/{\tt artdata} package to create simulated data sets
for the NTT and SOAR observations, which were subsequently processed
through our pipeline using the same detection and extraction
procedures as for the science images. Within {\tt artdata} we used the
{\tt mkobjects} task to draw artificial elliptical galaxies on science
fields as \citeauthor{deVauc48} profiles with the same noise
properties as the real data. We scaled the surface brightness of the
galaxy profiles according to their magnitude for a range of assumed
sizes ($1-7$~kpc, but converging to the typical image seeing at the
faint limits). Using this procedure we generated 20 and 25 artificial
elliptical galaxies for the NTT and SOAR datasets respectively over a
wide range of magnitudes ($19<i<27$) sampled at small ($\Delta_{\rm
  mag}=0.1$) steps. We repeated this 10 times for each dataset at each
step. At each magnitude step we matched the recovered galaxies in the
simulations with the input positions from the artificial catalogs. In
Figure~\ref{fig:recovery} we show the recovered fraction of galaxies
as a function of the input $i$-band magnitude for the NTT and SOAR
dataset, as well as the 80\% and 50\% levels for reference. From this
exercise we conclude that we reach 80\% completeness at $i=23.5$ and
$i=24.3$ for elliptical galaxies for the NTT and SOAR datasets
respectively. We emphasize that we followed the same observing
strategy (i.e. same filter set, dithering pattern and exposure times)
for both runs and therefore the relatively large difference (0.8~mag)
in the completeness limit is likely due to a combination of telescope
size (3.6-m vs 4.1-m), CCD sensitivity and telescope throughput.  The
re-aluminization of the SOAR primary mirror in November 2009, just
before our run, might have played a role in the higher sensitivity of
the SOAR imaging.

We can use the completeness limits estimated
from our simulations to determine how far in redshift we can ``see''
massive clusters.
For this, we compare the completeness limits of our observations to the
expected and observed (i.e. known) apparent magnitudes of galaxies {\em in}
clusters as a function of redshift. 
We estimated the expected apparent galaxy $i$-band magnitude as a
function of redshift using $L^*$ as defined for the population of red
galaxies by \cite{Blanton-03} at $z=0.1$ and allowing passive
evolution according to a solar metallicity \citet{BC03} $\tau=1.0$~Gyr
burst model formed at $z_f=5$. 
We show this in Figure~\ref{fig:mi_z} for a range of luminosities
($0.4L^*$, $L^*$ and $4L^*$) aimed at representing the cluster members
from the faint ones to the BCG. We also show as different gray levels
the 80\% completeness as determined by the simulations for the NTT and
SOAR run, which provides evidence that we can reach typical $L^*$
galaxies to $z\approx0.8$ for both datasets (and $z\approx1$ for the
SOAR data).
We also compare the depth of our imaging campaign to the observed
$i$-band apparent magnitudes of galaxies in three massive high
redshift ($z>1.2$) clusters for which we could find suitable data in
the literature for comparison. In Figure~\ref{fig:mi_z} we plot the
observed magnitude of the three brightest galaxies for the clusters
RDCS~1252.9$-$2927 \citep{Blakeslee-03,Rosati-04} at $z=1.237$,
XMMU~J2235.3$-$2557 \citep{Mullis-05,Rosati-09} at $z=1.393$ and
XMMXCS~J2215.9$-$1738 \citep{Stanford-06,Hilton-09} at $z=1.457$.  We
conclude that, with the possible exception of a cluster like
XMMXCS~J2215.9$-$1738 observed during the NTT run, we comfortably
detect luminous galaxies in clusters at high redshift. This provides
further evidence that our observation's integration times allow us to
confidently identify real SZE clusters up to $z\approx1.2-1.4$, 
below this redshift according to $\Lambda$CDM model $>99\%$ of all clusters more massive
$8\times10^{14}M_{\odot}$ should reside. In Section~\ref{sec:unconfirmed}
we further discuss the possibility of missing $z>1.2$ massive clusters in our
observations.

\subsection{Additional Imaging Data}

A number of the SZE cluster candidates were associated with previously
known clusters at lower redshifts. For a few of these we obtained new
NTT/SOAR $gri$ observations in order to calibrate our photometric
redshift and to test our identification procedure; however, for the
most part these were avoided and we relied on public archival imaging
and on data from other existing programs to compare with the SZE
sources.

AS0295 and AS0592 are included among the low redshift SZE clusters
reported by \cite{Hincks-09} and are the focus of a separate observing
program by our group (09B-0389, PI:Hughes) aimed at obtaining weak
lensing masses using deep $V$ and $R$ observations.  These were obtained
using the MOSAIC camera on the Blanco 4-m telescope on January 9, 2010
under photometric conditions.

\begin{figure*}
\centerline{
\includegraphics[width=3.5in]{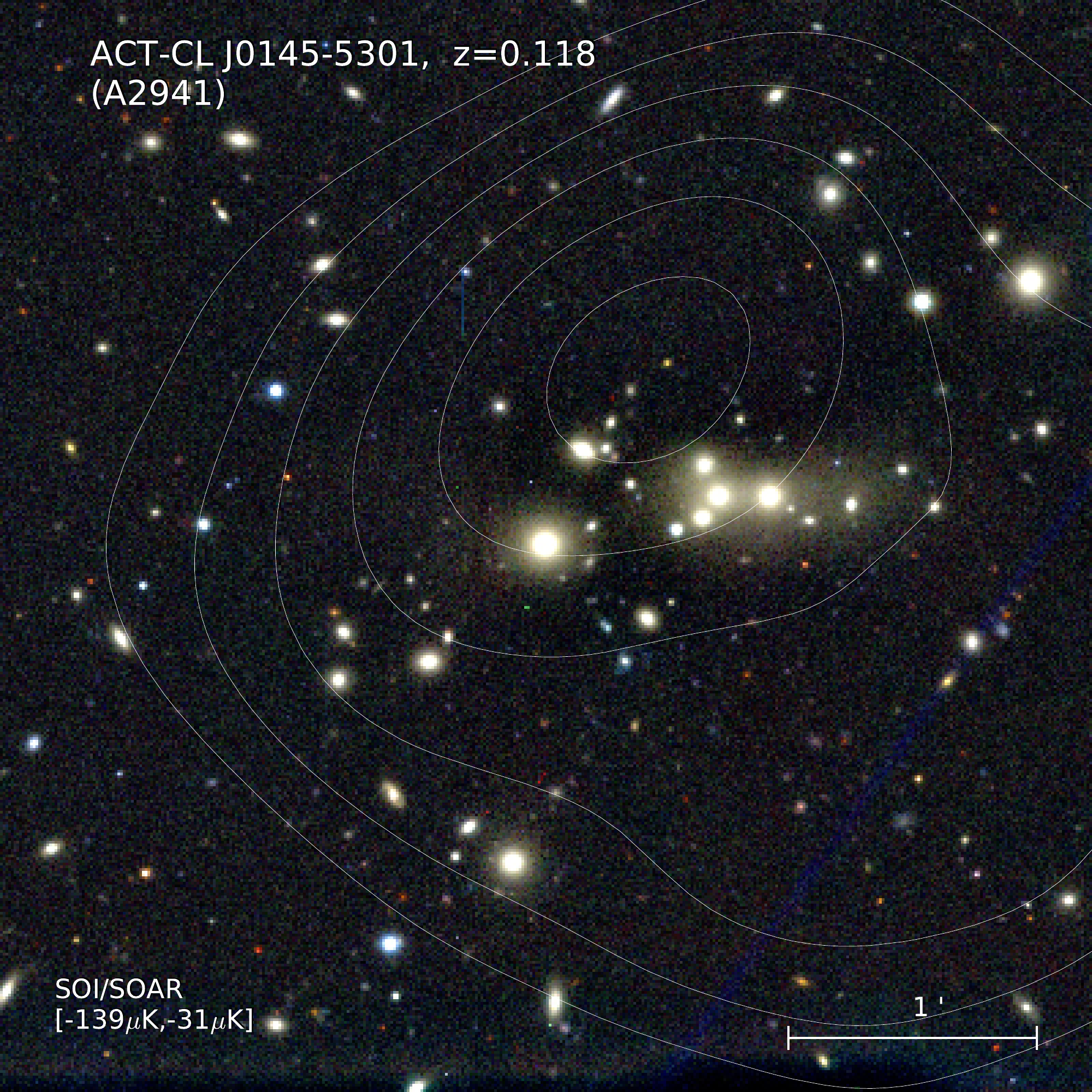}  
\includegraphics[width=3.5in]{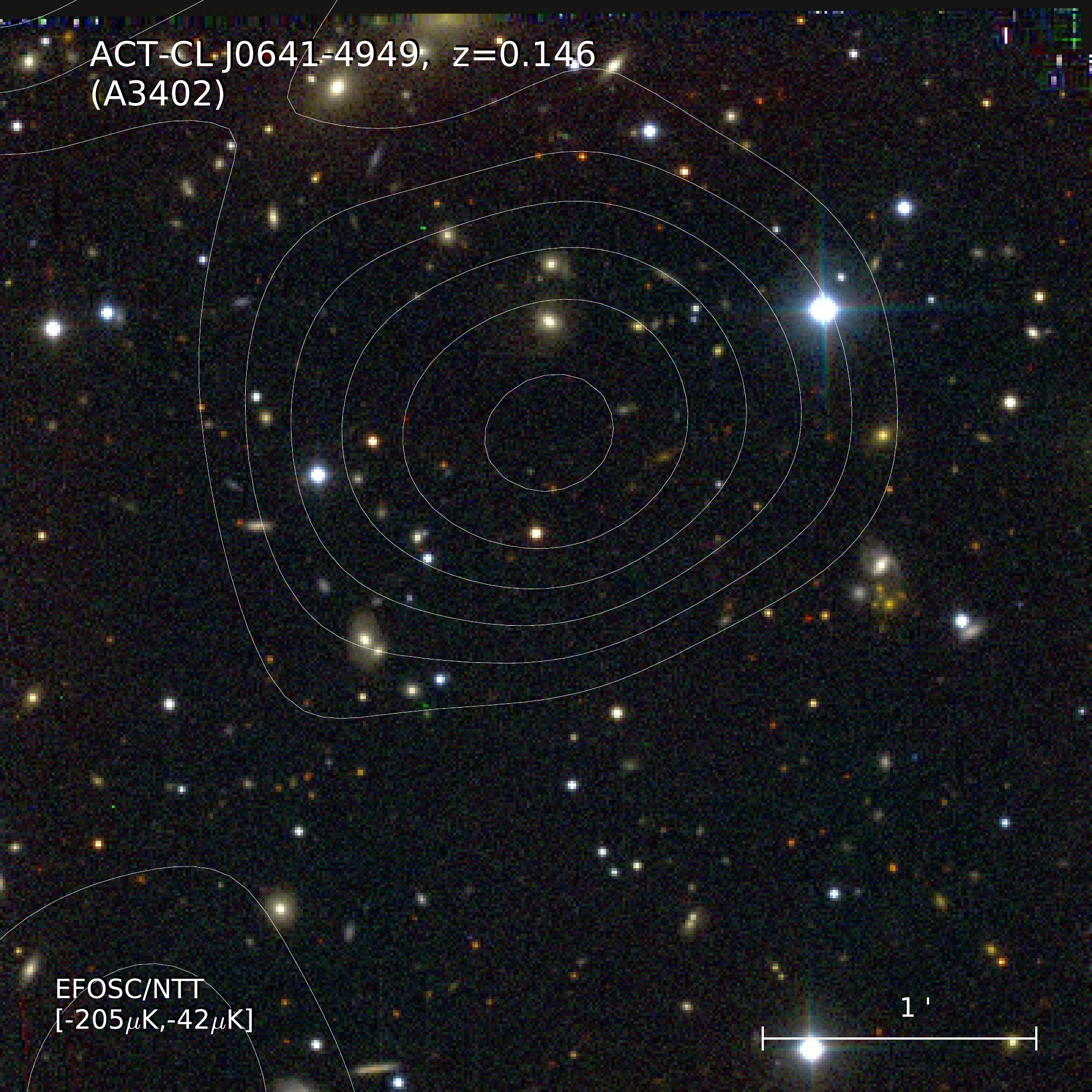}} 
\centerline{
\includegraphics[width=3.5in]{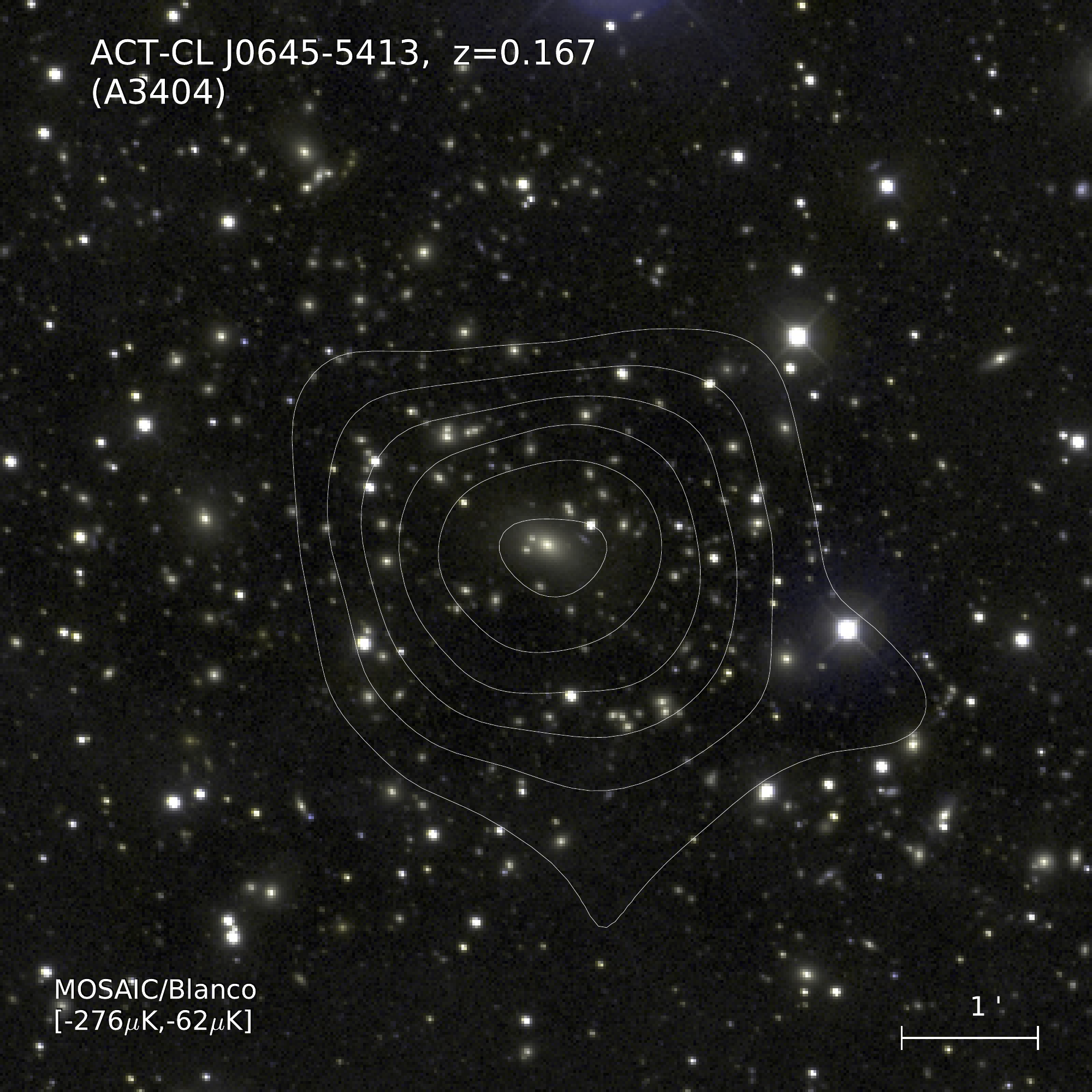}           
\includegraphics[width=3.5in]{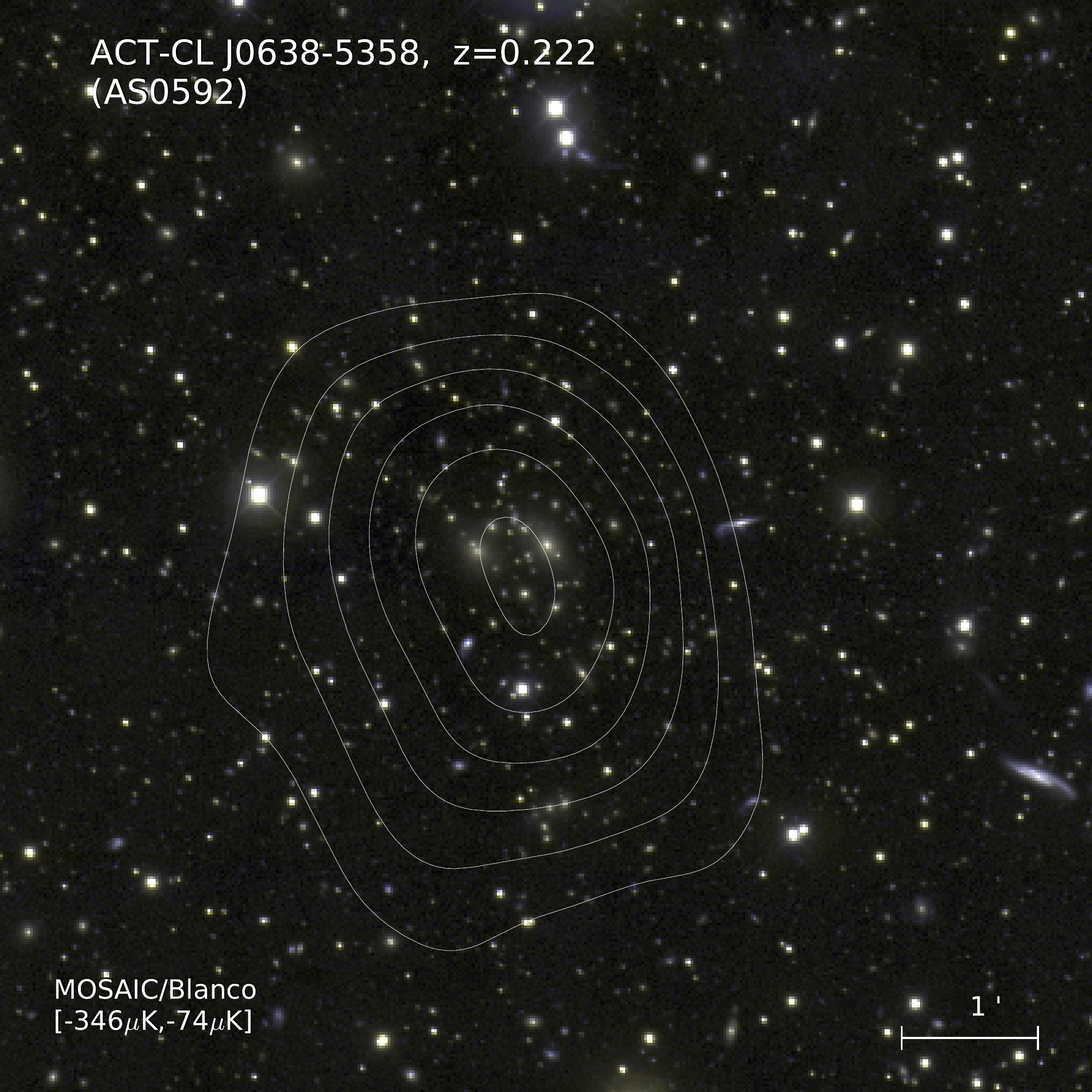}}         
\caption{Composite color images for ACT-CL~J0145-5301 (Abell 2941),
  ACT-CL~J0641-4949 (Abell 3402), ACT-CL~J0645-5413 (Abell 3404) and
  ACT-CL~J0638-5358 (Abell S00592). The horizontal bar shows the scale
  of the images, where north is up and east is left. White contours
  show the 148~GHz SZE maps with the minimum and maximum levels, in
  $\mu$K, displayed between brackets.}
\label{fig:CL1}
\end{figure*}

\begin{figure*}
\centerline{
\includegraphics[width=3.5in]{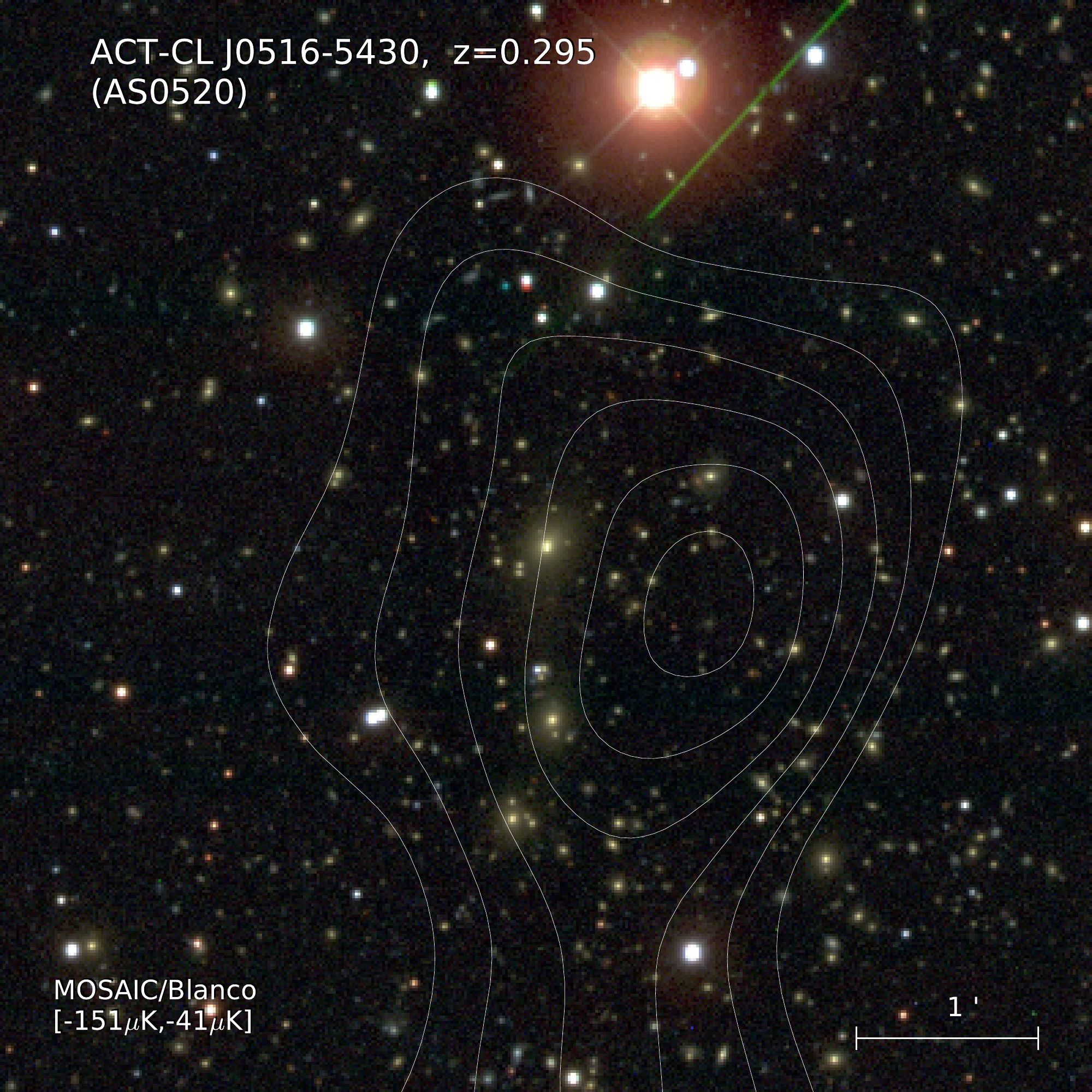}  
\includegraphics[width=3.5in]{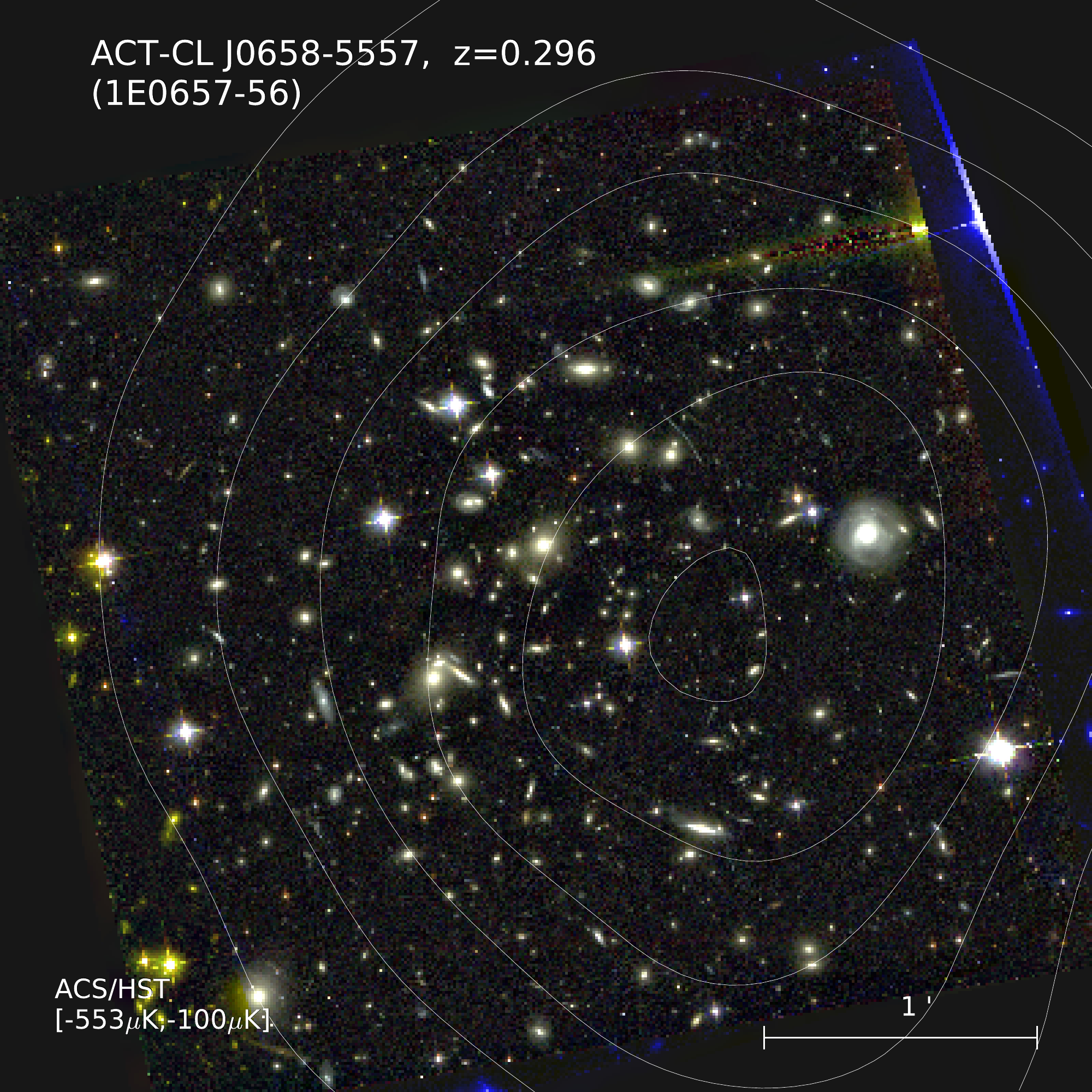}       
}
\centerline{
\includegraphics[width=3.5in]{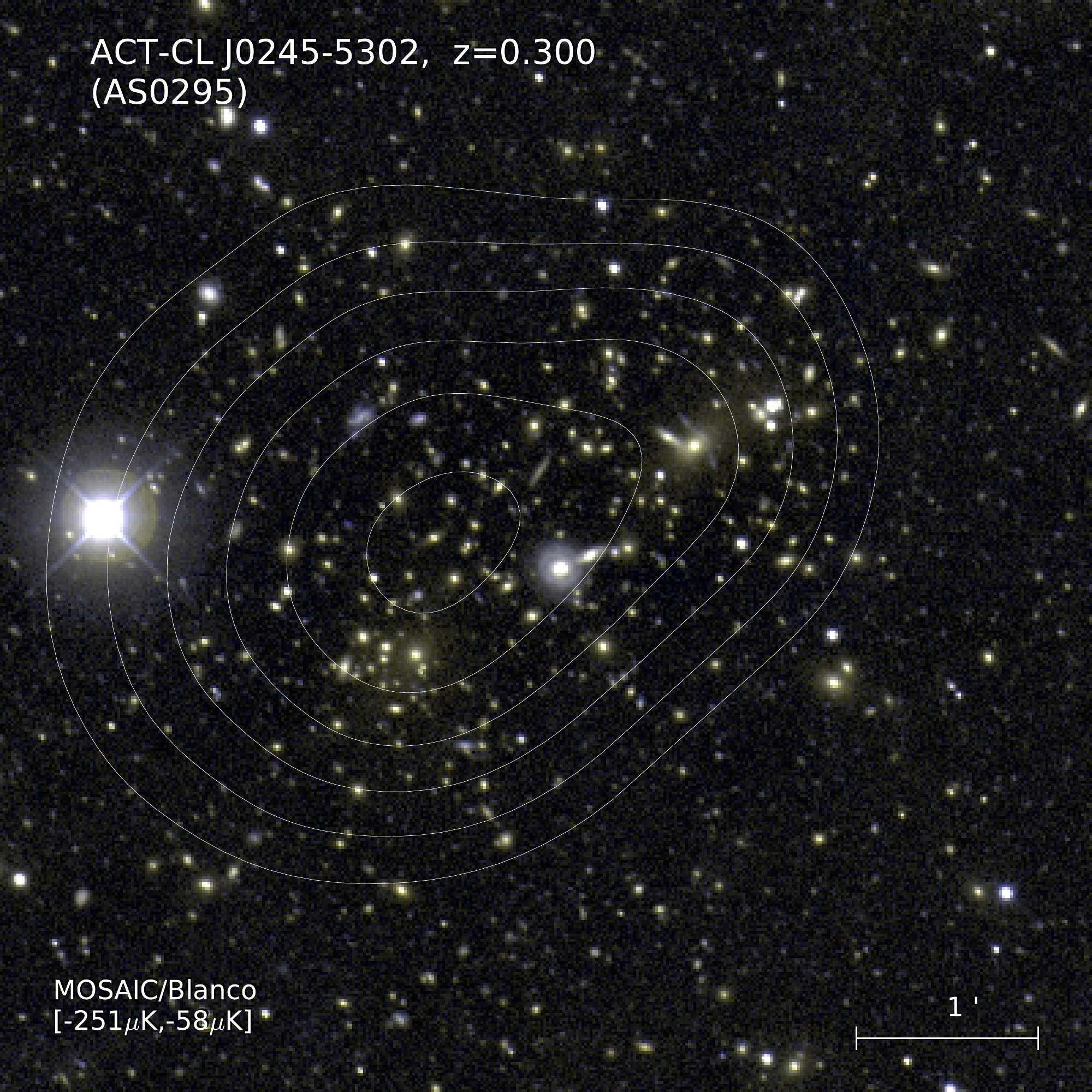}         
\includegraphics[width=3.5in]{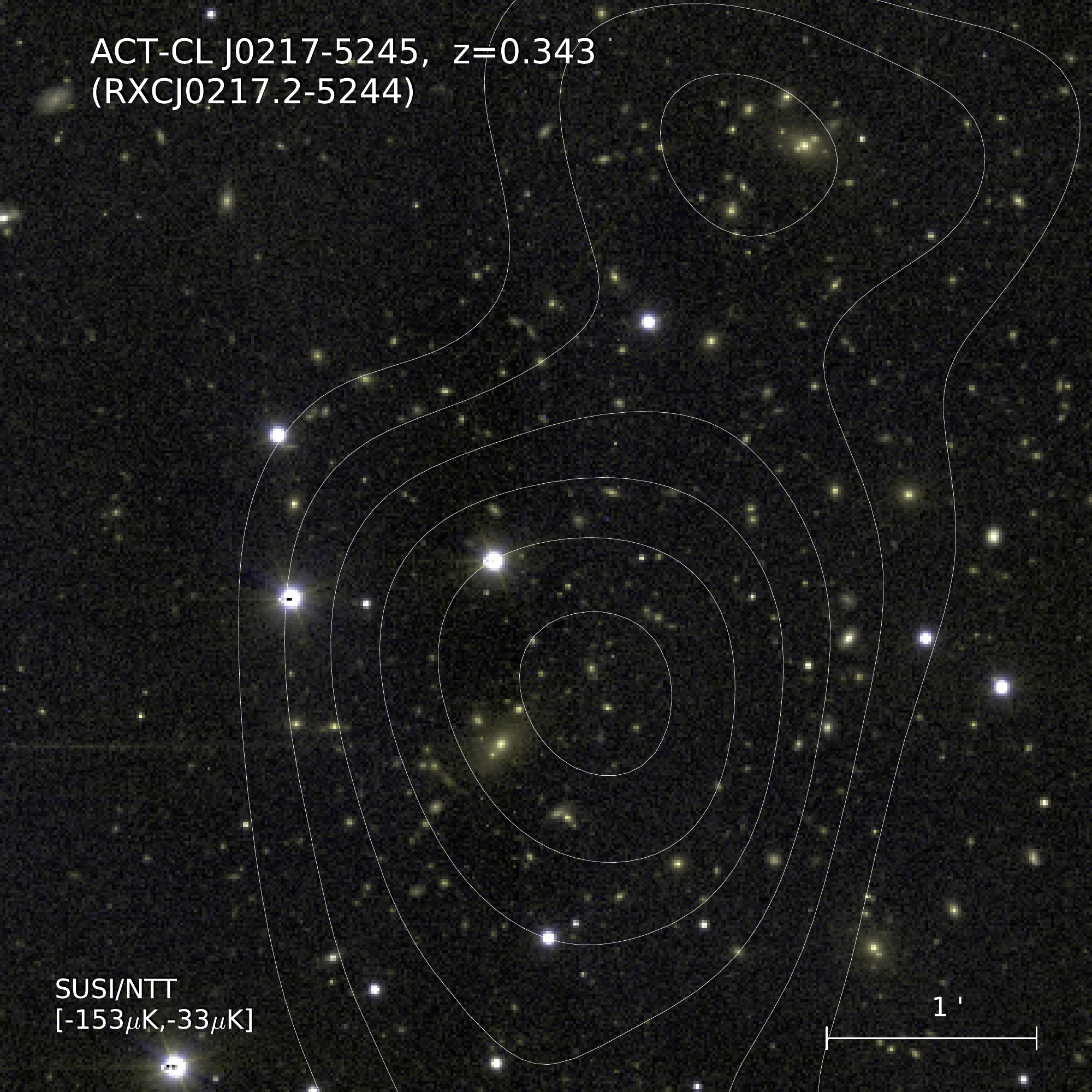}  
}
\caption{Composite color images for ACT-CL~J0516-5430 (Abell S0520),
  ACT-CL~J0658-5557 (Bullet), ACT-CL~J0245-5302 (Abell S0295) and
  ACT-CL~J0237-4939 (RXC~J0217.2-5244). The horizontal bar shows the
  scale of the images, where north is up and east is left.  White
  contours show the 148~GHz SZE maps with the minimum and maximum
  levels, in $\mu$K, displayed between brackets.}  
\label{fig:CL2}
\end{figure*}

\begin{figure*}
\centerline{
\includegraphics[width=3.5in]{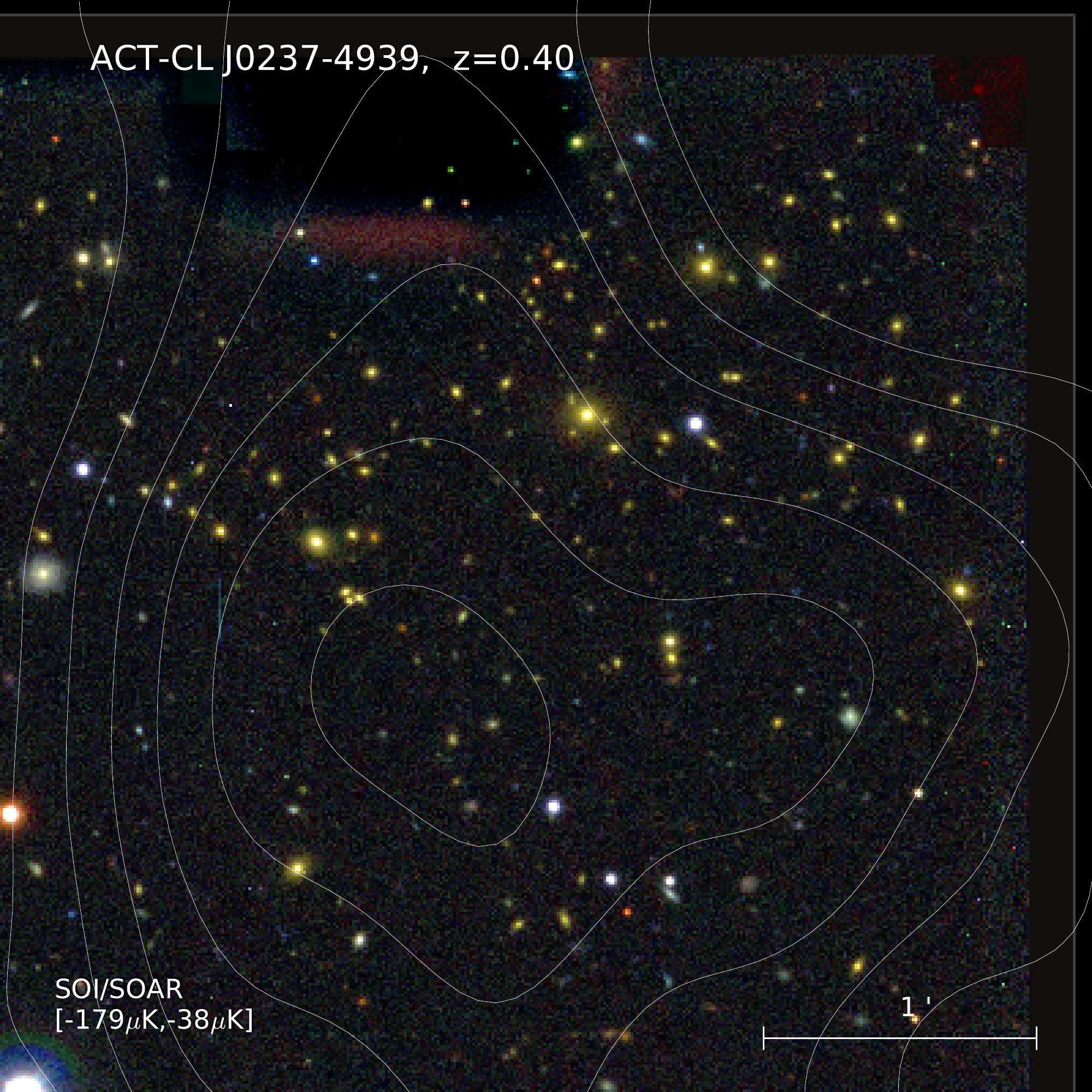}  
\includegraphics[width=3.5in]{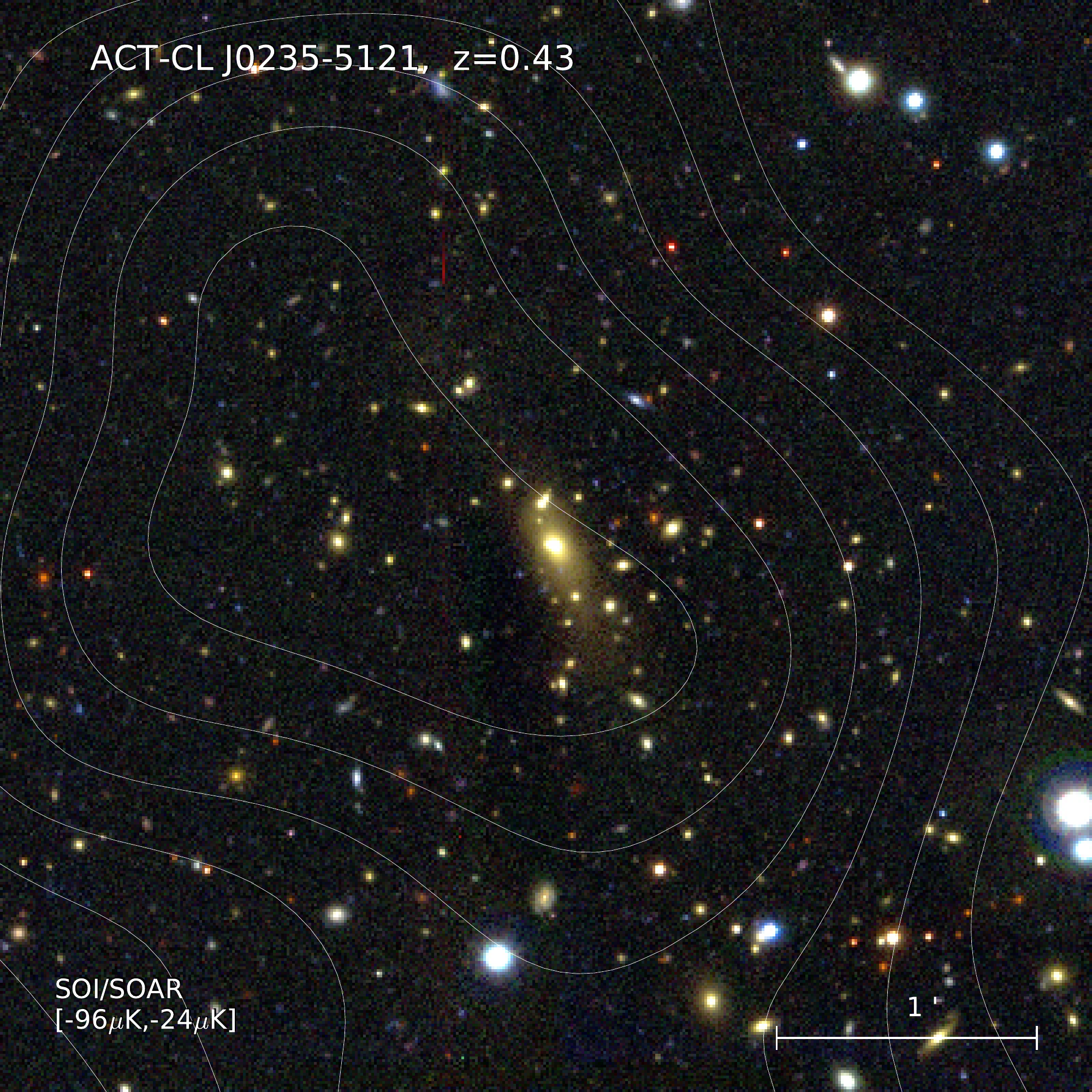}  
}
\centerline{
\includegraphics[width=3.5in]{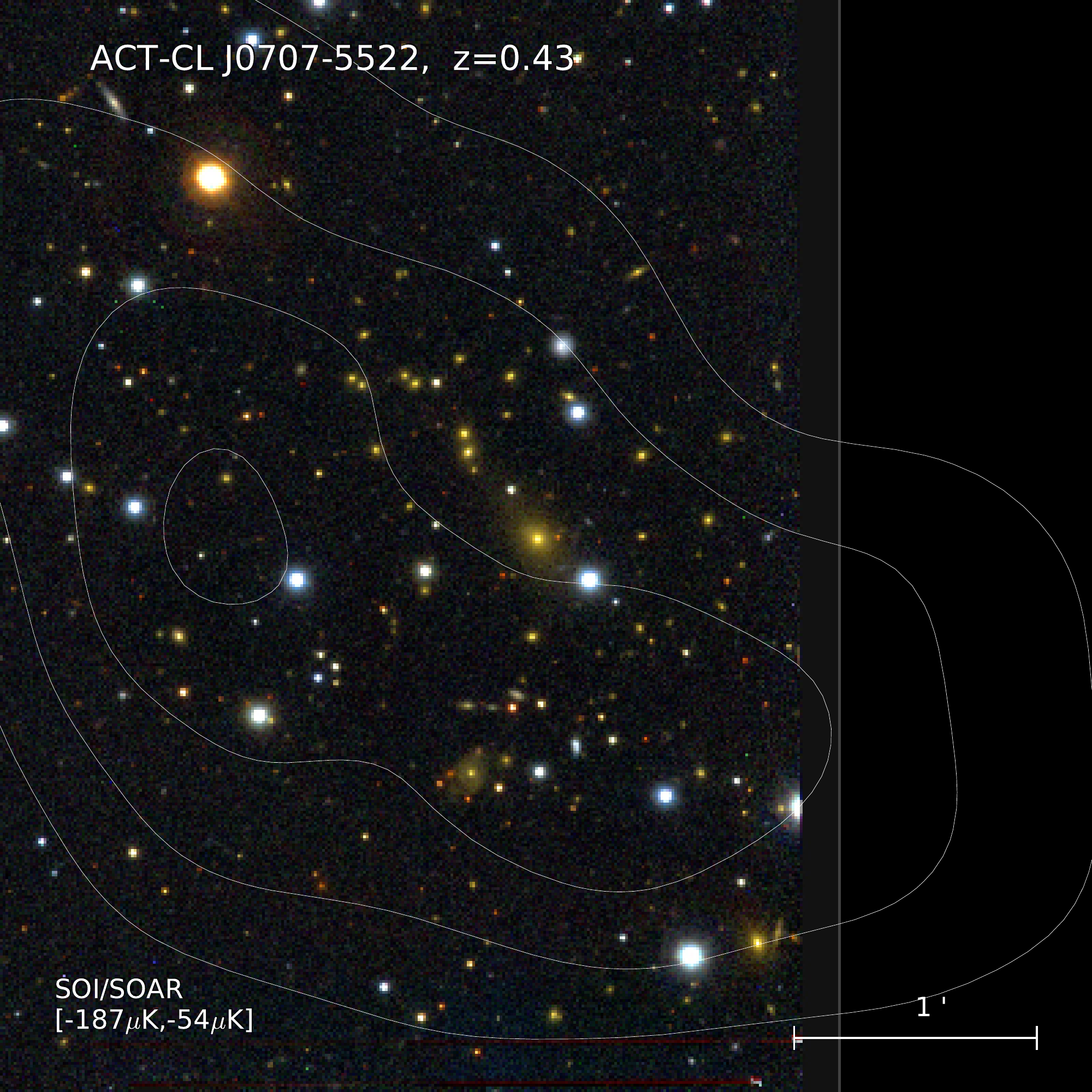}  
\includegraphics[width=3.5in]{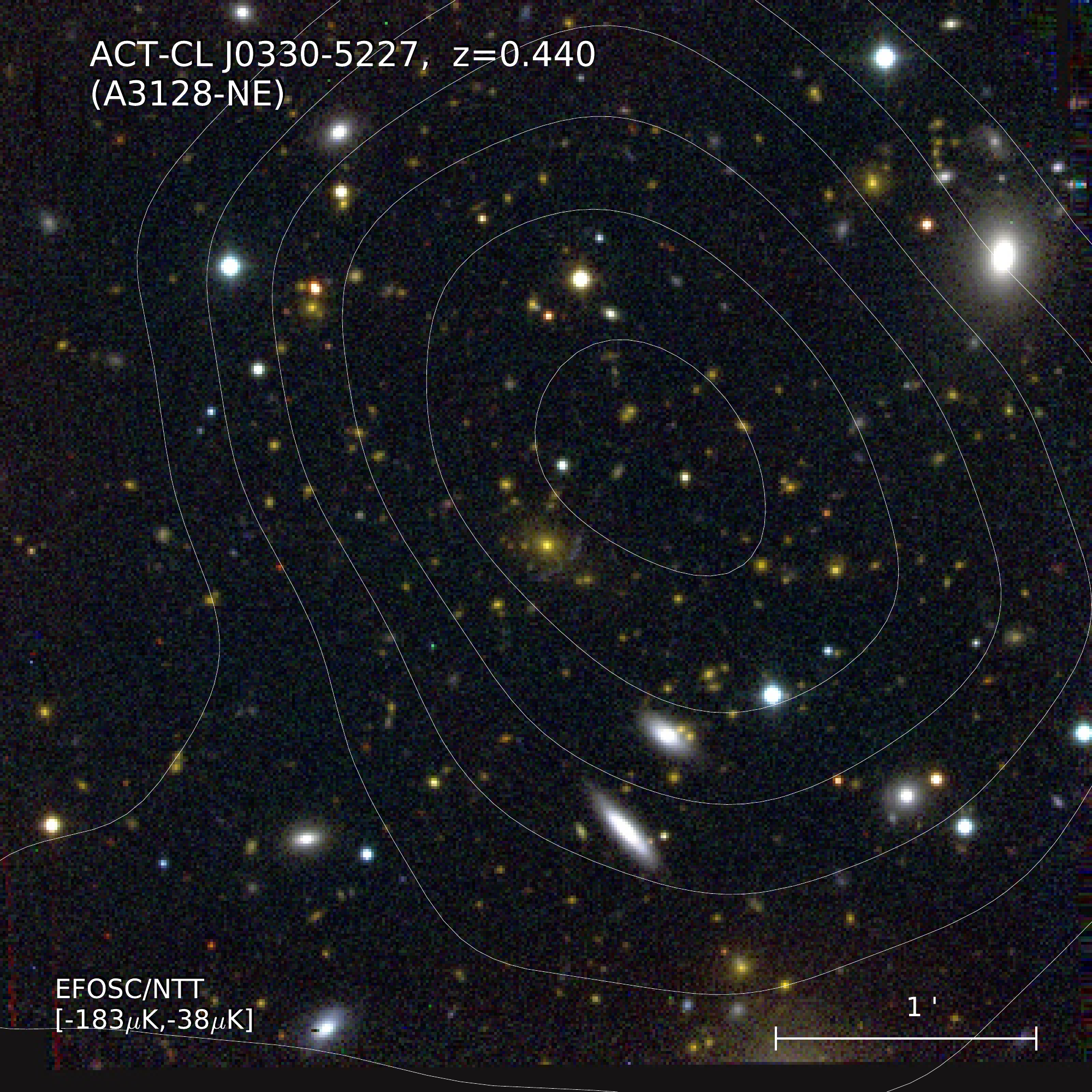}  
}
\caption{Composite color images for ACT-CL~J0237-4939,
  ACT-CL~J0707-5522, ACT-CL~J0235-5121 and ACT-CL~J0330-5227 (Abell
  3128 North-East). The horizontal bar shows the scale of the images,
  where north is up and east is left. White contours show the 148~GHz
  SZE maps with the minimum and maximum levels, in $\mu$K, displayed
  between brackets.}
\label{fig:CL3}
\end{figure*}

\begin{figure*}
\centerline{
\includegraphics[width=3.5in]{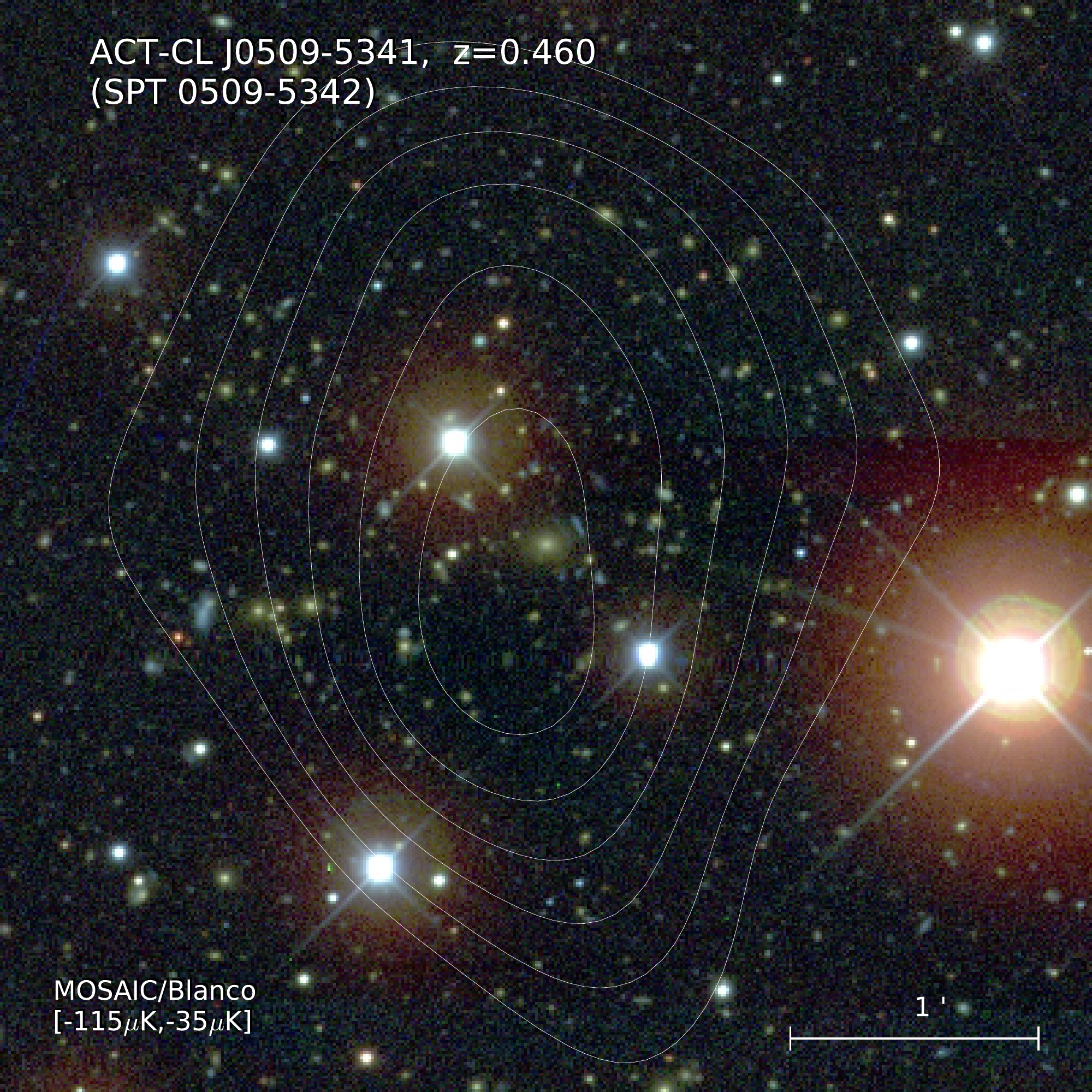}  
\includegraphics[width=3.5in]{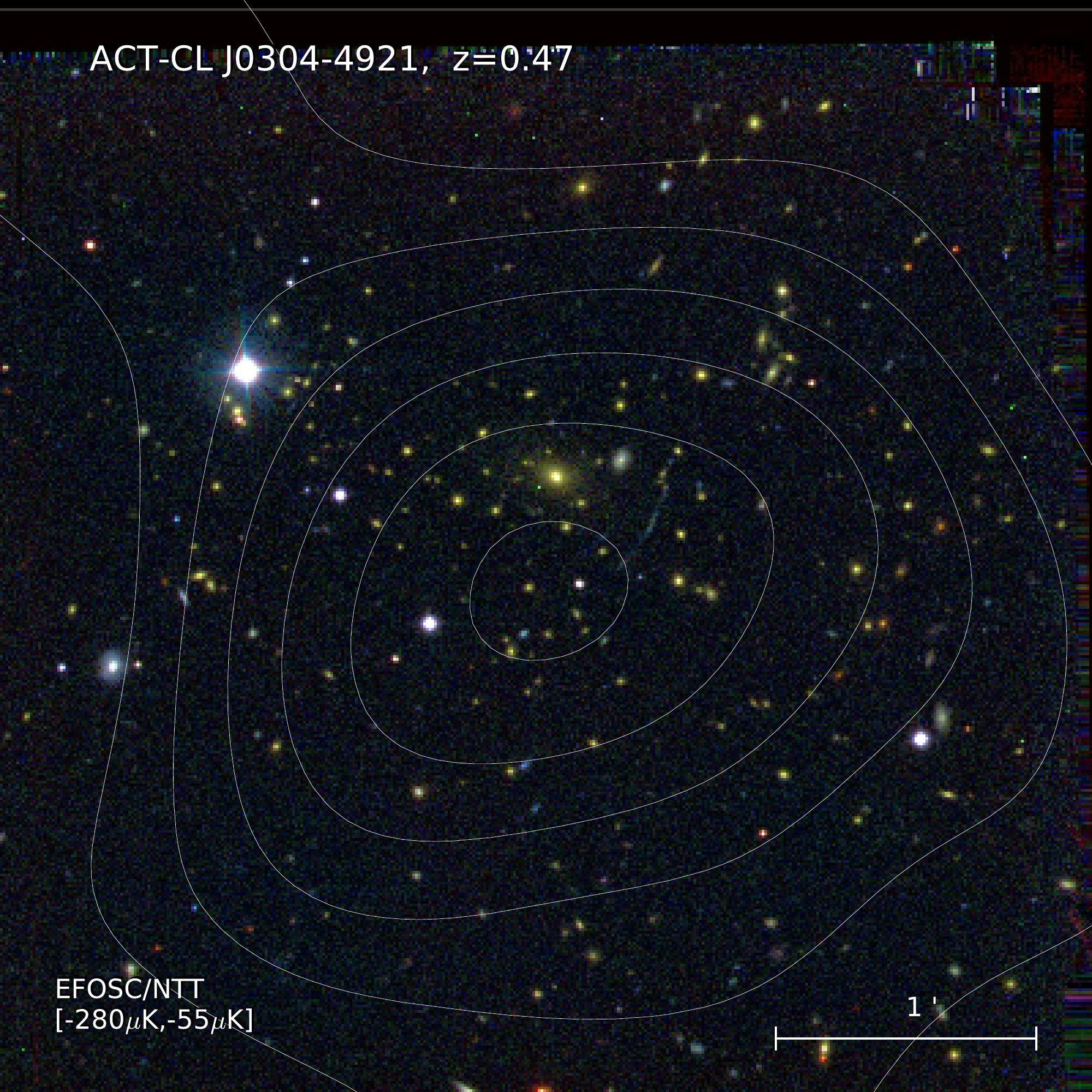}  
}
\centerline{
\includegraphics[width=3.5in]{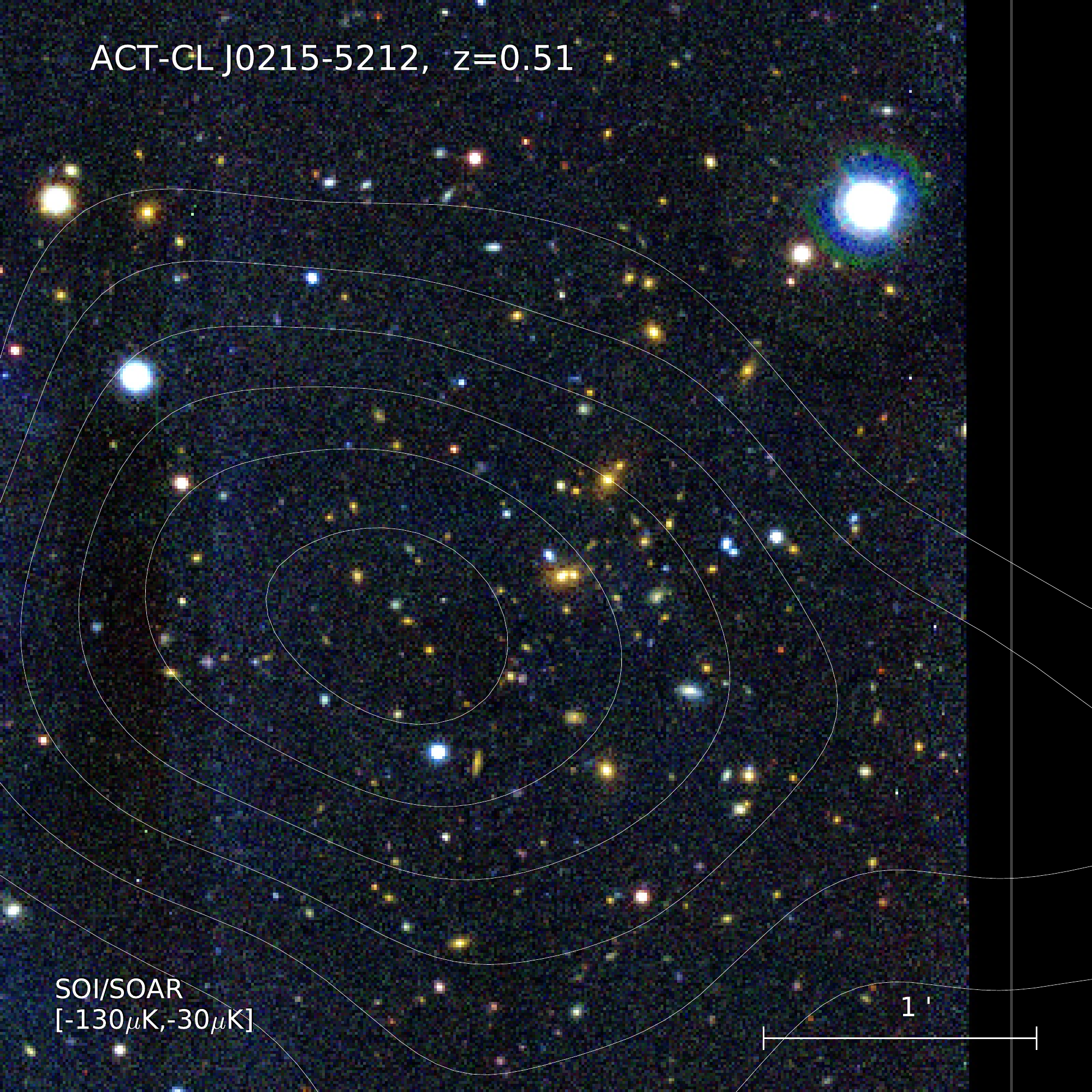}  
\includegraphics[width=3.5in]{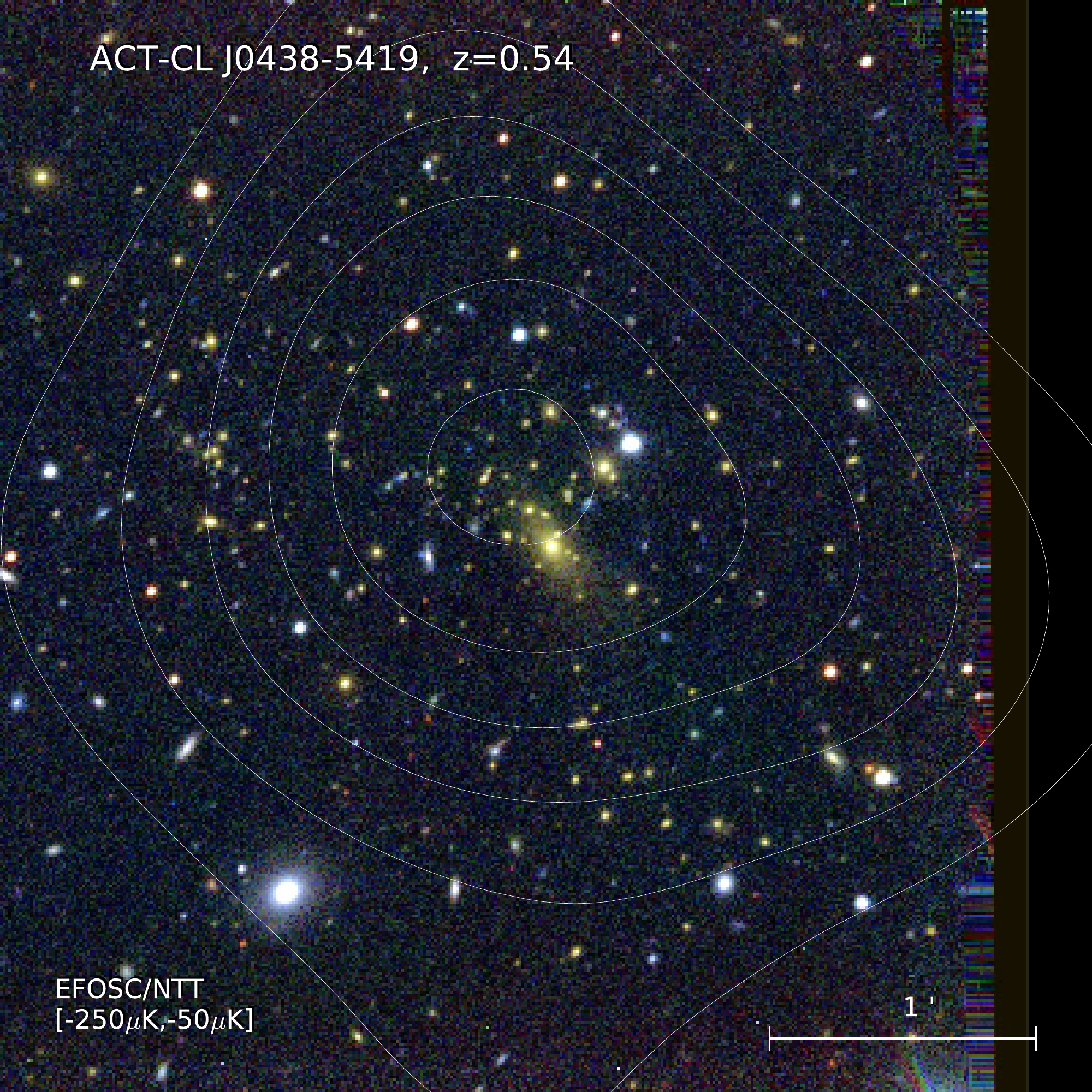}  
}
\caption{Composite color images for ACT-CL~J0509-5341
  (SPT-CL~J0509-5342), ACT-CL~J0304-4921 , ACT-CL~J0215-5212 and
  ACT-CL~J0438-5419. The horizontal bar shows the scale of the images,
  where north is up and east is left. White contours show the 148~GHz
  SZE maps with the minimum and maximum levels, in $\mu$K, displayed
  between brackets.}
\label{fig:CL4}
\end{figure*}

\begin{figure*}
\centerline{
\includegraphics[width=3.5in]{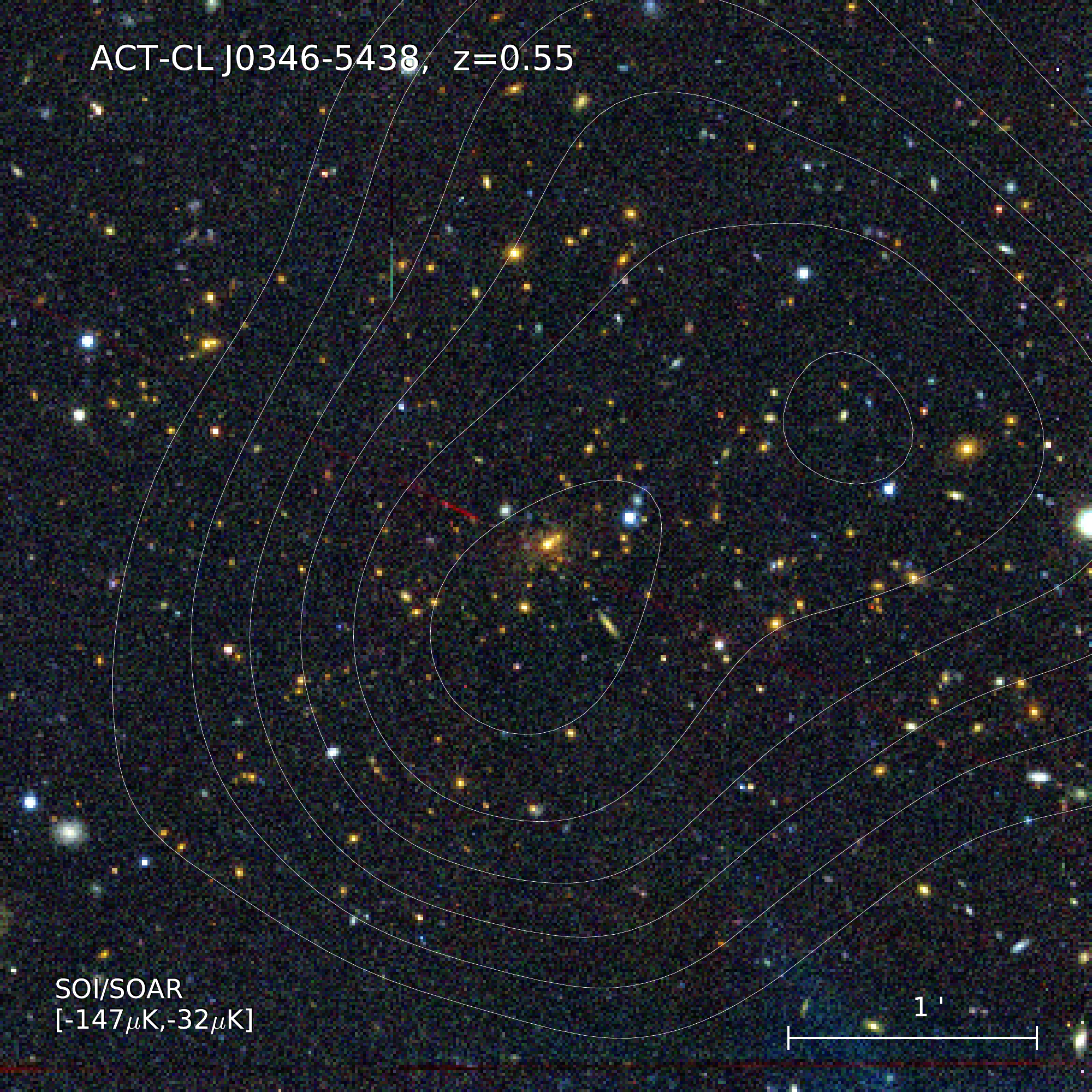}  
\includegraphics[width=3.5in]{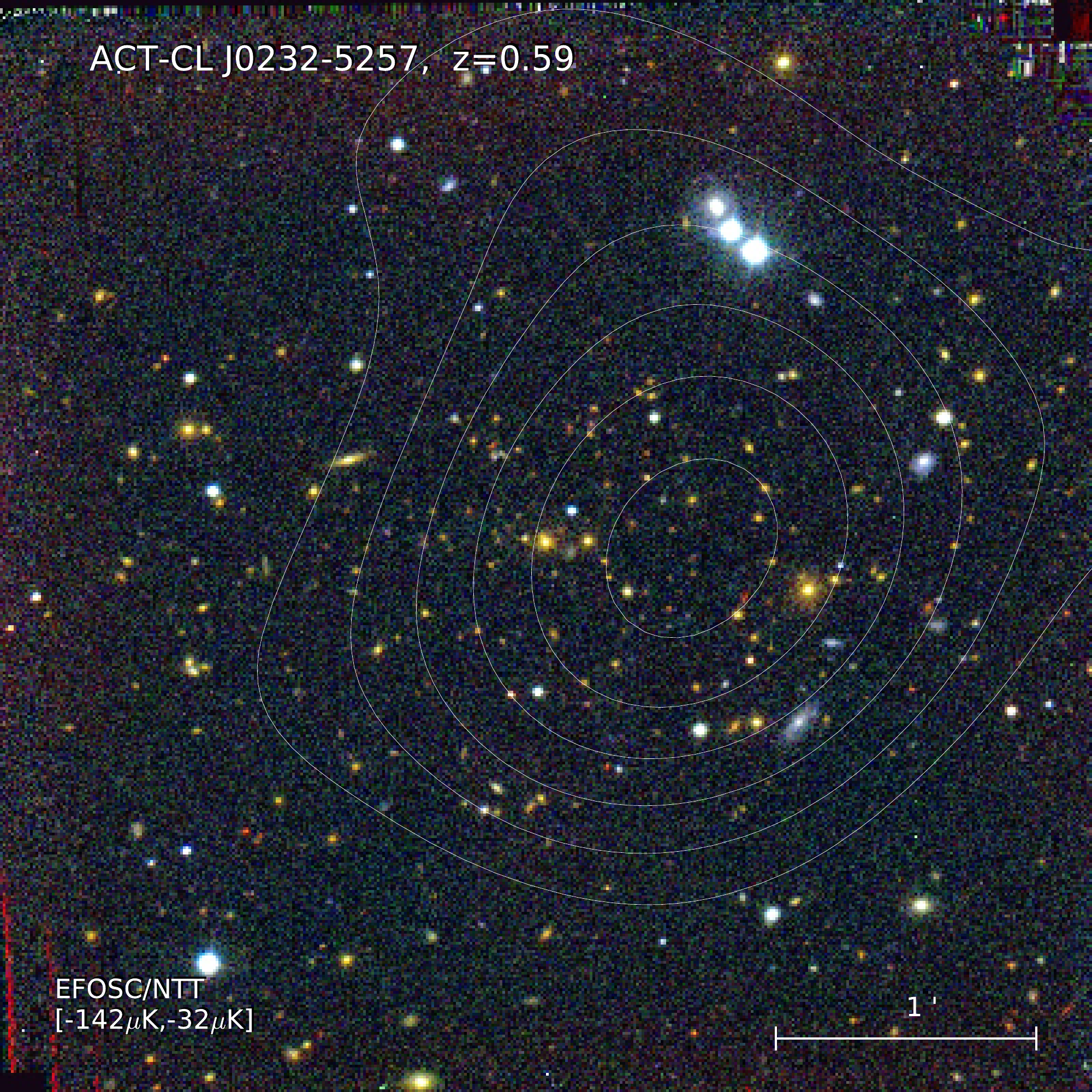}  
}
\centerline{
\includegraphics[width=3.5in]{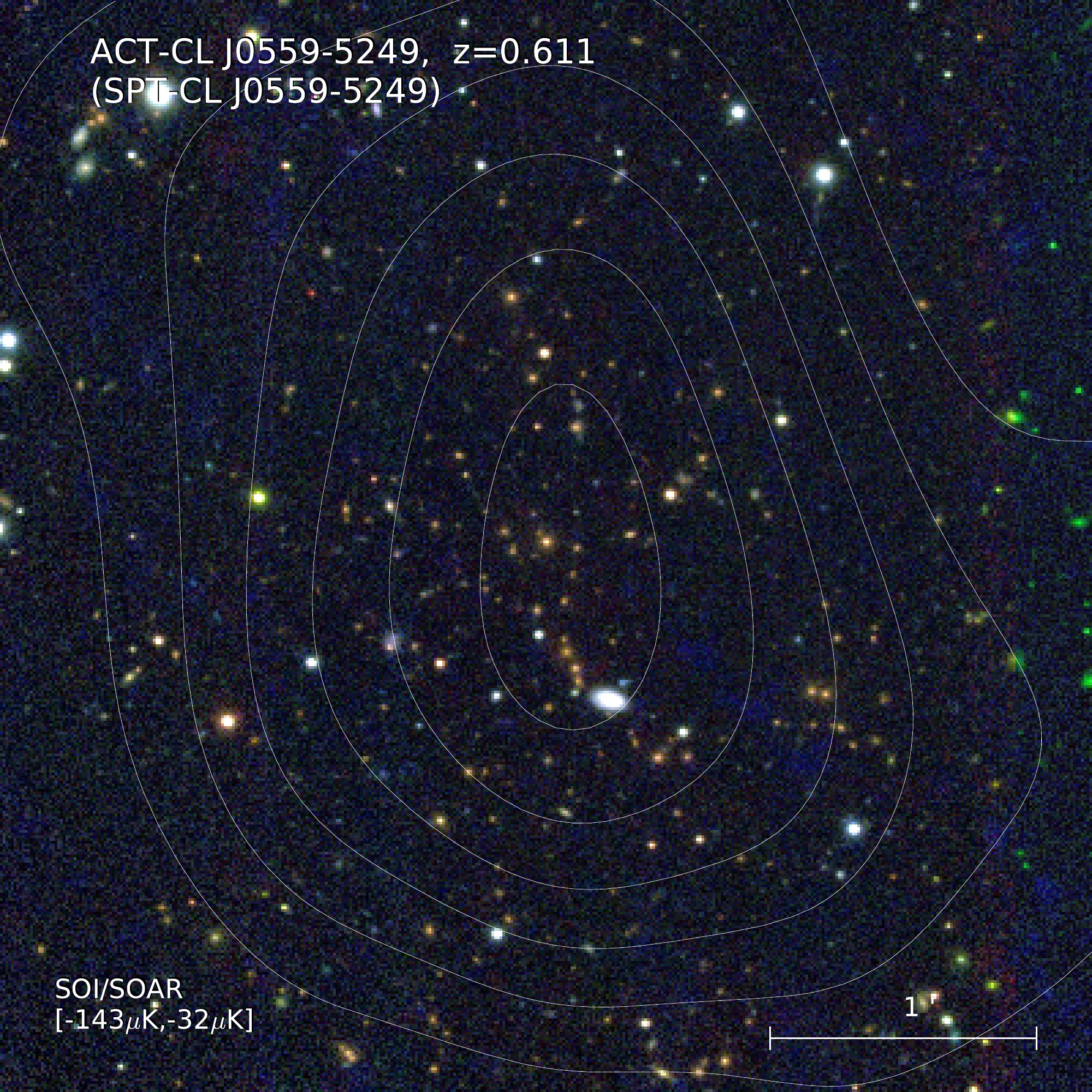}  
\includegraphics[width=3.5in]{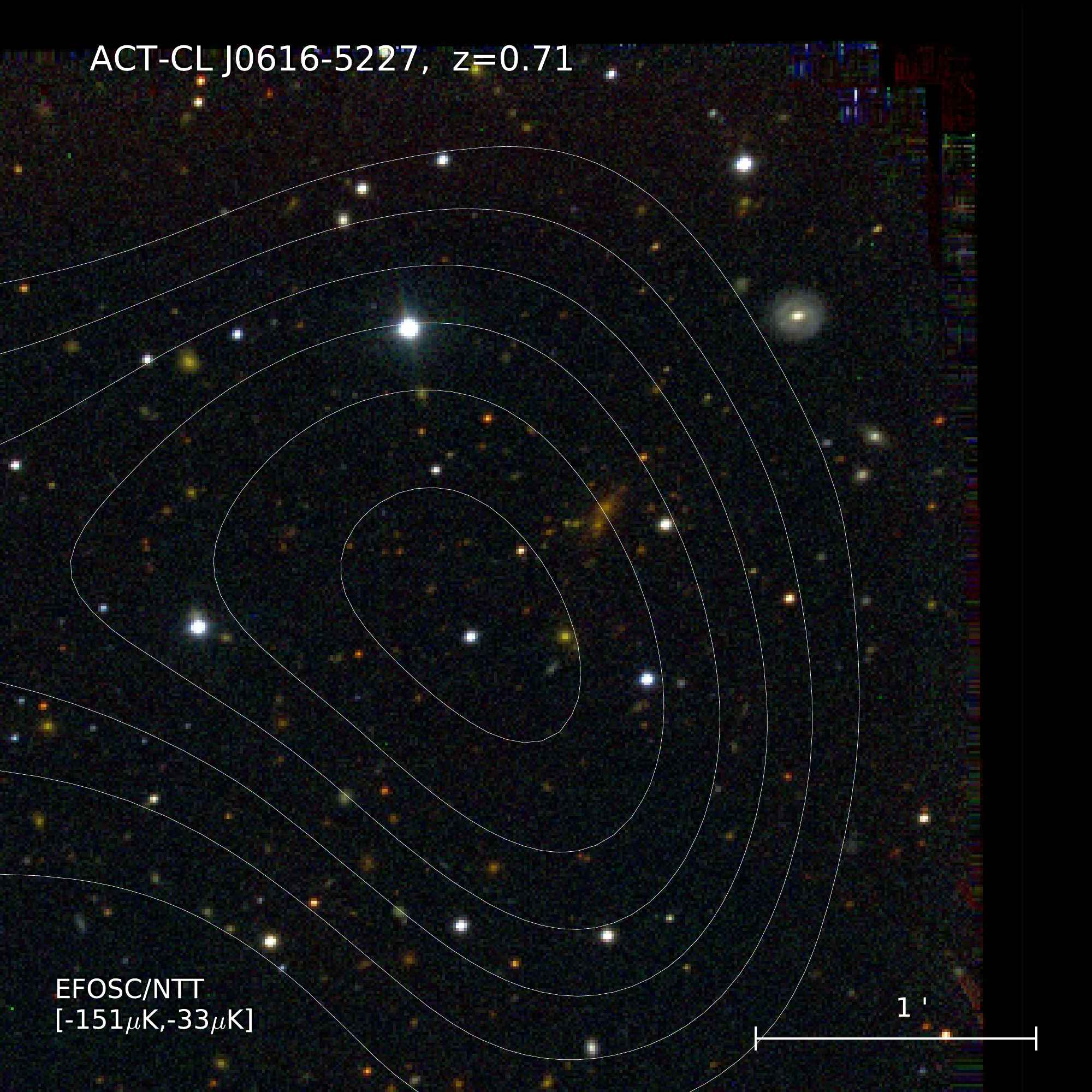}  
}
\caption{Composite color images for ACT-CL~J0346-5438,
  ACT-CL~J0232-5257, ACT-CL~J0559-5249 ( SPT-CL~J0559-5249) and
  ACT-CL~J0616-5227. The horizontal bar shows the scale of the images,
  where north is up and east is left. White contours show the 148~GHz
  SZE maps with the minimum and maximum levels, in $\mu$K, displayed
  between brackets.}
\label{fig:CL4}
\end{figure*}

\begin{figure*}
\centerline{
\includegraphics[width=3.5in]{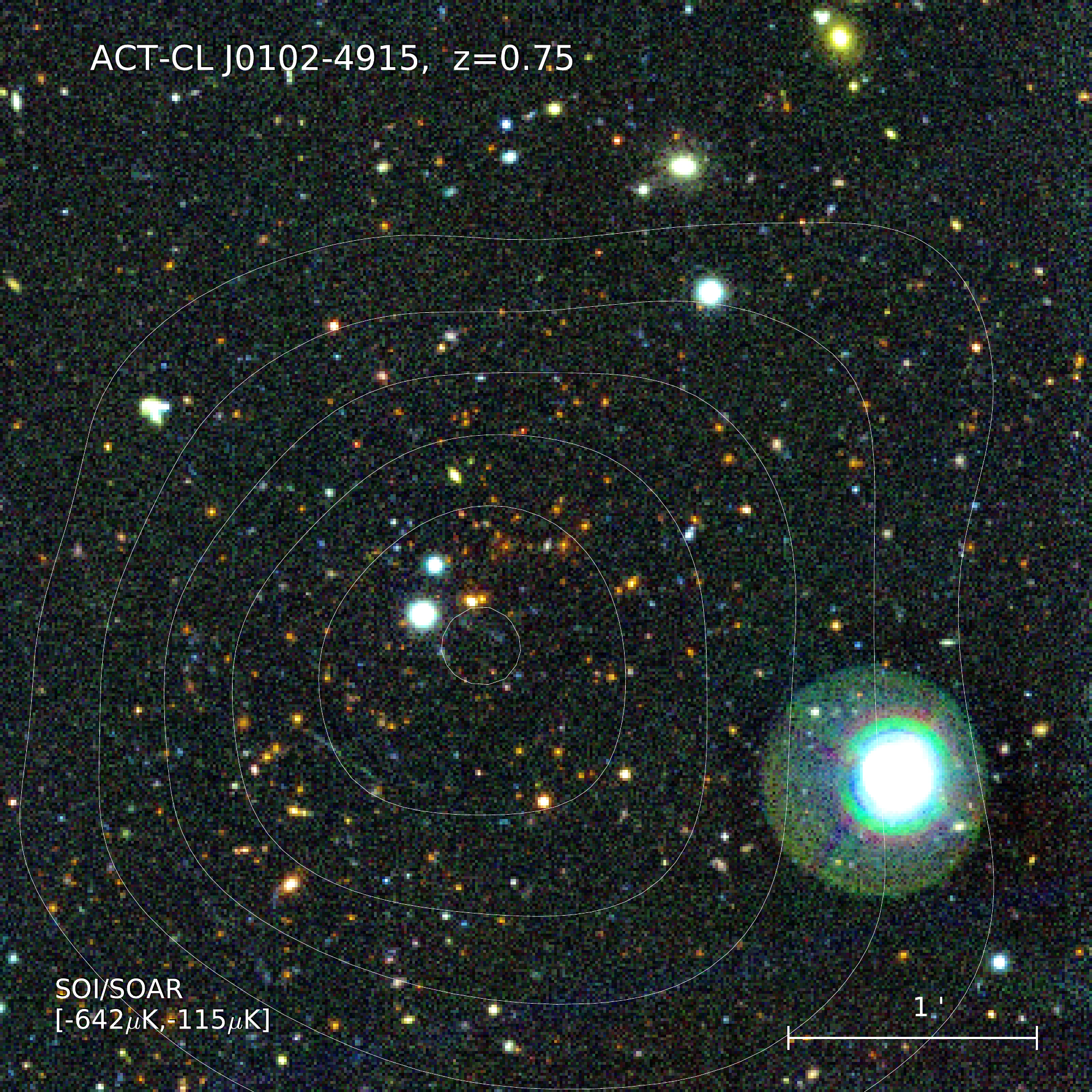}  
\includegraphics[width=3.5in]{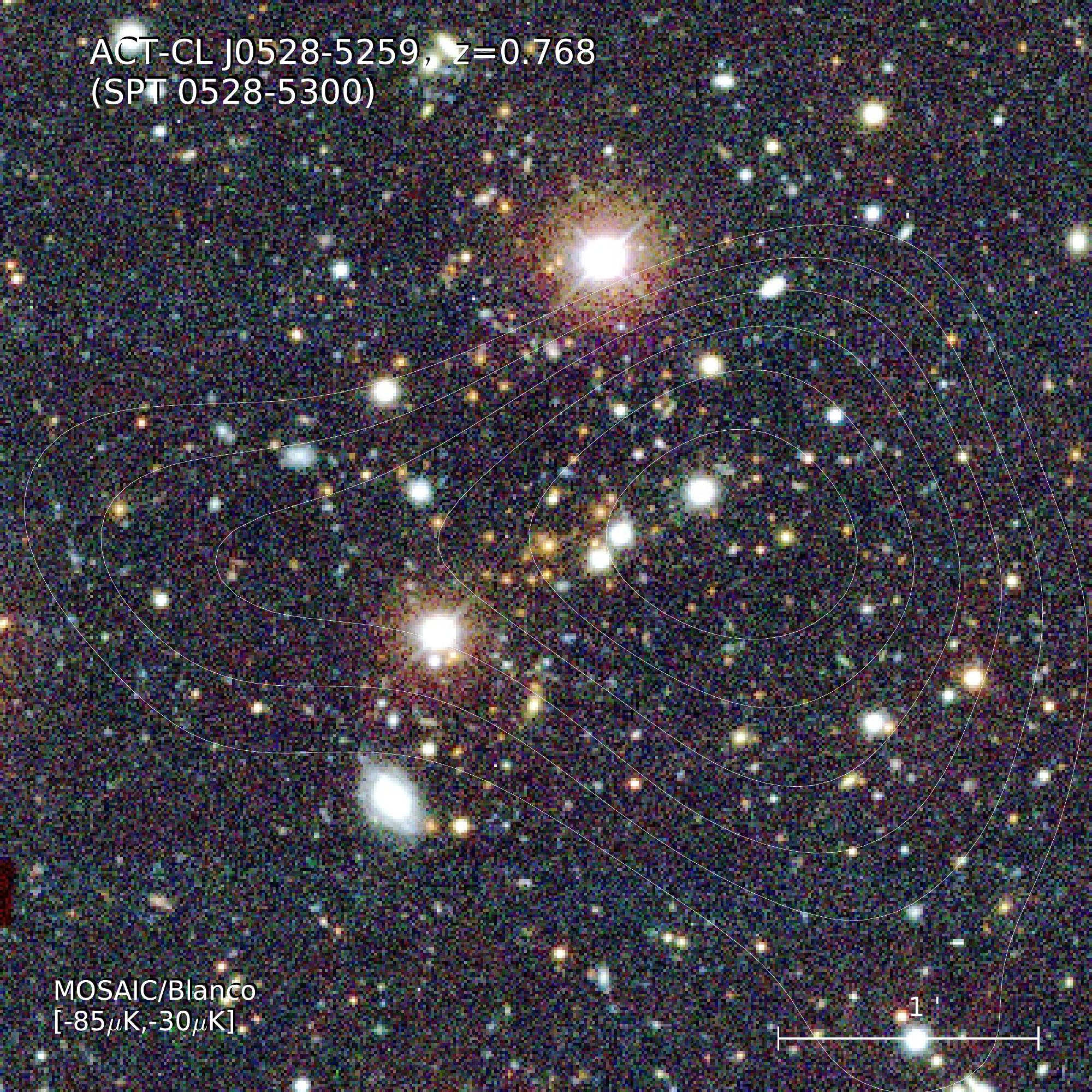}  
}
\caption{Composite color images for ACT-CL~J0102-4915 and
  ACT-CL~J0528-5259 (SPT-CL~J0528-5300). The horizontal bar shows the scale of the
  images, where north is up and east is left.  White contours show the 148~GHz
  SZE maps with the minimum and maximum levels, in $\mu$K, displayed
  between brackets.}
\label{fig:CL6}
\end{figure*}
\begin{figure}
\includegraphics[width=3.5in]{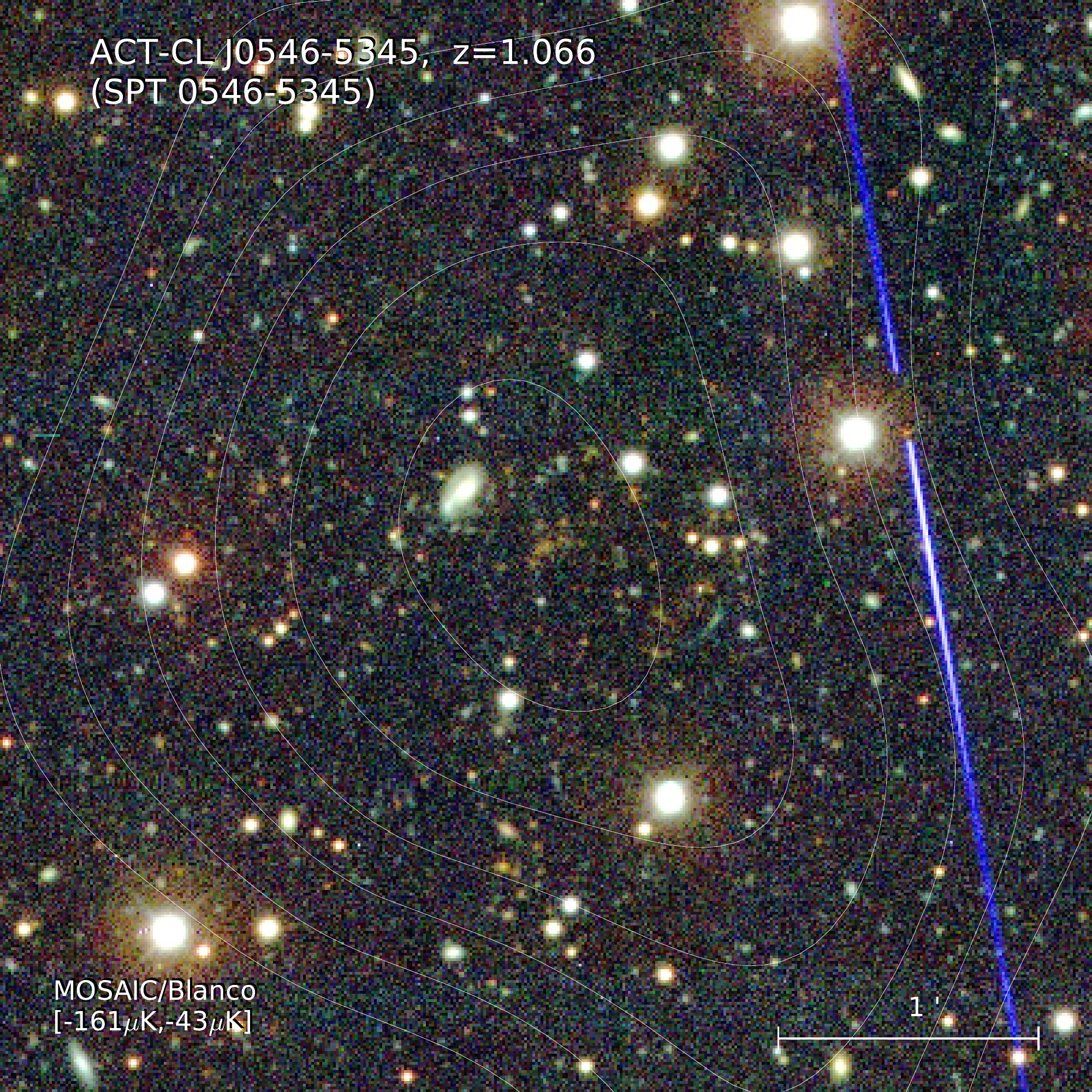}  
\caption{Composite color images for ACT-CL~J0546-5345
  (SPT-CL~J0546-5345). The horizontal bar shows the scale of the
  images, where north is up and east is left. White contours show the
  148~GHz SZE maps with the minimum and maximum levels, in $\mu$K,
  displayed between brackets.}
\label{fig:CL7}
\end{figure}

1E0657$-$56 (the Bullet Cluster) has been extensively observed at
different wavelengths. Here we show the optical observations of the
cluster central region taken with the Advanced Camera for Surveys
aboard the Hubble Space Telescope (HST) (GO:10200, PI:Jones and
GO:10863, PI:Gonzalez) using the F606W ($V$), F814W ($I$) and F850LP
($z$) filters which were taken from the HST
archive\footnote{http://archive.stsci.edu/}.

For RXCJ0217.2$-$5244 we used our own pipeline to process raw $V$
(720s) and $R$ (1200s) images obtained from the ESO
archive\footnote{http://archive.eso.org/} (70.A-0074-A, PI:Edge)
utilizing the SUSI camera on the NTT on Oct 7 and 12, 2002.

A fraction of the area covered by ACT overlaps with the 41~deg$^2$
$griz$ imaging from the CTIO 4-m telescope MOSAIC camera on the 05hr field
from the Southern Cosmology Survey \citep{SCSII}. We used these data to
search for optical identification of SZE cluster candidates in the
region to avoid re-targeting with the NTT and SOAR telescopes.

\section{Results}

\subsection{The ACT SZE Cluster Sample}

Our analysis has resulted in a sample of 23 optically-confirmed SZE
clusters selected from 455 square-degrees of ACT data between
$0.1<z<1.1$. Nine of these clusters, such as 1E0657$-$56 (the Bullet
Cluster), Abell S0592, and Abell 3404 are well known and appear at low
redshift ($z<0.4$); and four systems in our sample have been reported
as SZE-selected clusters by SPT in their area overlapping with the ACT
coverage. Three of these systems (ACT-CL~J0509$-$5341,
ACT-CL~J0528$-$5259, and ACT-CL~J0546$-$5345) were originally reported
as SZE-selected clusters by \cite{Stan09}, with optical and X-ray
properties subsequently studied by us \citep{Menanteau-Hughes-09}, and
ACT~J0559$-$5249 has been recently reported by \cite{High-10}. 
About two-thirds of the clusters we present here (14),
however, are new SZE-discovered systems (4 of which overlap with the SPT
cluster sample of \citealt{High-10}) previously unrecognized in the
optical or X-ray bands; ten of these are entirely new discoveries by ACT, with
photometric redshifts ranging from $z=0.4$ to $z=0.8$. All of the new
clusters are optically quite rich and a number show strong lensing
arcs (see section~\ref{sec:lensing}). 
We present the full list of clusters, positions and redshifts
(photometric in some cases) in Table~\ref{tab:clusters}.
In Figures~\ref{fig:CL1}-\ref{fig:CL7} we show the composite color
optical images of all clusters in our sample with the ACT 148\,GHz
contours (six equally-spaced between the minimum and maximum levels in
$\mu$K shown between brackets) overlaid on the images.

In the following sections we provide detailed information on
individual clusters in the sample in increasing redshift order.

\begin{figure*}
\centerline{\includegraphics[width=2.3in]{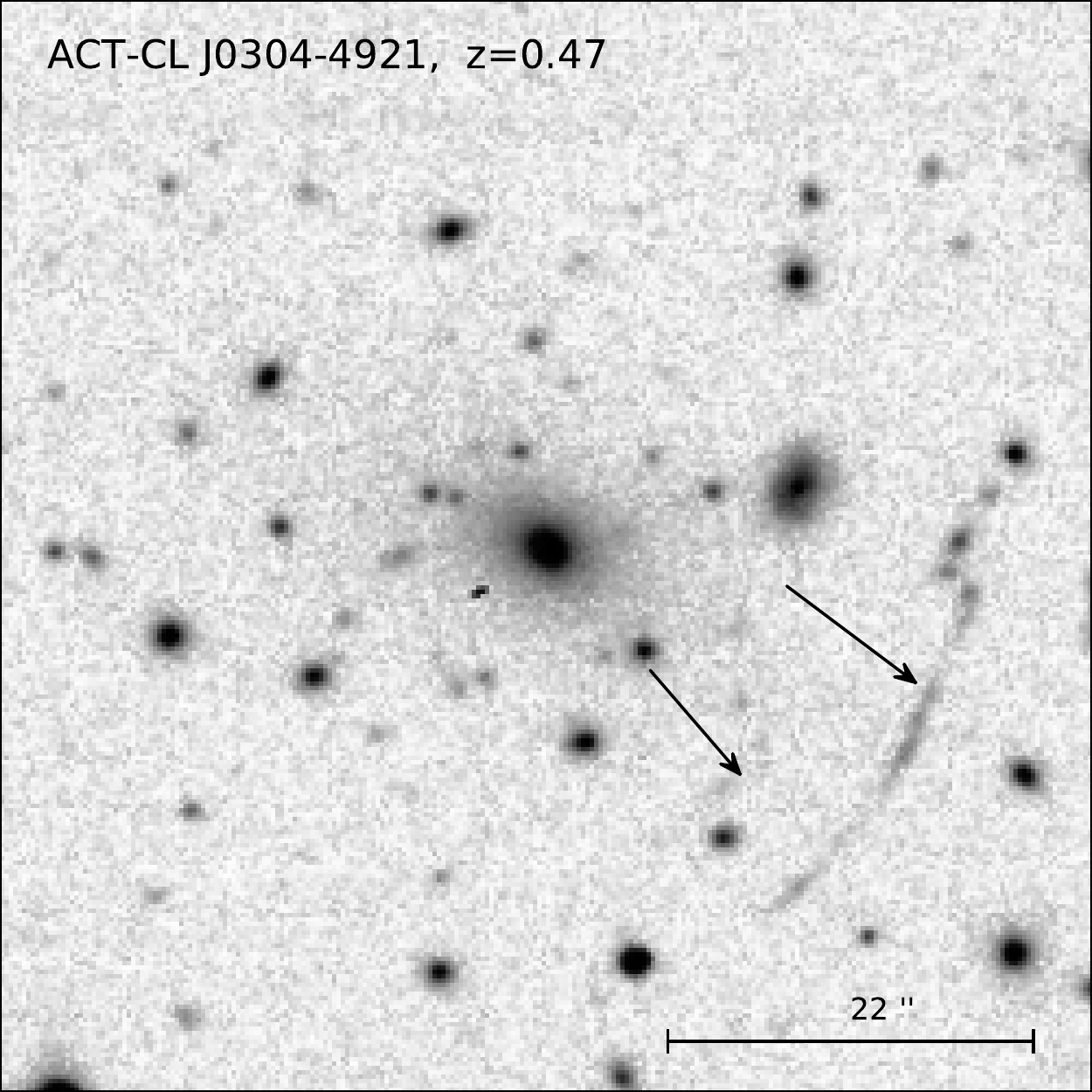}
            \includegraphics[width=2.3in]{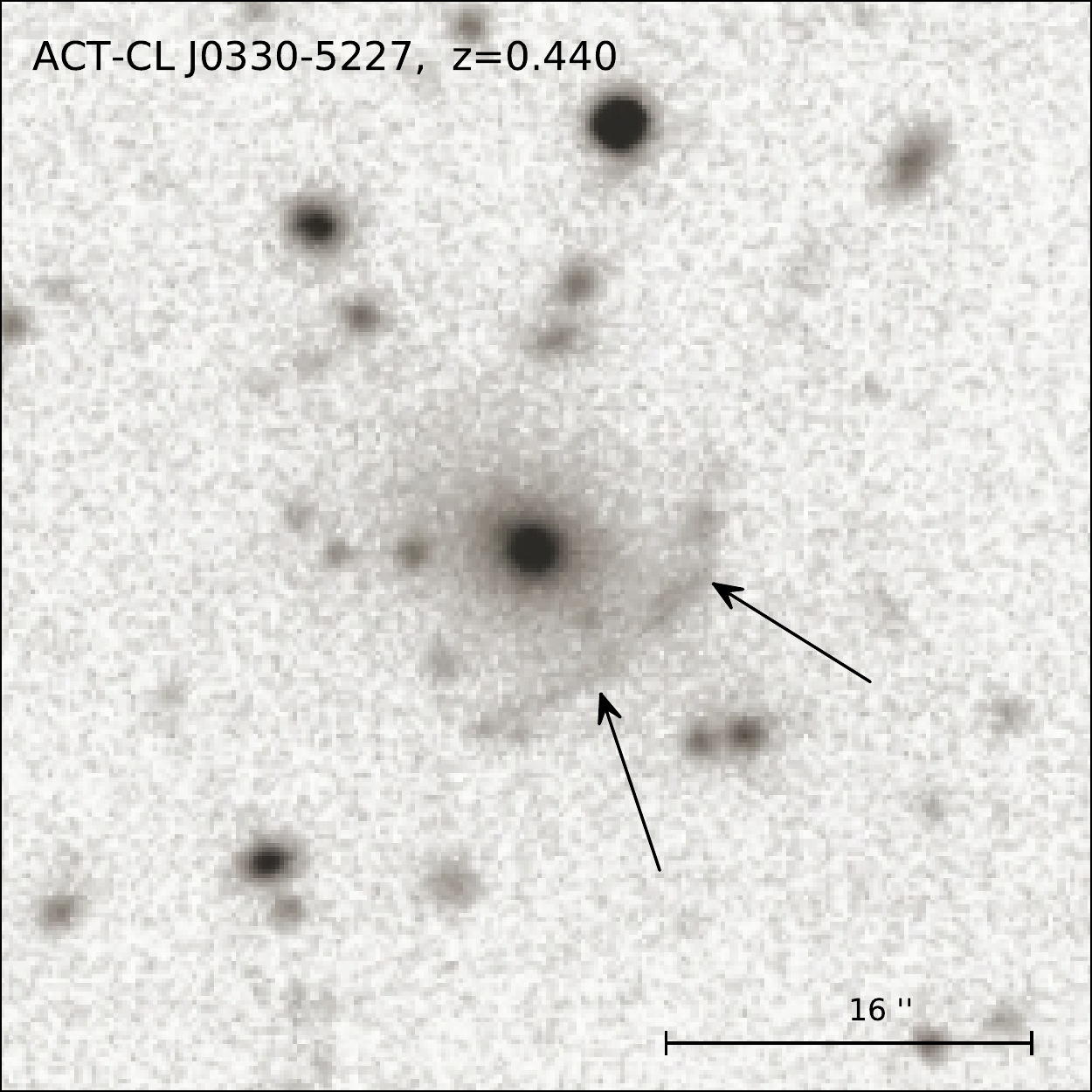}
            \includegraphics[width=2.3in]{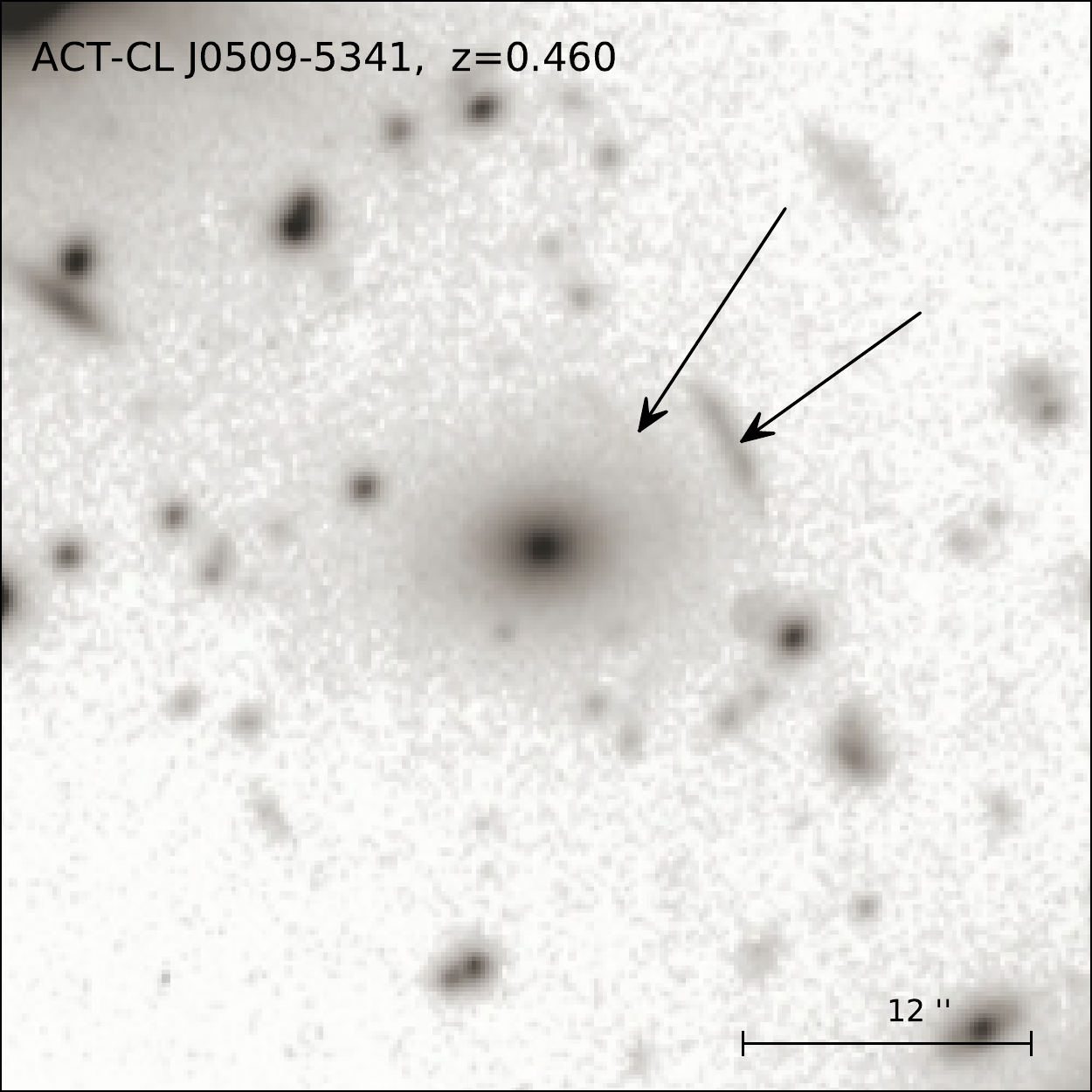}}
\caption{Close-up monochrome images of examples of strong lensing arcs in three
  clusters in the ACT sample.}
\label{fig:arcs}
\end{figure*}

\subsubsection{ACT-CL~J0145-5301 (Abell 2941)}

Abell 2941 is the lowest redshift cluster in our sample and it was
originally reported by \cite{Sersic-1974} as a compact cluster in his
list of peculiar galaxies, groups and clusters based on the Maksutov
plate collection of the Universidad de Chile's Observatorio
Astronomico. It was later included by \cite{ACO-89} as an optically
rich cluster, with R=2 according to the \cite{Abell-1958}
classification.  Abell 2941 has also been identified as a {\em ROSAT} All
Sky Survey (RASS) source \citep{Voges-99} and X-ray cluster,
RXCJ0145$-$5300 at $z=0.1183$ by \cite{deGrandi-99}.

\subsubsection{ACT-CL~J0641-4949 (Abell 3402)}

Abell~3402 is included in the catalog of rich optical clusters of
\cite{ACO-89} with R=0 richness.  We identify its BCG (by examining
the DSS images) as the 2MASS source 2MASS~J06413783$-$4946548
\citep{Skrutskie-02}.  The cluster has not been targeted
spectroscopically, however, the redshift of the BCG ($z=0.146$) has
been reported as part of the 6dF survey \citep{6dF}. It is worth
mentioning that despite its low redshift, Abell 3402 has not been
reported as an X-ray source by any survey yet (including the
RASS). Moreover, the relatively large distance between the SZE
centroid and the cluster BCG ($1.8\arcmin$) may suggest that this is
potentially a spurious association. Note that the BCG falls right at
the edge of the field of view shown in Fig.~4 (upper right panel).
Further spectroscopy and X-ray observations of the cluster will
provide definite information to determine whether this is a bona fide
association with the SZE detection or a case of a ``cluster behind a
cluster'' like ACT-CL~J0330$-$5227 (see section \ref{sec:A3128NE})

\subsubsection{ACT-CL~J0645-5413 (Abell 3404)}

Abell~3404 was first reported by \cite{ACO-89} as an optical cluster
with R=1 richness class and it was also detected as a RASS bright
source \citep{Voges-99}. The cluster redshift, $z=0.1661$, was
reported by \cite{deGrandi-99} and the system is also know by its
REFLEX designation RXCJ0645.4$-$5413 \citep{Bohringer-04}.

\subsubsection{ACT-CL~J0638-5358 (Abell S0592)}

Abell~S0592 was first reported by \cite{ACO-89} in their supplementary
table of southern clusters. These were clusters not rich enough or too
distant to satisfy the criteria to be included in the rich nearby main
cluster catalog. It was also later reported as a \rosat\ bright
source in the RASS \citep{Voges-99} and as the REFLEX X-ray cluster,
RXC~J0638.7$-$5358 \citep{Bohringer-04} at $z=0.2216$
\citep{deGrandi-99}. 
Abell~S0592 was first reported as
an SZE source by ACT (ACT-CL~J0638-5358) in \cite{Hincks-09}.

\subsubsection{ACT-CL~J0516-5430 (Abell S0520)}

Abell~S0520 is an optically-rich cluster reported by \cite{ACO-89} in
their supplementary catalog and it is located at $z=0.294$
\citep{Guzzo-99}. It has been reported as a \rosat\ bright source in
the RASS \citep{Voges-99} and it is known by its REFLEX designation
RXCJ0516.6$-$5430 \citep{Bohringer-04}. This cluster was also detected
as an SZE cluster by SPT \citep{Stan09} and
the results from several mass estimates (optical, X-ray and weak
lensing) suggest a value of $M\simgt5\times10^{14}\,M_{\odot}$
\citep{Zhang-06,Menanteau-Hughes-09,McInnes09}.

\subsubsection{ACT-CL~J0245-5302 (Abell S0295)}

The galaxy cluster Abell S0295 at $z=0.300$ \citep{Edge-94}, was
originally selected optically \citep{ACO-89} and was also detected in
the RASS \citep{Voges-99}. There has been a reported discovery of a
giant strong-lensing arc near the brightest cluster galaxy
\citep{Edge-94}, which we confirm in our imaging (see lower left panel
of Figure~\ref{fig:CL2}). There were also previous (unsuccessful) attempts at
detecting the SZE effect at 1.2-mm and 2-mm using SEST
\citep{Andreani-96}. The ACT detection of this cluster is highly
significant (it is ranked in the top 5 of the sample of 23).
Abell~S0295 was first reported as an ACT SZE source
(ACT-CL~J0638-5358) by \cite{Hincks-09}. {\it ASCA} observations
yielded an average temperature of $kT = 6.72\pm 1.09$~keV
\citep{Fukazawa-04}. Figure~\ref{fig:CL2} shows our deep, multicolor optical
imaging of this cluster from the CTIO 4-m telescope in January 2010.
The distribution of galaxies is highly elongated and matches well the
shape of the SZE contours; both suggest the idea that this is a
merging system.

\subsubsection{ACT-CL~J0217-5245 (RXC~J0217.2-5244)}

RXC~J0217.2$-$5244, at $z=0.343$, was first identified as an X-ray
cluster by the REFLEX survey \citep{Bohringer-04}.  Close comparison
of the SUSI/NTT archival imaging with the ACT 148GHz contours (see
Figure~\ref{fig:CL2}) shows a double-peaked morphology in the SZ maps
that matches the position of the cluster BCG and the second brightest
elliptical galaxy.

\subsubsection{ACT-CL~J0330-5227 (Abell 3128 North-East)}
\label{sec:A3128NE}

Abell 3128 North-East (NE) is a great example of the power of the SZE
to detect clusters regardless of redshift and only limited by
mass. Until \cite{Werner-07}, using \xmm\ data, revealed a more
distant ($z=0.440$) and more massive cluster superposed on the
northeastern component of Abell 3128, it was believed to be part of
the Horologium-Reticulum supercluster at $z=0.06$, although its X-ray
morphology is clearly double peaked with the two components separated
on the sky by some $12^\prime$ \citep{Rose-02}. \cite{Hincks-09}
confirmed the \cite{Werner-07} results that Abell 3128 (NE) is indeed a
separate higher redshift cluster, and a case of finding a cluster
behind a cluster. We do not detect with ACT a significant decrement
from the lower redshift southwestern component.
We show our optical EFOSC/NTT imaging of the cluster in
Figure~\ref{fig:CL3}. There is a strong lensing arc a few arcsecond
southwest of the central brightest cluster galaxy.

\subsubsection{ACT-CL~J0658-5557 (1ES0657$-$558)}

1ES0657$-$558 is the famous ``Bullet'' cluster, which we detect with
the highest significance of any cluster in the ACT sample. Previous
detections of the mm-band SZE signal from this cluster have been
reported by ACBAR \citep{Gomez-04}, APEX-SZ \citep{Halverson-09} and
SPT \citep{Plagge}.

\subsubsection{Overlap with the SPT sample}

Four clusters in our sample overlap with SZE discovered systems
previously reported by SPT. This presents an opportunity to compare
observed clusters properties using different observing strategies and
techniques (albeit sometimes the same archival data).

ACT-CL~J0509$-$5345, ACT-CL~J0547$-$5345 and ACT-CL~J0528$-$5399 were
first reported by \cite{Stan09} and their physical properties
(photometric redshifts, luminosities and mass estimates) reported in
\cite{Menanteau-Hughes-09}. They have been the focus of intense
multi-wavelength observations to establish their redshifts and
estimate their masses using different techniques. Our team has secured
multi-object spectroscopic redshifts for these systems using FORS2 on
the VLT during 2009B (084.A-0577, PI:Infante) as part of a dedicated
program aimed to calibrate the observable--cluster mass scaling
relations of SZE clusters at high redshift (see Infante
et~al. 2010). Two of these systems, J0509$-$5345 ($z=0.461$) and
J0528$-$5399 ($z=0.768$), have spectroscopic redshifts consistent with
those recently reported by \cite{High-10}. For ACT-CL~J0547$-$5345 we
have secured the highest spectroscopic redshift, $z=1.0678$ for an
SZE-selected clusters until now (Infante et al.~2010, \citealt{Brodwin-10}).

The spectroscopic redshift for ACT-CL~J0559$-$5249,
(SPT-CL~J0559$-$5249), $z=0.611$ has been reported by
\citealt{High-10} and provides a consistency check for our photometric
redshift for this system, $z=0.56\pm0.06$.

\subsection{Gravitational lensing features}
\label{sec:lensing}

A number of the clusters in our sample show gravitational lensing
features, indicative of massive systems. The excellent seeing
conditions of the 2009B runs used for this study, enable us to
unambiguously determine that 6 of the 23 clusters show strong lensing
arcs.  Three of these have already been reported in the literature:
ACT-CL~J0658$-$5557 (1E0657-56/Bullet), ACT-CL~J0245$-$5302 (AS0295),
and ACT-CL~J0546$-$5345.  Arcs in the other three are reported here
for the first time: ACT-CL~J0330$-$5227 (Abell 3128-NE),
ACT-CL~J0509-5341, and ACT-CL~J0304$-$4921. In Figure~\ref{fig:arcs}
we show close-up monochromatic images of the clusters with newly
identified strong lensing arcs in the sample.

\subsection{Photometric Redshifts of the Clusters}

\renewcommand{\arraystretch}{1.250}
\begin{deluxetable*}{cccccc}
\tablecaption{Cluster Photometric Redshifts}
\tablehead{
\colhead{ACT Descriptor} & 
\colhead{$z_{\rm cluster}$} & 
\colhead{$z_{\rm BCG}$}    &
\colhead{$z_{\rm spec}$}    &
\colhead{Tel.} &
\colhead{Filters} 
}
\startdata
ACT-CL~J0232-5257  & $0.59 \pm 0.07$  & $0.59^{+0.56}_{-0.63}$ & \nodata & EFOSC/NTT & $g,r,i$\\
ACT-CL~J0304-4921  & $0.47 \pm 0.05$  & $0.42^{+0.41}_{-0.51}$ & \nodata & EFOSC/NTT & $g,r,i$\\    
ACT-CL~J0330-5227  & $0.52 \pm 0.07$  & $0.48^{+0.42}_{-0.59}$ & 0.440   & EFOSC/NTT & $g,r,i$\\    
ACT-CL~J0438-5419  & $0.54 \pm 0.05$  & $0.54^{+0.44}_{-0.56}$ & \nodata & EFOSC/NTT & $g,r,i$\\
ACT-CL~J0616-5227  & $0.71 \pm 0.10$  & $0.71^{+0.67}_{-0.78}$ & \nodata & EFOSC/NTT & $g,r,i$\\
ACT-CL~J0641-4949  & $0.18 \pm 0.07$  & $0.16^{+0.14}_{-0.21}$ & 0.146   & EFOSC/NTT & $g,r,i$\\
ACT-CL~J0645-5413  & $0.19 \pm 0.04$  & $0.19^{+0.16}_{-0.22}$ & 0.167   & EFOSC/NTT & $g,r,i$\\
ACT-CL~J0102-4915  & $0.75 \pm 0.04$  & $0.78^{+0.76}_{-0.81}$ & \nodata & SOI/SOAR  & $g,r,i,z$\\ 
ACT-CL~J0145-5301  & $0.12 \pm 0.06$  & $0.13^{+0.11}_{-0.18}$ & 0.118   & SOI/SOAR  & $g,r,i$\\ 
ACT-CL~J0215-5212  & $0.51 \pm 0.05$  & $0.52^{+0.50}_{-0.55}$ & \nodata & SOI/SOAR  & $g,r,i$\\ 
ACT-CL~J0235-5121  & $0.43 \pm 0.06$  & $0.42^{+0.31}_{-0.49}$ & \nodata & SOI/SOAR  & $g,r,i$\\ 
ACT-CL~J0237-4939  & $0.40 \pm 0.05$  & $0.36^{+0.35}_{-0.43}$ & \nodata & SOI/SOAR  & $g,r,i$\\ 
ACT-CL~J0346-5438  & $0.55 \pm 0.05$  & $0.58^{+0.54}_{-0.64}$ & \nodata & SOI/SOAR  & $g,r,i$\\
ACT-CL~J0559-5249  & $0.56 \pm 0.06$  & $0.56^{+0.55}_{-0.64}$ & 0.611   & SOI/SOAR  & $g,r,i,z$\\
ACT-CL~J0707-5522  & $0.43 \pm 0.07$  & $0.44^{+0.33}_{-0.52}$ & \nodata & SOI/SOAR  & $g,r,i$\\
ACT-CL~J0509-5341  & $0.36 \pm 0.13$  & $0.35^{+0.35}_{-0.50}$ & 0.461   & MOSAIC/Blanco & $g,r,i$\\
ACT-CL~J0516-5430  & $0.25 \pm 0.08$  & $0.26^{+0.26}_{-0.47}$ & 0.294   & MOSAIC/Blanco & $g,r,i$\\
ACT-CL~J0528-5259  & $0.69 \pm 0.05$  & $0.69^{+0.67}_{-0.72}$ & 0.768   & MOSAIC/Blanco & $g,r,i$\\
ACT-CL~J0546-5345  & $0.93 \pm 0.13$  & $0.95^{+0.89}_{-1.04}$ & 1.066   & MOSAIC/Blanco & $g,r,i$\\
\enddata
\label{tab:photo-z}
\end{deluxetable*}
\renewcommand{\arraystretch}{1.}

\begin{figure}
\centerline{\includegraphics[width=3.2in]{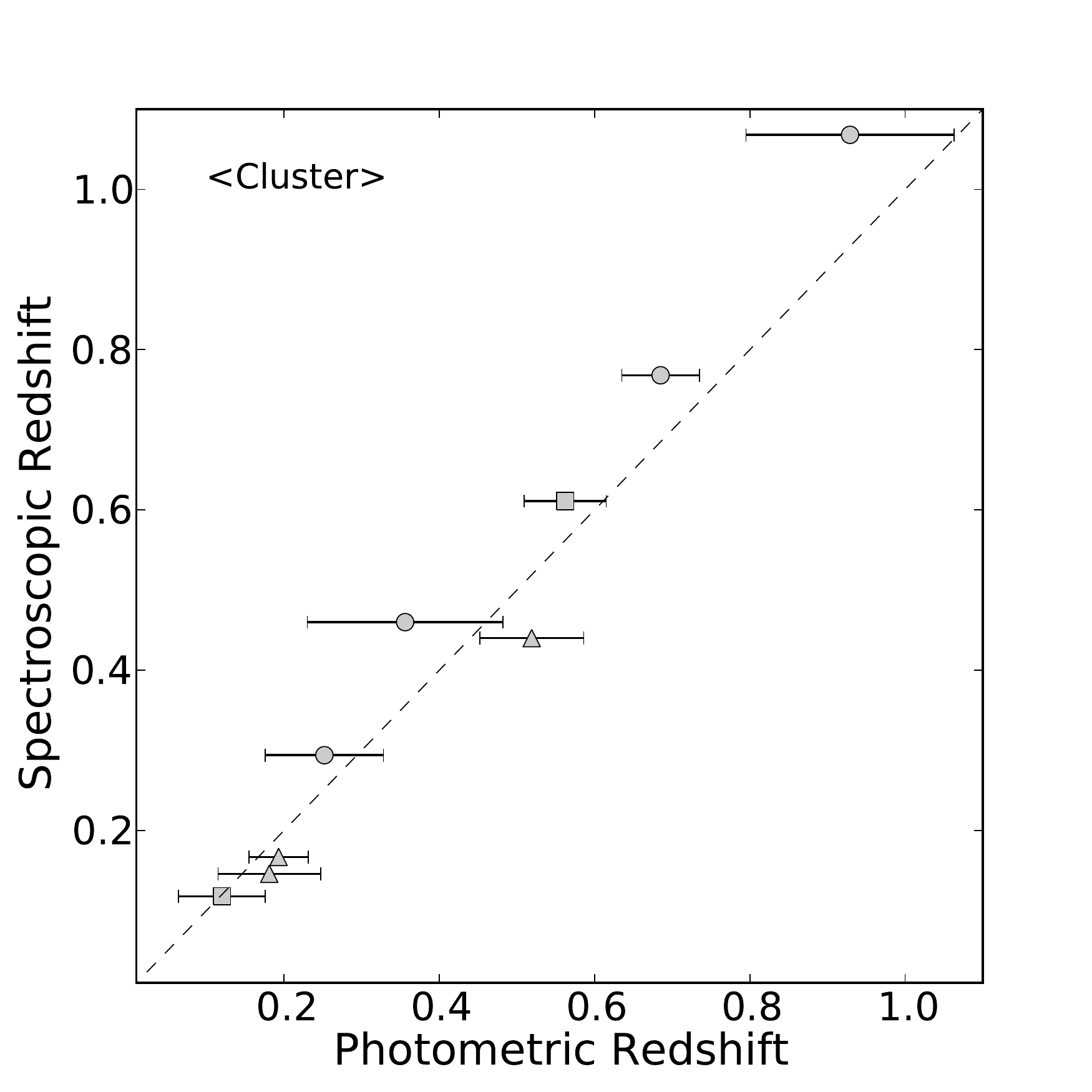}}  
\centerline{\includegraphics[width=3.2in]{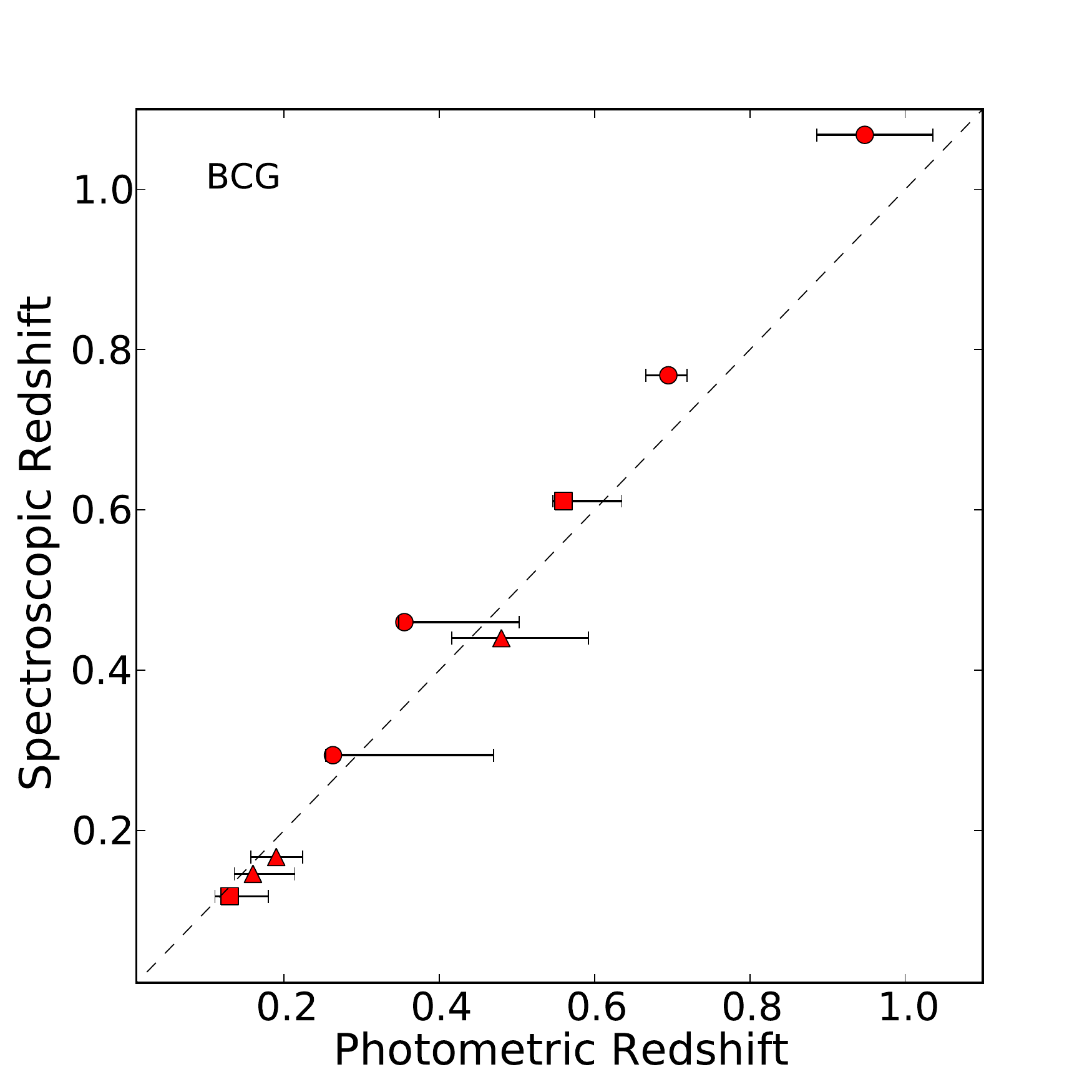}} 
\caption{Spectroscopic redshift versus photometric redshift for the
  sub sample of 8 ACT clusters with known redshifts. Circles,
  triangles and squares represent imaging data from the MOSAIC/Blanco,
  EFOSC/NTT and SOI/SOAR telescopes. Error bars represent 68\% C.L.
  uncertainties on the cluster photometric redshift. }
 \label{fig:photo-z}
\end{figure}

Ten systems in our sample of ACT SZE clusters are completely new
systems, for which the only optical data available are from our NTT
and SOAR imaging collected during 2009B. The main objective of our
observations was to confirm SZE clusters via optical identification as
real systems; and in the case that observations were carried under
photometric conditions --- which was true for both NTT and SOAR
runs --- to compute crude photometric redshift estimates using the
$gri$ (and $z$ in some cases) imaging for the observed
clusters. Therefore we can provide only photometric redshift
estimates for these new systems (see Table~\ref{tab:clusters}).  A
small subsample of the targeted clusters (9) has spectroscopic
information available, which we can use to perform some basic
comparison between the photometric and spectroscopic
redshifts. Unfortunately, the imaging of this subsample is quite
heterogeneous and comes from different datasets (4 from CTIO 4-m, 3
from NTT and 2 from SOAR) making it difficult to characterize the
photometric redshift errors for the sample.
In Figure~\ref{fig:photo-z} we show the comparison between
spectroscopic and photometric redshift for the mean cluster value (top
panel) and the BCG (bottom panel) for the subsample of nine
clusters. We calculate the mean photometric redshift for each cluster,
$z_{\rm cluster}$, with a procedure similat to that used in \cite{SCSII}.
This consists of iteratively selecting cluster member galaxies
using a $3\sigma$ median sigma-clipping algorithm  within a local
color-magnitude relation (using all available colors)
defined by galaxies that:
a) were photometrically classified as E or E/S0s according to their
BPZ-fitted SED type, b) are within a projected radius of $500$~kpc
from the BCG and c) are within the redshift interval $|z-z_{\rm cluster}| = |\Delta
z|=0.05$.
The uncertainty on this redshift
estimate was taken to be the weighted rms of the individual galaxies
chosen as members.  The uncertainties associated with the BCG
photometric redshifts were estimated from the BCG $p_{\rm BPZ}(z)$
function.  These probability functions are non-Gaussian as revealed by
the asymmetric error bars in Figure~\ref{fig:photo-z} (lower panel).

In Table~\ref{tab:photo-z} we provide the mean cluster and BCG
photometric redshift estimates and associated 68\% C.L.\ uncertanties.
Despite the relatively large errors on the photometric redshifts
(driven by the limited number of bands used) they are still useful for
providing cosmological constraints (see Sehgal et al.~2010).

\subsection{False Positive Rate}

\begin{figure}
\centerline{\includegraphics[width=3.9in]{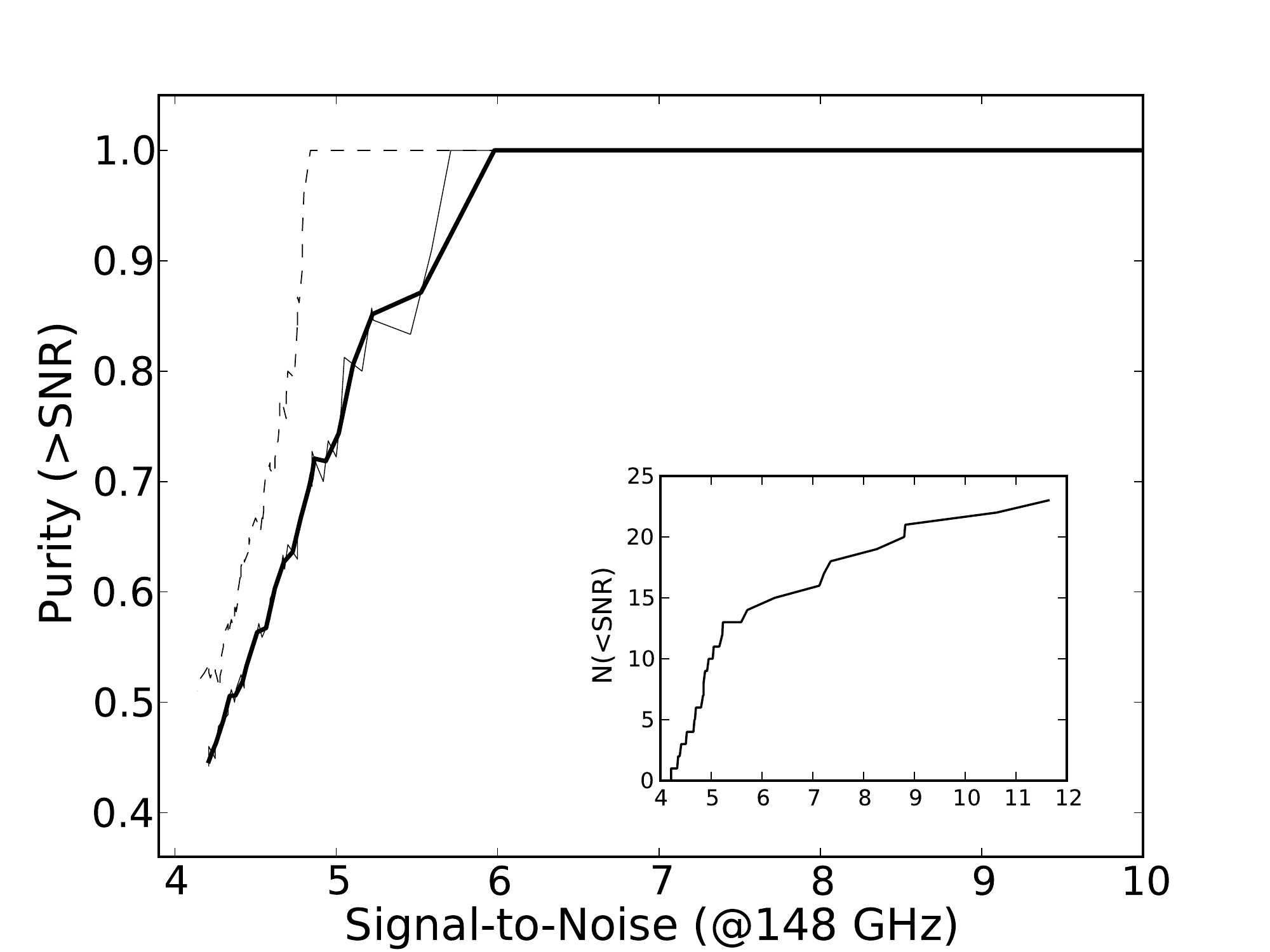}}
\caption{The SZE sample purity greater than a given SNR in the ACT 148
  GHz maps. The purity is defined as the ratio of the number of
  confirmed clusters to the number of observed clusters. The thick
  solid line represents the purity binned in n=2 events, while the
  thin line represents no binning. The dashed line represent the
  fraction of candidates from the sample greater than a given SNR that
  were actually observed. The inset plot shows the cluster cumulative
  distribution as a function of SNR for the optically-confirmed
  cluster sample.}
\label{fig:purity}
\end{figure}

We expect fewer than $0.010$ false positives based on targeting 49
regions of the sky and assuming an expected cluster number density of
0.06~deg$^{-2}$ and a positional offset between the BCG and the SZ
target centroid of $2^\prime$. This assumes a random association with
a cluster at or above the mass threshold of our sample. We would
expect to find one random association for a cluster density of
$5.96$~deg$^{-2}$, which corresponds to a mass threshold of
$1.7\times10^{14}\,M_\odot$ assuming a distribution of massive haloes
using the \cite{Tinker-08} mass function. Abell~3402 might be in this
situation, which would explain the large positional offset between its
BCG and peak SZ decrement and its lack of significant X-ray emission.

\subsection{The Purity of the Sample}
\label{sec:purity}

The extraction methodology of SZE cluster candidates evolved during
the 2009B season and as a result the candidates observed in the first
run (at the NTT) came from a list of targets that was later revised
once the selection method was more stable (see
Section~\ref{sec:sz}). The revised sample for our second run (at SOAR)
in December 2009 contained all of the high signal detections observed
on the NTT run. This later list, from which we targeted 49 systems, is
therefore the most uniformly observed as a function of signal-to-noise
and the one best suited to estimate the purity of SZE candidates.
We utilized the Blanco archival data in the 05hr field to supplement
our NTT and SOAR imaging, although no new clusters were found (beyond
those previously found by SPT). 
By purity here we mean the ratio of optically-confirmed clusters to SZE detections.
In Figure~\ref{fig:purity} we show the purity from the sample of 49
targeted systems as a function of signal-to-noise as the thick and
thin solid line types (these correspond to different event binning).  
Since we did not observe every cluster candidate at signal-to-noise
ratios below 5, for reference we also show, using dashed lines, the
fraction of candidates greater than a given signal-to-noise that were
actually imaged in this program.

We achieve 100\% purity for signal-to-noise ratios greater than 5.7
where there are 10 clusters.  This drops to a purity value of 80\% for
a SNR of 5.1.  We targeted all of the SZ candidates down to a
signal-to-noise of 4.8 and for this sample the purity is 71\% (17/24).
Below this SNR value the purity of the SZ candidates drops sharply.
Only six clusters were optically confirmed out of 25 candidates
targeted (24\% purity) over the SNR range 4.2--4.8.  However, as we
were unable to target all of the SZ candidates in this SNR range,
assigning a precise estimation of the purity is not possible.  Another
estimate for the purity comes from conservatively assuming that none
of the untargeted SZE candidates is a real system, in which case the
overall purity would be $\simeq9\%$ in the signal-to-noise ratio range
between 4.2 and 4.8.

An important conclusion of this analysis is that within our high SNR
sample ($>$5.7), which was fully observed, we report no clusters above
a redshift of 1.1.  As we show next, this distribution is fully
consistent with expectations of $\Lambda$CDM which are based on
Gaussian fluctuations in the primordial matter density distribution.
Deviations from Gaussianity can perturb the expected number of massive
clusters at high redshift.  Thus the large area surveyed by ACT and
the well-established redshift distribution of the sample, can be used
to put important constraints on the level of non-Gaussianity, defined
by the dimensionless coupling parameter, $f_{\rm NL}$
\citep[see][]{KS-01,LV-08}.  This will be the focus of a future study.

\subsection{Mass Threshold of the Cluster Sample}

\begin{figure}
\centerline{\includegraphics[width=3.9in]{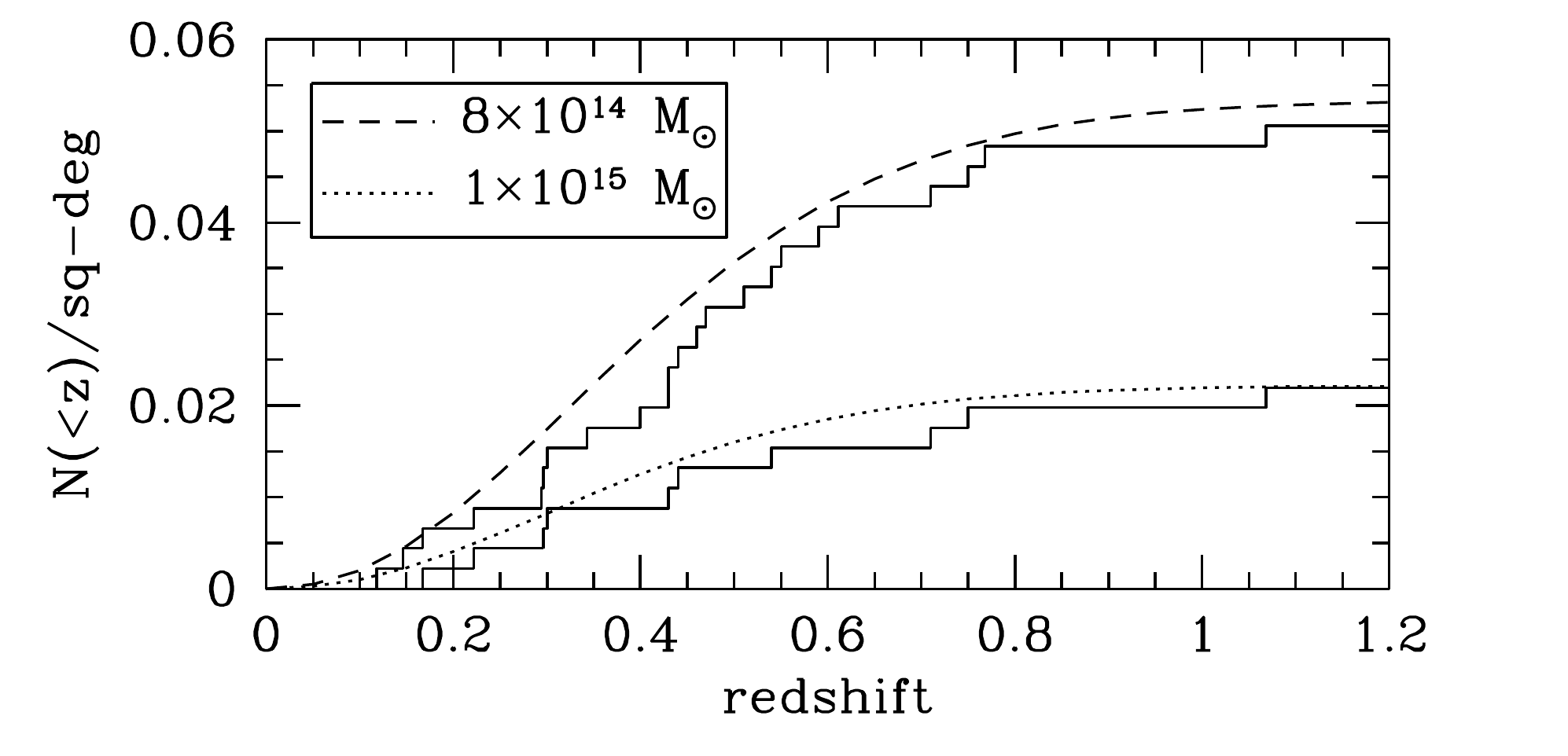}}
\caption{Cumulative distribution of ACT SZE clusters per square degree
  as a function of redshift. The top histogram is for the full sample
  of 23 clusters; the bottom one is for the high significance
  subsample of 10 clusters. The curves show the best fit model
  distributions for each. }
\label{fig:dndz}
\end{figure}

Here we describe how we estimate the approximate mass limit of the
observed sample.  We binned the number of clusters as a function of
redshift in bins of $\Delta z=0.2$ and fitted to the redshift distribution of
massive haloes using the \cite{Tinker-08} mass function and
cosmological parameters from \cite{Komatsu09}, assuming that all 
(no) clusters above (below) the threshold were detected.  The number
counts were assumed to be Poisson-distributed in each bin and the
appropriate likelihood function was calculated and used as the
figure-of-merit function for the fits.
In Figure~\ref{fig:dndz} we show the cumulative number distribution of
optically-confirmed ACT SZ clusters (the histograms) per square degree
as a function of redshift.  We make two assumptions here: a) that the
area surveyed is 455~deg$^2$ even though all of this was not covered
to the same depth, and b) that the sample is complete out to a
redshift of 1.2.  The upper histogram is for the entire sample of
clusters from Table~2, while the bottom histogram is for the subset of
clusters above a SNR of 5.7 which has 100\% purity.  We overplot
curves that give the expected distribution for different mass
thresholds based on structure formation in a $\Lambda$CDM model
\citep{Tinker-08}. The full sample of 23 is well characterized by a
mass threshold of $8\times10^{14}\,M_\odot$ and the 10 highest
significance clusters, where all ACT SZ candidate clusters were
optically confirmed, are consistent with a mass threshold of
$1\times10^{15}\,M_\odot$.  These mass thresholds are in agreement
with their X-ray luminosities (Fig.~\ref{fig:fx-z}) and with
simulations given the current ACT survey noise (Sehgal et al.~2010).
Thus there is consistency between the SZ, optical, and X-ray
observations with the model. Note that if the actual area were
significantly smaller than 455~deg$^2$ or clusters were missed, the
mass thresholds would decrease. Therefore the mass threshold values
quoted here can be considered strong upper limits.

\subsection{Unconfirmed High Signal-to-Noise Candidates}
\label{sec:unconfirmed}

Beyond the reported sample of 23 clusters (Table~\ref{tab:clusters}),
there are in addition three SZE cluster candidates (not reported) at
relatively high SNR ($>$5.1) for which we do not find an optical
counterpart using the procedure described above.  One of these we
observed at the NTT and then again for twice as long at SOAR, where we
also included the $z$-band. Although there are potential BCGs in the
field of view (based on our low accuracy photometric redshifts), none
is accompanied by a red sequence containing more than $\simeq$ 15
galaxies within a projected radius of 500~kpc. This value is
consistent with the average surface density of galaxies at $z\simeq1$
and additionally is far below the typical observed richness of the
confirmed clusters ($\simeq$ 60 members) in the optical. Thus we can
confidently conclude that these three candidates are not optical
clusters in the redshift range $z\simless 1.2$ with optical properties
like those of the confirmed sample (Table~\ref{tab:clusters}).  An
obvious possibility is that these are high redshift clusters, beyond
the reach of our observational program.  However, for the
\citet{Tinker-08} distribution of massive haloes in the $\Lambda$CDM
cosmology, we expect only 0.6 or 0.1 clusters (for $M=6$ and $8 \times
10^{14}\, M_\odot$, respectively) beyond a redshift of 1.2 over the
area surveyed. Future near-IR or warm phase {\it Spitzer} IRAC imaging
is required to determine whether these are high-redshift massive
clusters.

\section{Cluster X-ray Properties}

\begin{figure*}
\centerline{\includegraphics[width=7in]{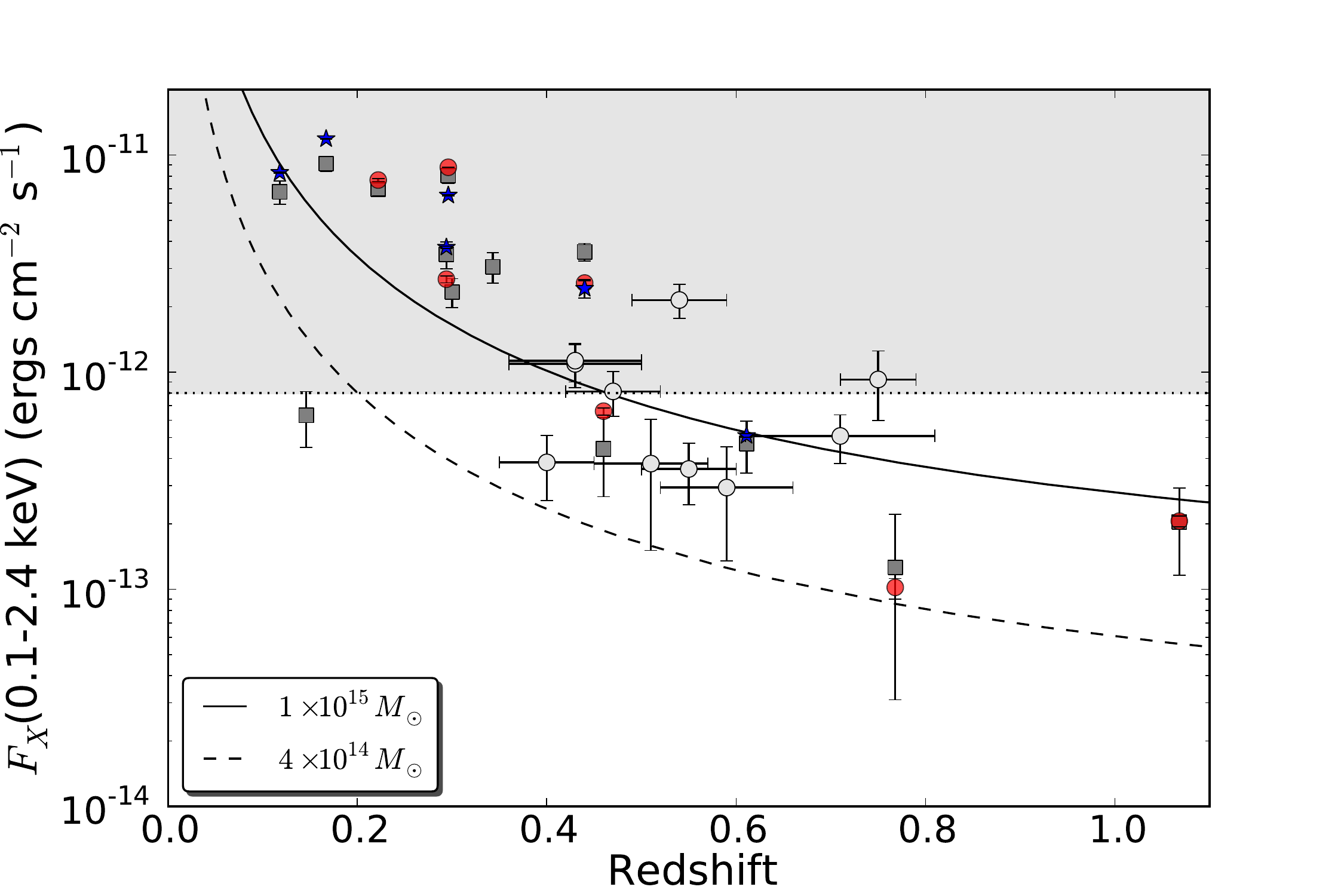}}
\caption{The soft X-ray flux (in the observed band of 0.1--2.4~keV) as a
  function of redshift for ACT SZE-selected clusters. The filled
  squares are known clusters and the open circles are RASS fluxes for
  seven new ACT clusters with only photometric redshifts.  Red circles
  and blue stars represent flux values from \chandra\ and
  \xmm\ observations respectively. The solid and dashed curves
  represent the expected X-ray fluxes for clusters with masses
  $M_{200}=4\times10^{14}\,M_\odot$ and $1\times10^{15}\,M_\odot$,
  respectively, using scaling relations from \cite{Vikhlinin09a}. The
  gray area at the top corresponds approximately to the regime of the
  RASS bright source catalog \citep{Voges-99}.  }
\label{fig:fx-z}
\end{figure*}

In order to estimate masses we turn to X-ray observations, using
archival datasets from \rosat, \chandra\ and \xmm, and measure X-ray
luminosities to serve as coarse mass proxies. Here we describe the
X-ray data reduction and analysis and present the X-ray properties of
the ACT clusters.

\subsection{\rosat\ Observations}

We searched for X-ray counterparts to the ACT clusters using the RASS
data following the same procedure as in \citet{Menanteau-Hughes-09}.
The raw X-ray photon event lists and exposure maps were downloaded
from the MPE {\em ROSAT}
Archive\footnote{ftp://ftp.xray.mpe.mpg.de/rosat/archive/} and were
queried with our own custom software.  At the position of each
cluster, RASS count rates in the $0.5-2$~keV band (corresponding to PI
channels 52--201) were extracted from within radii of 3$^\prime$ to
9$^\prime$ (chosen appropriately for each cluster) for the source
emission and from within a surrounding annulus (12$^\prime$ to
30$^\prime$ inner and outer radii) for the background
emission. Obvious point sources were excluded from both source and
background regions.  The background-subtracted count rates were
converted to X-ray luminosity (in the 0.1--2.4 keV band) assuming a
thermal spectrum ($kT=5$~keV) and the Galactic column density of
neutral hydrogen appropriate to the source position, using data from
the Leiden/Argentine/HI Bonn survey \citep{kalberla05}. In
Table~\ref{tab:RASS} we show the X-ray fluxes and luminosities for
all ACT clusters, regardless of the significance of the RASS
detection.

\subsection{Chandra Observations}

All available archival \chandra\ data for clusters in our sample are
used in this analysis.  The combined exposure times of the observations for
each cluster are shown in Table~\ref{tab:xspec_chandra}.  Data are
reduced using CIAO version 4.2 and calibration data base version
4.2.0.   The data are processed starting with the level 1 events data,
removing the cosmic ray afterglow correction, and generating a new bad
pixel file that accounts for hot pixels and cosmic ray afterglows.
Using the newly generated bad pixel file, the charge transfer
inefficiency correction, time-dependent gain adjustment, and other
standard corrections are applied to the data.  The data are filtered
for \asca\ grades 0, 2, 3, 4, 6 and status=0 events and the good time
interval data provided with the observations are applied.  Periods of
high background count rate are excised using an iterative procedure
involving creating light curves in background regions with 259 s bins
(following the ACIS ``Blank-Sky'' Background File reduction), and
excising time intervals that are in excess of 3 $\sigma$ (=rms) from
the mean background count rate.  This sigma clipping procedure is
iterated until all remaining data lie within 3 $\sigma$ of the mean.
All observations are in the ACIS-I configuration and background
regions are chosen among the chips free from cluster emission, at
approximately the same distance from the readout as the cluster.  This
is to account for a gradient in the front-illuminated chip response
seen when investigating deep blank-sky exposures
\citep{bonamente2004}.

Images are created by binning the data by a factor of four, resulting
in a pixel size of 1.97\arcsec.  Exposure maps are constructed for
each observation at an energy of 1 keV.  Images and exposure maps are
combined in cases where there are multiple observations of the same
cluster.  A wavelet-based source detector is used on the combined
images and exposure maps to generate a list of potential
point sources.  The list is examined by eye, removing bogus or suspect
detections, and then used as the basis for our point source mask.

Spectra are extracted in regions that are 3$\sigma$ above the
background level, where $\sigma$ is the uncertainty in the background
measurement.  Surface brightness profiles are constructed using
5$\arcsec$ bins and are used to estimate the background surface
brightness and the uncertainty in the background.  The radius
corresponding to 3$\sigma$ above the background, $\theta_{3\sigma}$,
is computed and used for the spectral extraction, accounting for the
point source mask.  Spectra are extracted and responses calculated for
each individual observation.  All of the spectra for a given cluster
are then fit simultaneously.

ACIS quiescent backgrounds are used to construct background spectra
for each of the observations.  Appropriate quiescent background files are
reprojected to match each observation and background spectra are
extracted in the same $\theta_{3\sigma}$ regions used for the cluster
spectra. The background spectra are then normalized to match the
count rate of the cluster spectra in the 9.5-12.0~keV range, where the
\chandra\ response is negligable and the particle background
dominates \citep[e.g.,][]{vikhlinin2005}.

XSPEC \citep{arnaud1996} is used to fit the X-ray spectrum assuming a
thermal plasma emission model \citep[using the so-called ``Mekal''
  model;][]{mewe1985}, multiplied by a Galactic absorption model, with
solar abundances of \citet{asplund2009} and cross sections of
\citet{balucinska1992} including an updated He cross section
\citep{yan1998}.  For the absorption component we fixed the column
density of neutral hydrogen to the Galactic value toward the cluster
position (obtained in the same way as for the RASS data).  Spectra are
grouped such that there are a minimum of 20 counts per bin and the
$\chi^2$ statistic is used for the fit.  All spectra for a given
cluster are fit simultaneously to the same plasma model, including
linking the normalizations between data sets.  The fit is limited to
photons within the energy range 0.3-9.0 keV.  To estimate fluxes in
the 0.1-2.4 keV \rosat\ band, the energy range is extended below
\chandra's nominal energy range.  Fluxes, $F_X$, and luminosities,
$L_X$, in the 0.1-2.4 keV band are then computed from the best fit
model.  Uncertainties in the fluxes and luminosities are estimated by
varying the temperature, $T_e$, abundance, $Z$, and normalization of
the model using the extremities of the 68\% confidence regions for the
parameters. The results are summarized in
Table~\ref{tab:xspec_chandra}, which shows the redshift ($z$), total
effective exposure time ($t_{exp}$), column density ($N_H$), the
spectral extraction radius in angular ($\theta_{3\sigma}$) and
physical ($R_{3\sigma}$) units, spectral parameters from the fit
($T_e$ and $Z$), and the resulting flux and luminosity ($F_X$ and
$L_X$) for each cluster.

\begin{deluxetable*}{clcrrrrr}
\tablecaption{\rosat\ Cluster Properties}
\tablewidth{0pt}
\tablehead{
  \colhead{} & 
  \colhead{} & 
  \colhead{$t_{exp}$} &
  \colhead{$N_H$} &
  \colhead{$\theta$} &
  \colhead{$R$} &
  \colhead{$F_X$\tablenotemark{a}} &
  \colhead{$L_X$\tablenotemark{b}}
  \\
  \colhead{ACT Descriptor} & 
  \colhead{$z$} & 
  \colhead{(s)} & 
  \colhead{($10^{20}$ cm$^{-2}$)} &
  \colhead{(arcmin)} &
  \colhead{($h_{70}^{-1}$ kpc)} &
  \colhead{(0.1$-$2.4 keV)} &
  \colhead{(0.1$-$2.4 keV)}
}
\startdata
ACT-CL-J0145-5301 &   0.118  &   247 &   2.67 & 9.0  & 1151  & $ 67.59\pm8.39$ & $  2.30\pm0.29$ \\ 
ACT-CL-J0641-4949 &   0.146  &   859 &   3.37 & 6.0  &  920  & $  6.32\pm1.82$ & $  0.34\pm0.10$ \\ 
ACT-CL-J0645-5413 &   0.167  &   496 &   5.60 & 9.0  & 1542  & $ 91.17\pm7.02$ & $  6.48\pm0.50$ \\ 
ACT-CL-J0638-5358 &   0.222  &   834 &   6.06 & 8.0  & 1717  & $ 69.57\pm4.74$ & $  9.13\pm0.62$ \\ 
ACT-CL-J0516-5430 &   0.294  &   380 &   2.05 & 7.0  & 1844  & $ 34.83\pm4.90$ & $  8.43\pm1.19$ \\ 
ACT-CL-J0658-5557 &   0.296  &   470 &   4.90 & 7.0  & 1853  & $ 80.26\pm6.53$ & $ 19.72\pm1.61$ \\ 
ACT-CL-J0245-5302 &   0.300  &   621 &   3.15 & 7.0  & 1870  & $ 23.39\pm3.54$ & $  5.92\pm0.90$ \\ 
ACT-CL-J0217-5245 &   0.343  &   357 &   2.68 & 7.0  & 2047  & $ 30.56\pm4.86$ & $ 10.40\pm1.65$ \\ 
ACT-CL-J0237-4939 &   0.40   &   778 &   2.55 & 3.5  & 1128  & $  3.84\pm1.28$ & $  1.84\pm0.61$ \\ 
ACT-CL-J0235-5121 &   0.43   &   460 &   3.03 & 3.5  & 1178  & $ 10.91\pm2.44$ & $  6.14\pm1.37$ \\ 
ACT-CL-J0707-5522 &   0.43   &   620 &   6.37 & 3.5  & 1178  & $ 11.29\pm2.28$ & $  6.36\pm1.28$ \\ 
ACT-CL-J0330-5227 &   0.440  &   746 &   1.44 & 6.0  & 2046  & $ 35.77\pm3.35$ & $ 21.20\pm1.98$ \\ 
ACT-CL-J0509-5341 &   0.461  &   418 &   1.46 & 3.5  & 1224  & $  4.43\pm1.76$ & $  2.90\pm1.15$ \\ 
ACT-CL-J0304-4921 &   0.47   &   699 &   1.77 & 5.0  & 1770  & $  8.15\pm1.89$ & $  5.61\pm1.30$ \\ 
ACT-CL-J0215-5212 &   0.51   &   369 &   2.49 & 5.0  & 1850  & $  3.79\pm2.28$ & $  3.13\pm1.88$ \\ 
ACT-CL-J0438-5419 &   0.54   &   308 &   1.02 & 3.5  & 1334  & $ 21.48\pm3.83$ & $ 20.21\pm3.60$ \\ 
ACT-CL-J0346-5438 &   0.55   &   774 &   1.44 & 3.0  & 1154  & $  3.58\pm1.13$ & $  3.52\pm1.11$ \\ 
ACT-CL-J0232-5257 &   0.59   &   425 &   2.84 & 3.5  & 1393  & $  2.94\pm1.59$ & $  3.39\pm1.83$ \\ 
ACT-CL-J0559-5249 &   0.611  &   995 &   5.06 & 4.0  & 1618  & $  4.69\pm1.27$ & $  5.85\pm1.58$ \\ 
ACT-CL-J0616-5227 &   0.71   &  1223 &   4.24 & 4.0  & 1724  & $  5.08\pm1.28$ & $  8.92\pm2.24$ \\ 
ACT-CL-J0102-4915 &   0.75   &   227 &   1.72 & 4.0  & 1761  & $  9.25\pm3.27$ & $ 18.39\pm6.49$ \\ 
ACT-CL-J0528-5259 &   0.768  &   718 &   3.24 & 3.0  & 1332  & $  1.26\pm0.95$ & $  2.64\pm2.00$ \\ 
ACT-CL-J0546-5345 &   1.066  &  1092 &   6.78 & 3.0  & 1462  & $  2.04\pm0.88$ & $  9.14\pm3.94$ 
\enddata
\tablenotetext{a}{Units are $\times 10^{-13}$ erg s$^{-1}$ cm$^{-2}$.}
\tablenotetext{b}{Units are $\times 10^{44}$ erg s$^{-1}$.}
\label{tab:RASS}
\end{deluxetable*}

\subsection{XMM-Newton Observations}

For the {\em XMM-Newton} observations we started with the
pipeline-processed data supplied as part of the standard processing.
A combined background-subtracted and exposure-corrected image was
created for each cluster in the 0.5--2 keV band by merging data from
the three EPIC cameras: PN, MOS1 and MOS2. These were used to define
the optimal radii for spectral extraction, initially taken as the
radius where the estimated cluster emission fell below the background
level. These source regions were further refined by increasing the
extraction radii by factors of 10\% until the extracted flux values
converged to an accuracy of a few percent.  The regions, in nearly all
cases, were elliptical.  Background spectra came from a surrounding
annular region with sufficient area to obtain good photon statistics.
Event lists were filtered to remove time periods of high count rate
background flares and to include only good event patterns ($\leq$ 12
for MOS and $\leq$ 4 for the PN).  Other standard filters to remove
bad pixels and chip edges, for example, were applied.  We used
standard {\em XMM-Newton} software tools to calculate the instrumental
response functions for each cluster (treating the data from each
camera independently). The cluster image itself was used for the
weighting to generate the response functions.  The same absorbed
thermal plasma model as described above was used to fit the PN, MOS1,
and MOS2 data, again using XSPEC. The individual spectral data sets
were fitted separately with linked parameters. X-ray fluxes and
luminosities were determined using the best-fitted spectrum.  The
uncertainties on these include the 1$\sigma$ range on temperature and
overall normalization.  Results are given in Table~\ref{tab:xmm}.

\subsection{Summary of X-ray Properties}

The soft X-ray fluxes as a function of redshift are shown in
Figure~\ref{fig:fx-z}. We plot all datasets: RASS in gray,
\chandra\ in red, and \xmm\ in blue.  We indicate the region (at the
high-end flux) that approximately corresponds to the flux limit of the
RASS bright source catalog \citep{Voges-99}. In cases of multiple
X-ray datasets the agreement among the observed fluxes is good, with
the possible exception of ACT-CL-J0330-5227, where the emission from
the foreground cluster contaminates the RASS flux estimate.

For reference we show curves of the expected (observed-frame) X-ray
fluxes for clusters with assumed masses of
$M_{200}=4\times10^{14}M_\odot$ (dashed) and $1\times10^{15}M_\odot$
(solid) using the X-ray luminosity versus mass scaling relation in
\cite{Vikhlinin09a}.  We use their eqn.~22, which includes an
empirically determined redshift evolution.  We convert their X-ray
band (emitted: 0.5--2 keV band) to ours (observed: 0.1--2.4 keV)
assuming a thermal spectrum at the estimated cluster temperature
determined using the mass-temperature relation also from
\cite{Vikhlinin09a}.  This too has a redshift dependence, so the
estimated temperatures vary in the ranges 2.9--5.0 keV and 5.2--9.0
keV over $0.0<z<1.1$ for the two mass values we plot.  We use
conversion factors assuming the redshift-averaged temperatures of 4
keV and 7 keV, since the difference in conversion factor over the
temperature ranges is only a few percent, negligible on the scale of
Figure~\ref{fig:fx-z}.  The mass values in \cite{Vikhlinin09a} are
defined with respect to an overdensity of 500 times the {\it critical}
density of the Universe at the cluster redshift, while in this paper
we define cluster masses with respect to an overdensity of 200 times
the {\it average} density at the cluster redshift.  We calculate the
conversion factor assuming an NFW halo mass profile with the power-law
relation between concentration and mass determined by \cite{Duffy08}
from large cosmological $N$-body simulations.  This mass conversion
factor is approximately 1.8 averaged over redshift, varies from 2.2 to
1.7 over $0.0<z<1.1$, and depends only weakly on cluster mass (few
percent) for the 2 values plotted here.

The X-ray fluxes of the ACT SZE-selected clusters scatter about the
$1\times10^{15}\,M_\odot$ curve, validating the mass threshold we
estimated above from the cluster number counts.  The most extreme
outlier (on the low flux side) is Abell 3402, indicating a
considerably lower mass for this system, assuming (as we have) that
the SZE candidate is at the redshift of Abell 3402.  For those
clusters with archival \chandra\ or \xmm\ data our measured
temperatures are typically high $kT >5$ keV and in good agreement with
results from the recent SPT cluster X-ray analysis \citep{SPT-x}.  The
masses inferred from the X-ray temperatures are consistent with the
mass curves shown in Figure~\ref{fig:fx-z}.

\begin{deluxetable*}{lccccccccc}
\tablewidth{0pt}
\tablecolumns{10}
\tablecaption{\chandra\ Cluster Properties \label{tab:xspec_chandra}}
\tablehead{
  \colhead{} 
  & \colhead{} 
  & \colhead{$t_{exp}$} 
  & \colhead{$N_H$}
  & \colhead{$\theta_{3\sigma}$}
  & \colhead{$R_{3\sigma}$}
  & \colhead{$T_e$} 
  & \colhead{$Z$}
  & \colhead{$F_X$\tablenotemark{a}} 
  & \colhead{$L_X$\tablenotemark{b}}
\\
  \colhead{ACT Descriptor} 
  & \colhead{$z$} 
  & \colhead{(ks)}
  & \colhead{($10^{20}$ cm$^{-2}$)}
  & \colhead{(arcmin)} 
  & \colhead{($h_{70}^{-1}$ kpc)} 
  & \colhead{(keV)} 
  & \colhead{($Z_\odot$)} 
  & \colhead{(0.1$-$2.4 keV)} 
  & \colhead{(0.1$-$2.4 keV)}
}
\startdata
ACT-CL J0638-5358 &  $0.222$ & $\phn 19.9$    & $6.06$ & $5.20$ & $1116$      
                  & $\phn 8.51 ^{+0.36} _{-0.34}$ 
                  & $0.46 ^{+0.09} _{-0.09}$ 
                  & $76.64      ^{+1.02}  _{-1.02}$ 
                  & $\phn 9.87  ^{+0.13}  _{-0.13}$ \\
ACT-CL J0516-5430 &  $0.294$ & $\phn\phn 9.5$ & $2.05$ & $4.05$ & $1068$ 
                  & $\phn 9.59 ^{+1.35} _{-1.05}$ 
                  & $0.49 ^{+0.25} _{-0.24} $                  
                  & $26.79      ^{+1.00}  _{-0.93}$ 
                  & $\phn 6.30  ^{+0.23}  _{-0.22}$ \\
ACT-CL J0658-5557 &  $0.296$ & $589.4$        & $4.90$ & $5.92$ & $1568$      
                  & $    13.56 ^{+0.14} _{-0.14}$ 
                  & $0.36 ^{+0.02} _{-0.02}$ 
                  & $87.75       ^{+0.22}  _{-0.22}$ 
                  & $20.69      ^{+0.05}  _{-0.05}$ \\
ACT-CL J0330-5227 &  $0.440$ & $\phn 19.3$    & $1.44$ & $3.60$ & $1229$      
                  & $\phn 4.32 ^{+0.21} _{-0.19}$ 
                  & $0.11       ^{+0.07}  _{-0.07}$ 
                  & $25.72      ^{+0.62}  _{-0.60}$ 
                  & $15.50      ^{+0.37}  _{-0.36}$ \\
ACT-CL J0509-5341 &  $0.460$ & $\phn 28.8$    & $1.46$ & $1.95$ & $\phn 683$      
                  & $\phn 9.39 ^{+1.43} _{-1.07}$ 
                  & $0.35 ^{+0.22} _{-0.23}$ 
                  & $\phn 6.61  ^{+0.26}  _{-0.24}$ 
                  & $\phn 4.14  ^{+0.16}  _{-0.15}$ \\
ACT-CL J0528-5259 &  $0.768$ & $\phn 27.2$    & $3.24$ & $1.04$ & $\phn 463$  
                  & $\phn 4.67 ^{+1.45} _{-0.99}$ 
                  & $0.30 $                  
                  & $\phn 1.02  ^{+0.12}  _{-0.10}$ 
                  & $\phn 2.15  ^{+0.24}  _{-0.22}$ \\
ACT-CL J0546-5345 &  $1.068$ & $\phn 57.1$    & $6.78$ & $1.24$ & $\phn 605$  
                  & $\phn 8.54 ^{+1.38} _{-1.05}$ 
                  & $0.33 ^{+0.29} _{-0.29}$ 
                  & $\phn 2.06  ^{+0.12}  _{-0.11}$ 
                  & $\phn 8.43  ^{+0.47}  _{-0.44}$ \\
\enddata
\tablenotetext{a}{Units are $\times 10^{-13}$ erg s$^{-1}$ cm$^{-2}$.}
\tablenotetext{b}{Units are $\times 10^{44}$ erg s$^{-1}$.}
\end{deluxetable*}

\begin{deluxetable*}{lcccccccccrrrr}
\tablewidth{0pt}
\tablecolumns{12}
\tablecaption{\xmm\ Cluster Properties}
\tablehead{
  \colhead{} & 
  \colhead{} & 
  \colhead{$t_{exp}^{\rm MOS}$,$t_{exp}^{\rm PN}$\tablenotemark{a}} &
  \colhead{$N_H$} &
  \colhead{$\theta_{\rm minor}$,$\theta_{\rm major}$} &
  \colhead{$R_{mean}$\tablenotemark{b}} &
  \colhead{$T_e$} & 
  \colhead{$Z$} &
  \colhead{$F_X$\tablenotemark{c}} &
  \colhead{$L_X$\tablenotemark{d}}
  \\
  \colhead{ACT Descriptor} & 
  \colhead{$z$} & 
  \colhead{(ks)} & 
  \colhead{($10^{20}$ cm$^{-2}$)} &
  \colhead{(arcmin)} &
  \colhead{($h_{70}^{-1}$ kpc)} &
  \colhead{(keV)} & \colhead{($Z_\odot$)} &
  \colhead{(0.1$-$2.4 keV)} &
  \colhead{(0.1$-$2.4 keV)}
}
\startdata
ACT-CL~J0145-5301 & 0.118 & 37,22 & 2.67 & 6.5,10.8 & 1031 & $ 5.60\pm0.08$ & $0.30\pm0.02$ & $ 82.96\pm0.52$ & $ 2.83\pm0.02$ \\
ACT-CL~J0645-5413 & 0.167 & 44,28 & 5.60 & 5.4,8.9  & 1188 & $ 7.49\pm0.09$ & $0.24\pm0.01$ & $118.70\pm0.40$ & $ 8.36\pm0.03$ \\
ACT-CL~J0516-5430 & 0.294 & 11,8  & 2.05 & 6.2,8.0  & 1856 & $ 7.44\pm0.38$ & $0.21\pm0.05$ & $ 37.57\pm0.47$ & $ 8.93\pm0.11$ \\
ACT-CL~J0658-5557 & 0.296 & 33,21 & 4.90 & 3.9,5.1  & 1181 & $10.80\pm0.22$ & $0.22\pm0.03$ & $ 65.22\pm0.39$ & $15.51\pm0.09$ \\
ACT-CL~J0330-5227 & 0.440 & 71,57 & 1.44 & 4.0,4.0  & 1365 & $ 5.46\pm0.27$ & 0.30          & $ 24.28\pm2.34$ & $14.60\pm1.40$ \\
ACT-CL~J0559-5249 & 0.611 & 20,14 & 5.06 & 2.0,3.5  & 1070 & $ 8.09\pm0.75$ & 0.30          & $  5.10\pm0.10$ & $ 6.06\pm0.12$ \\
\enddata
\tablenotetext{a}{The quoted MOS exposure time is the average of the two MOS cameras.}
\tablenotetext{b}{Geometric mean of the major and minor axes.}
\tablenotetext{c}{Units are $\times 10^{-13}$ erg s$^{-1}$ cm$^{-2}$.}
\tablenotetext{d}{Units are $\times 10^{44}$ erg s$^{-1}$.}
\label{tab:xmm}
\end{deluxetable*}

\section{Conclusions}

After many years of intense effort, we have presented the first
optically-confirmed sample of 23 SZE-selected clusters of galaxies
from ACT over a cosmologically significant volume. Using a combination
of archival and targeted multi-color observations using 4-m telescopes
we have confirmed the presence of a BCG and a rich red sequence of
galaxies in 14 new SZE clusters (4 overlapping with previous ACT and
SPT detections), of which 10 are presented here for the first
time. Using a careful selection method of SZE candidates combined with
the results from our observations we have determined the purity of the
ACT SZE sample to be 100\% for SNR~$>5.7$, 80\% for SNR~$>5.1$, and
71\% for SNR~$>4.8$.  For lower SNR values we did not target all SZE
candidates; still it is clear that the purity of the lower
significance candidates has fallen substantially to between
$\simeq$9\% and 24\%.

By comparing the number distribution of clusters of galaxies as a
function of redshift with the expectations from $\Lambda$CDM
simulation, we estimate a typical lower mass limit of
$8.0\times10^{14}\,M_{\odot}$ for our full sample of 23.  Our highest
significance subsample of 10 ACT SZE cluster candidates (above a SNR
of 5.7) were all confirmed optically is consistent with a mass
threshold of $1.0\times10^{15}\,M_{\odot}$. We present the results of
analysis of archival X-ray data from \rosat, \chandra, and \xmm.
The X-ray data yield uniformly high
temperatures ($kT \simeq 7 - 17$ keV) and luminosities ($L_X \simeq
10^{44} - 10^{45}$ erg s$^{-1}$), which are broadly consistent with
the mass thresholds inferred from the cluster number counts as a
function of redshift.  Further work on measuring cluster masses
using X-ray data and other techniques (e.g., galaxy dynamics and weak
lensing) is underway.

While the $\Lambda$CDM model predicts that there should be very few
clusters with masses greater than $8 \times 10^{14}\, M_\odot$ at $z >
1.2$, (i.e., 0.1 clusters with our current area and mass limit or
0.4\% of the sample) early dark energy models
\citep{Bartelmann-06,Alam-10} and models with significant primordial
non-Gaussianity \citep{Cayon-10} predict significantly more high
redshift clusters. Moreover, in quintessence models, the number of
high redshift clusters is also a sensitive function of $w$, the dark
energy equation of state \citep{Alimi-10}. Therefore, the detection of
even a single massive high redshift cluster would challenge our
current standard paradigm.  However our ability to investigate this
question here is constrained somewhat by the upper redshift limit set
by the depth of our optical program.  Future follow-up IR observations
of high significance SZE candidates that are currently undetected in
our optical imaging can probe the existence of $z > 1.2$ massive
systems and provide a simple yet powerful test of $\Lambda$CDM.  Even
in the event of a null result (i.e., none of the high significance SZE
decrements turns out to contain a cluster), the ACT cluster sample
has the potential to set strong constraints on alternative dark energy
models.

\acknowledgments

The observations on which this paper were based represent the marriage
of two different communities (CMB and optical) in multiple countries
working for a common goal. In particular, the optical observations
were coordinated and led by Felipe Barrientos and Leopoldo Infante
(Pontificia Universidad Cat\'olica de Chile) and John P.~Hughes and
Felipe Menanteau (Rutgers University).
This work was supported by the U.S. National Science Foundation
through awards AST-0408698 for the ACT project, and PHY-0355328,
AST-0707731 and PIRE-0507768 (award number OISE-0530095). The PIRE
program made possible exchanges between Chile, South Africa, Spain and
the US that enabled this research program. 
Funding was also provided by Princeton
University and the University of Pennsylvania. 
We also acknowledge support from NASA/XMM grants NNX08AX55G and
NNX08AX72G to Rutgers University.
Computations were performed on the GPC supercomputer at the SciNet HPC
Consortium. SciNet is funded by: the Canada Foundation for Innovation
under the auspices of Compute Canada; the Government of Ontario;
Ontario Research Fund – Research Excellence; and the University of
Toronto.
This research is partially funded by ``Centro de Astrof\'{\i}sica FONDAP''
15010003, Centro BASAL-CATA and by FONDECYT under proyecto 1085286.
The observers (FM, JPH, JG, LI) would like to thank the La Silla, CTIO
and SOAR staff for their support during the runs.
The SOAR Telescope is a joint project of: Conselho Nacional de
Pesquisas Cientificas e Tecnologicas CNPq-Brazil, The University of
North Carolina at Chapel Hill, Michigan State University, and the
National Optical Astronomy Observatory.

\end{document}